\documentclass{dmathesis}[12pt]

\usepackage{graphicx}
\usepackage{amssymb}
\usepackage{multicol}
\usepackage{url}
\usepackage{latexsym}
\usepackage{amsmath}
\usepackage{booktabs}
\usepackage{relsize}
\usepackage{epsfig}
\usepackage{ccaption}
\usepackage{enumerate}
\usepackage{setspace}
\usepackage{xcolor}
\usepackage[percent]{overpic}

\newcommand{\be}{\begin{equation}}
\newcommand{\ee}{\end{equation}}
\newcommand{\bea}{\begin{eqnarray}}
\newcommand{\eea}{\end{eqnarray}}
\newcommand{\nn}{\nonumber\\}

\def\d{d}
\def\ln{\mathrm{ln}}
\def\ds{\displaystyle}
\def\pd{\partial}
\def\Area{\mathrm{Area}}
\def\za{z_\ast}

\def\tr{\mathrm{tr}}

\def\m{\mu}
\def\n{\nu}

\def\a{\alpha}
\def\b{\beta}

\def\t{\tau}
\def\Om{\Omega}
\def\th{\theta}
\def\p{\phi}
\def\D{\Delta}

\def\L{\mathcal{L}}

\def\e{\varepsilon}
\def\G{\Gamma}
\def\g{\gamma}
\def\la{\lambda}
\def\La{\Lambda}
\def\s{\sigma}

\def\A{\mathcal{A}}
\def\B{\mathcal{B}}
\def\H{\mathcal{H}}
\def\O{\mathcal{O}}
\def\N{\mathcal{N}}
\def\M{\mathcal{M}}
\def\R{\mathbb{R}}
\def\Z{\mathbb{Z}}

\def\thus{\Rightarrow}
\def\tend{\rightarrow}
\usepackage{fancyhdr}
\usepackage{epsfig}
\usepackage{cite}
\usepackage{graphicx}
\usepackage{amsmath}
\usepackage{theorem}
\usepackage{amssymb}
\usepackage{latexsym}
\usepackage{epic}

\newcounter{ind}

{\theorembodyfont{\rmfamily}}
{\theorembodyfont{\rmfamily}}
{\theorembodyfont{\rmfamily}}
{\theorembodyfont{\rmfamily}}
{\theorembodyfont{\rmfamily}}
{\theorembodyfont{\rmfamily}\newtheorem{Con}{{\textbf Conjecture}}[chapter]}

\begin{document}
\pagenumbering{roman}

\setcounter{page}{1}

\newpage

\thispagestyle{empty}
\begin{center}
  \vspace*{1cm}
  {\Huge \bf Gauge/Gravity Duality: Recovering the Bulk from the Boundary using AdS/CFT}

  \vspace*{2cm}
  {\LARGE\bf Samuel Bilson}

  \vfill

  {\Large A Thesis presented for the degree of\\
         [1mm] Doctor of Philosophy}
  \vspace*{0.9cm}
  
   \begin{center}
   \includegraphics[height=1.1in,width=2.5in]{logo1}
   \end{center}

  {\large Centre for Particle Theory\\
          [-3mm] Department of Mathematical Sciences\\
          [-3mm] Durham University\\
          [-3mm] England\\
          [1mm]  November 2010}

\end{center}

\newpage
\thispagestyle{empty}
\begin{center}
 \vspace*{2cm}
  \textit{\LARGE {Dedicated to}}\\ 
 Mum and Dad
\end{center}

\newpage
\thispagestyle{empty}
\addcontentsline{toc}{chapter}{\numberline{}Abstract}
\begin{center}
  \textbf{\Large Gauge/Gravity Duality: Recovering the Bulk from the Boundary using AdS/CFT}

  \vspace*{1cm}
  \textbf{\large Samuel Bilson}

  \vspace*{0.5cm}
  {\large Submitted for the degree of Doctor of Philosophy\\ November 2010}

  \vspace*{1cm}
  \textbf{\large Abstract}
\end{center}
Motivated by the holographic principle, within the context of the AdS/CFT Correspondence in the large t'Hooft limit, we investigate how the geometry of certain highly symmetric bulk spacetimes can be recovered given information of physical quantities in the dual boundary CFT. In particular, we use the existence of bulk-cone singularities (relating the location of operator insertion points of singular boundary correlation functions to the endpoints of boundary-to boundary null geodesics in the bulk spacetime) and the holographic entanglement entropy proposal (relating the entanglement entropy of certain subsystems on the boundary to the area of static minimal surfaces) to recover the bulk metric.

Using null and zero-energy spacelike boundary-to-boundary geodesic probes, we show that for classes of static, spherically symmetric, asymptotically AdS spacetimes, one can find analytic expressions for extracting the metric functions given boundary data. We find that if the spacetime admits null circular orbits, the bulk geometry can only be recovered from the boundary, down to the radius of null circular orbits. We illustrate this for various analytic and numerical boundary functions of endpoint separation of null and spacelike geodesics. We then extend our analysis to higher dimensional minimal surface probes within a class of static, planar symmetric, asymptotically AdS spacetimes. We again find analytic and perturbative expressions for the metric function in terms of the entanglement entropy of straight belt and circular disk subsystems of the boundary theory respectively. Finally, we discuss how such extractions can be generalised.

\chapter*{Declaration}
\addcontentsline{toc}{chapter}{\numberline{}Declaration}
The work in this thesis is based on research carried out at the
Centre for Particle Theory, the Department of Mathematical Sciences, Durham University, England.  No part of this thesis has been
submitted elsewhere for any other degree or qualification and it all
my own work unless referenced to the contrary in the text.

Chapters 1 and 6 are the author's own work. Chapter 2 is a review chapter of relevant background material. Chapters 3 and 4 contain the author's original work based on \cite{Bilson}. Chapter 5 is taken from the author's paper \cite{Bilson2}.

\vspace{2in}
\noindent \textbf{Copyright \copyright\; 2010 by Samuel Bilson}.\\
``The copyright of this thesis rests with the author.  No quotations
from it should be published without the author's prior written consent
and information derived from it should be acknowledged''.

\chapter*{Acknowledgements}
\addcontentsline{toc}{chapter}{\numberline{}Acknowledgements}
A PhD is a long battle between self-motivation and procrastination. That battle would not have been won, manifesting itself in this thesis, without the help of those around me. As such, I would like to thank those who have helped me when the completion of this thesis seemed very distant.\\
First of all, I would like to thank Veronika Hubeny for all her helpful comments and for steering me back on course when nothing seemed to work. I would like to thank all my office mates, for making my office a fun place to work and giving me an extra motivation to wake up in the morning. Special thanks goes to my flatmate Andy. He has been an amazing friend, and I have never had such a laugh sharing a house. I wish you all the best in life and I know that postdoc is out there somewhere! I also want to give special thanks to Jamie, who helped check my thesis. I also know, if it wasn't for him, I wouldn't have had half as much fun at Ustinov playing pool, poker, and generally getting hammered! Mentions should also go to Jonny, Gurdeep, Dave, Shane, Josie, Eimear, Peter, Luke, Ian and all my pool friends. Without Ian, I would never have had as much fun playing in the DUPL. I think after 4 years, I have finally learnt how to hold a cue!      

\tableofcontents
\listoffigures
\clearpage


\pagenumbering{arabic}
\setcounter{page}{1}

\chapter{Introduction}
\label{Intro}

\numberwithin{equation}{chapter}
\textit{This chapter is a review of background material written in the author's own words.}

\section{The Search for Quantum Gravity}

In the field of theoretical physics, we search for fundamental physical theories which provide us with a conceptual structure that we develop and use in order to organise and understand the world, and make predictions about it. Theories of this type have been applied throughout history to help us develop the physics and technology that we now take for granted. In the modern era, there have been two crucial theories which have extensively exhibited such properties: General Relativity (GR) and Quantum Field Theory (QFT)\footnote{General Relativity is covered extensively in \cite{wald}, whereas \cite{Peskin:1995ev} gives a good introduction to all aspects of Quantum Field Theory.}.

In 1905\cite{SR}, Albert Einstein proposed his theory of Special Relativity (SR), to explain the constancy of the speed of light required by Maxwell's field equations and remove the unwanted ether. This introduced an apparent contradiction with Newton's theory of gravitation\cite{Newton}. Newton described gravity as an instantaneous ``action-at-a-distance'' force. This conflicted with the fundamental tenet of SR, where an upper bound exists on how quickly information can travel, namely the speed of light $c$. Einstein resolved this contradiction by proposing his Theory of General Relativity\cite{GR}. He showed that gravity, is in fact, just geometry. More specifically, spacetime is a dynamical object which is warped by a gravitational field. This is a radical departure from the Newtonian notion of absolute time. The fact that spacetime is warped by gravitational fields is encapsulated by Einstein's field equations
\be
G_{\m\n}=8\pi G_NT_{\m\n}.
\ee
$G_{\m\n}$ is called the Einstein tensor, and since it contains second order terms in the derivative of the metric $g_{\m\n}$, provides information of the curvature of the spacetime. $T_{\m\n}$ is called the stress-energy tensor and provides information about the matter/energy distribution. $G_N$ is Newton's constant.

GR has proved to stand up to all experimental observations thus far, from cosmological distances to millimetre scales. It has also made predictions (the classical tests of GR) which were later observed, including Eddington's observation of the bending of light around the Sun, the anomalous precession of the perihelion of Mercury and the gravitational red shift of light. Applications of GR have lead to relativistic astrophysics, cosmology and GPS technology. The wealth of experimental data and verifiable predictions allows us to call GR a fundamental theory of nature.

In the 1920's, quantum theory was developed in an attempt to understand the strange phenomena occurring at the atomic scale. Through multiple experiments around the turn of the $20^{\text{th}}$ century, it was found that phenomena at small scales from a fraction of a millimetre down to $10^{-19}$ metres are probabilistic in nature. Heisenberg relation
\be
\D x\D p\geq\dfrac{\hbar}{2},
\ee
where $2\pi\hbar=h=6.62\times 10^{-34}\,\text{m}^2\text{kgs}^{-1}$ is Planck's constant, is a realisation of the fact that the position ($x$) and momentum ($p$) of a particle cannot be measured simultaneously and are thus manifestly uncertain. In quantum mechanics, matter content of a system is encapsulated by a wave function, $\Psi$, whose dynamics is described by Schr\"{o}dinger's equation, 
\be
i\hbar\dfrac{\pd}{\pd t}\Psi=\hat{H}\Psi,
\ee
where $\hat{H}=-\frac{\hbar^2}{2m}\nabla_{\vec{x}}^2+V(\vec{x})$ is the Hamiltonian operator.

Due to work by Dirac and others in the late 1920's and 1930's, quantum mechanics was made Lorentz invariant and thus combined with SR to form quantum field theory (QFT). QFT was applied to describe the electromagnetic force (QED), and then the weak and strong nuclear forces (QCD). This has helped explain many observations at the small scale. The best example comes from QED which has been verified by agreement in precision tests of the fine structure constant to within ten parts in a billion. Applications of QFT have led to atomic physics, nuclear physics, particle physics, condensed matter physics, semiconductors, lasers, computers and quantum optics. The wealth of experimental data and verifiable predictions allows us to call QFT a fundamental theory of nature.

No known fundamental interactions fall outside the frameworks of GR and QFT. None the less, these two pillars of modern physics contradict each other. QFT is formulated using a fixed non-dynamical background, whereas GR is formulated on a dynamical spacetime. QFT requires that dynamical fields be quantised, whereas GR requires the fields to be deterministic and sit on a smooth Riemannian manifold. With all their immense empirical successes, QFT and GR have left us with an understanding of the physical world which is unclear and badly fragmented. With this in mind, many prominent theoretical physicists including Dirac, Feynman, Weinberg, DeWitt, Wheeler, Penrose, Hawking, t'Hooft and notable others sought a resolution to this conflict in the form of a theory of quantum gravity (QG). But, despite some 80 years of active research, no one has yet formulated a consistent and complete QG. In fact, it took until 1986\cite{GRQFT} to show that GR cannot be quantised using conventional quantum field theoretical techniques. This begs the question, why search for such a theory?

The philosophical answer is the unification of physics. Throughout history, there have been many examples of contradictions between known theories, leading to the discovery of a more general theory which provided a new conceptual picture of the world. For example, the contradiction between Maxwell's theory of light and Newtonian mechanics lead to the discovery of special relativity. The contradiction between quantum mechanics and special relativity lead to quantum field theory. The contradiction between Newtonian gravity and special relativity lead to general relativity. The hope is that a theory of quantum gravity will lead to a deeper understanding of the universe as we know it and its applications will lead to new physics and technology. QG also may provide the answers to conceptual questions persisting in existing theories. GR predicts singularities, but an initial singularity in cosmological models causes problems. The hope is that QG might determine those initial conditions. 

QG is also motivated from inconsistencies in QFT. It has been suggested that the divergences which appear in perturbation theory at the ultraviolet (UV) scale arise from our ignorance of a more fundamental theory, which might provide an automatic cut-off at the Planck scale (the scale at which QG effects become important)\cite{landau}. As a final motivation, we would like to develop QG to understand the physics of massive but small phenomena, where both gravitational and quantum effect become important. This is most notable in the study of black holes, for which there are many unresolved issues.

Motivating QG is one thing, it is quite another to formulate it consistently\footnote{For a more complete picture of the current status of QG, refer to \cite{QG1,QG2}}. The two main approaches which have borne most fruit in recent times have been quantum geometry and string theory. Quantum geometry or loop quantum gravity (LQG) is regarded as a non-perturbative candidate for QG via the quantisation of GR in the canonical formalism. Whereas string theory, which will be described in a lot more detail in the \S\ref{stringth}, is a more ambitious project in that it attempts not only to recover QG, but to describe all the fundamental interactions of nature. In this sense, string theory can be considered as a ``theory of everything''. It achieves this by quantisation of the perturbative string, and so provides a perturbative formulation of QG\footnote{However, as will be noted in \S\ref{stringth} and \S\ref{adscftint}, recent progress has lead to a non-perturbative description of string theory, which will prove fundamental to the work presented in this thesis.}. Although there are no experimental constraints on QG outside the regimes of validity of GR and QFT, both LQG\cite{LQGBH} and string theory\cite{STBH} have recovered one of the major theoretical constraints on QG, which is the Bekenstein-Hawking entropy $S_{\text{BH}}$ of a black hole with horizon area $A$\cite{Bek1,Bek2,Bek3}
\be
\label{BHent}
S_{\text{BH}}=\dfrac{A}{4G\hbar}.
\ee
Despite the progress towards QG, both approaches come with disadvantages. LQG still has not found a consistent way of imposing the Hamiltonian constraint to control the dynamics of the theory, whereas imposing consistency in string theory requires us to live in a fanciful world of extra dimensions with multiple unobserved particles. Some have suggested this a symptom of a more fundamental problem, which is our lack of understanding of what exactly the quantisation of spacetime really means. Both theories may still prove successful, or maybe even an amalgamation of the two\cite{smolin}, but these problems have lead many to consider other, more extreme alternatives. These include causal set theory\cite{QGcs}, twistor theory\cite{twistor}, the null surface formalism\cite{QGns} and non-commutative geometry\cite{noncom}

\section{String Theory}
\label{stringth}

As we have already noted, string theory (or superstring theory) is one of the main candidates for a consistent and complete theory of quantum gravity. In this section we explore the history of its creation and development, its basic features, and its relevant applications.

String theory describes relativistic one dimensional objects which can either be open (where the string has end points) or closed (where the string forms a loop). These strings sweep out a worldsheet in a target spacetime similar to particles sweeping out world-lines. Since string theory is a relativistic quantum field theory including gravity, it must contain the three respective fundamental constants $c, \hbar$ and $G$. Thus strings have a characteristic length scale $l_s$, which can be estimated by the Planck length
\be
l_P=\left(\dfrac{\hbar G}{c^3}\right)^{3/2}=1.6\times 10^{-33}\mathrm{cm}.
\ee
Since the Planck scale is way beyond what any present particle detector can probe, strings would look like point particles to current accelerators, and is thus consistent with current observations.

String theory was formulated in the 1960's by Veneziano\cite{venez} as an attempt to describe the strong interactions between large numbers of mesons and hadrons. Excitement in this approach grew as the S-matrix scattering amplitudes matched those found in meson scattering experiments. However, by 1973, it was shown that the correct quantum field theory for strong interactions was quantum chromodynamics (QCD), and so interest waned. In the following years, it was shown that the strong force, along with the electromagnetic and weak forces (which had already been unified into Glashow, Salam and Weinberg's electroweak theory by 1967\cite{EW}) could be combined to form what is known as the ``Standard Model'' of elementary particles. 

The standard model is a gauge theory (a theory which is invariant under gauge transformations) with gauge group $SU(3)\times SU(2)\times U(1)$. The gauge fields combine to represent the force carrying bosons of the model. These include the photon for electromagnetic force, the W and Z bosons for the weak nuclear force, and the gluon for the strong nuclear force. $SU(3)$ is the gauge group of QCD with three colours of quarks, and $SU(2)\times U(1)$ is the gauge group of electroweak theory. The standard model, combined with GR, describes all the fundamental interactions of particle physics, and is consistent with virtually all physics down to scales probed by particle accelerators (approximately $10^{-16}$cm)\footnote{At the time of writing, there is a particle accelerator at CERN called the Large Hadron Collider (LHC) which is probing to smaller scales. The hope is to find the one remaining undetected particle in the standard model, the Higgs boson, which is required by the theory to give masses to the fermions.}. 

Unfortunately, this theory is not complete. If one tries to combine gravity with quantum theory, one ends up with a non-renormalisable quantum field theory, which hints at new physics at very high energy scales. If one analyses quantum gravity as a quantum field theory, where the graviton interaction has a coupling constant proportional to Newton's constant $G_N$, one finds that the perturbation theory is divergent, and breaks down at energy scales greater than the Planck mass $M_P=G_N^{1/2}=1.22\times 10^{19}$ GeV. This lead many theorists in the mid 1970's to find a new direction to combine gravity (or its conjectured spin-2, force carrying boson, the graviton $g_{\m\n}$) with the rest of particle theory. It was string theory that provided the answer, where in 1974 it was found that a spin-2 particle which interacts like a graviton was one of the massless states within the string spectrum\cite{grav1,grav2}. This lead to the first formulation of what is now known as bosonic string theory. This discovery was subsequently followed by the fact that fermions could be combined to such a theory via the requirement that the theory be supersymmetric. This led to the first example of a supersymmetric string theory, or ``superstring'' theory for short.  Since string theory naturally includes GR, it was therefore proposed as a unified theory of all the fundamental forces of nature.

For all the successes of superstring theory, there were still mathematical inconsistencies, including many divergent quantities. It was only until 1985 that these problems were resolved, using what is known as Green-Schwarz anomaly cancellation\cite{GS}. The main consequence of this was the limitation of the number of consistent superstring theories to five, and the restriction that all these theories must live in ten spacetime dimensions (9 space and 1 time). This is what is now known as the ``first superstring revolution''. The five consistent superstring theories are known as Type I (a theory of unoriented strings), Type IIA and IIB (theories of oriented strings, with the difference being that IIB is chiral where as IIA is not) and the heterotic string theories $SO(32)$ and $E_8\times E_8$. If superstring theory was to be proposed as a fully fledged ``theory of everything'', the apparent contradiction between strings living in a 10-dimensional spacetime, and our 4-dimensional observed reality, was resolved through the notion of ``compactification''. By compactifying six of the nine spatial directions on a small enough space so that it could not be resolved with current microscopes, one could make $9+1$ dimensional space look $3+1$ dimensional. This was achieved consistently via the use of ``Calabi-Yau'' manifolds\cite{CY}.

\begin{figure}
\centering
\includegraphics[height=3in,width=3in]{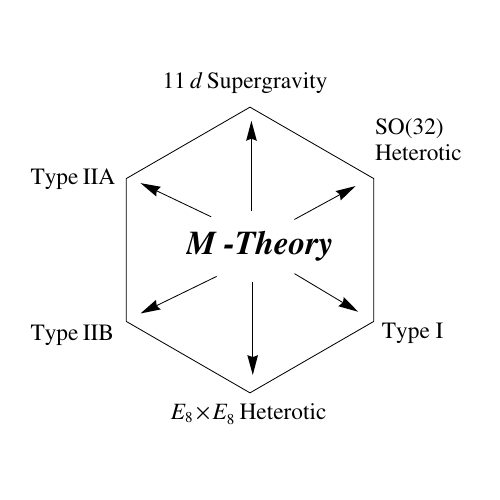}
\caption{\label{Mtheory}This is a schematic diagram expressing the fact that M-theory is related to all five superstring theories and 11$D$ supergravity}
\end{figure}

Up until 1995, this was the status of string theory. The classical theory was well understood with the classical equations of motion resembling GR. Quantum string perturbation theory was also well understood. However, a non-perturbative description of string theory was needed to explain quantum gravitational effects at strong coupling. This requirement lead to the ``second superstring revolution'', where in 1995, Witten suggested a candidate for a non-perturbative description of string theory. The proposal was that the five existing superstring theories were just certain limits of an eleven dimensional theory, known as ``M-Theory''\cite{M1}. This was motivated by the existence of dualities between the ten dimensional superstring theories and eleven dimensional supergravity, which is the low energy limit of M-theory. Eleven dimensions is the maximum number of spacetime dimensions to which one can formulate a consistent supersymmetric theory of gravity\cite{nahm}. The central idea of string duality is that the strong coupling limit of one string theory is equivalent to another theory in the weak coupling limit. Via a process known as ``T-duality'', which inverts the size of a compactified direction of spacetime along which the string propagates, and ``S-duality'', which inverts the string coupling constant, all five 10-dimensional superstring theories can be related to one another. M-theory is then realised as the strong coupling limit of Type IIA or $E_8\times E_8$ heterotic superstring theories (see figure \ref{Mtheory}).
Since the string dualities relate weak and strong coupling, fundamental perturbative string states in one theory are related to non-perturbative solitonic objects in another. These non-perturbative objects exist in M-theory in the form of ``$p$-branes'', where $p$ is the number of spatial extensions. A special class of such branes, discovered by Polchinski\cite{dbranes}, were found by T-dualising the Neumann boundary conditions of strings to give Dirichlet boundary conditions. These are known as Dirichlet $p$-branes or ``D-branes''. The stringy description of such objects are given as a topological defect on which the ends of open strings are constrained to move. It is this discovery of D-branes which has lead to the rapid growth in the understanding of the non-perturbative aspects of string theory. It was the use of D-branes in counting microstates of a black hole that lead to a microscopic explanation of black hole entropy\cite{STBH}, and it was the non-relativistic dynamics of D-branes that allowed the probing of distances smaller than the Planck length\cite{dbsd}. D-branes can be directly applied in brane-world scenarios in which the universe is modelled as one big D-brane\cite{RS}. Finally, D-branes have lead to the most important discovery in theoretical physics in recent times, namely Maldecena's AdS/CFT correspondence, which we will describe in the next section.
 
\section{The AdS/CFT Correspondence}
\label{adscftint}

It was mentioned in \S\ref{stringth} that it was the study of D-branes which led to Maldacena's AdS/CFT conjecture\cite{Maldacena,MAGOO,Witten1,SYM1}. This is because open string states which act as gauge fields correspond to fluctuations of D-brane geometry (see \S\ref{dbranes} for more detail). Thus one can describe D-branes using gauge theory. Equivalently, D-branes warp the geometry of the spacetime. So they can also be described in terms of a gravitational theory. It was this relationship or ``duality'' between gauge theory and gravity in the context of D-branes which lead Maldacena to his conjecture.

Maldecena's AdS/CFT correspondence states that
\begin{Con}
\textit{Type IIB superstring theory with} $AdS_5\times S^5$ \textit{boundary conditions is equivalent to} $\mathcal{N}=4$, \textit{SU(N) Super-Yang-Mills theory (SYM) in} $3+1$ \textit{dimensions}. 
\end{Con}
$\mathcal{N}=4$, SYM in four dimensions is a non-Abelian supersymmetric gauge theory with conformal symmetry. Anti-de Sitter, or $AdS$ space, is a maximally symmetric solution of Einstein's equations with negative cosmological constant. This implies that the AdS/CFT correspondence is a duality between a four dimensional gauge theory and a five dimensional gravitational theory. In some sense, it is useful to think of the gauge theory as ``living on the boundary'' of the bulk AdS spacetime. Infrared effects in AdS space correspond to ultraviolet effects in the boundary theory, known as the UV/IR connection. The fact that the physics in the bulk $AdS$ space can be described by a field theory of one less dimension is one example of the ``holographic principle''. In a quantum gravity theory, all physics within some volume can be described in terms of some theory on the boundary which has one bit of information per Planck area. This holographic bound is the physical interpretation of the UV/IR connection\cite{Hol5}.

The idea of a holographic interpretation of a quantum gravity theory in terms of a boundary gauge theory lead Polchinski and Horowitz\cite{adscft2} and others to propose a ``gauge/gravity duality''. This asserts that
\begin{Con}
\textit{Hidden within every non-Abelian gauge theory, even within the weak and strong nuclear interactions, is a theory of quantum gravity}.
\end{Con}
Such a duality can be motivated without the notion of string theory, although one finds string theory is hidden within this description.

The fact that gauge theories and string theories are related is not unusual. It had already been noted by t'Hooft\cite{largeN} in 1973 that the large $N$ limit (where $N$ is the number of colours) of certain gauge theories at strong coupling correspond to worldsheet string perturbation theory at weak coupling. It is this strong/weak coupling duality in AdS/CFT that allows us to study the non-perturbative regime of string theory by looking at weakly coupled gauge theories.

An immediate application of AdS/CFT was proposed by Witten\cite{Witten1}, where he found a one-to-one correspondence between operators in the field theory and fields propagating in AdS space by equating the generating functional of correlation functions in the CFT to the full partition function of string theory. This allows computation of correlation functions in the gauge theory in terms of supergravity Feynman diagrams. This is an explicit manifestation of the AdS/CFT correspondence and provides us with a dictionary from which one can relate observables in the bulk to observables on the boundary. In particular, since AdS/CFT relates weak and strong coupling, bulk physics with high curvature is related to weakly coupled field theories where one can perform perturbation theory. This has been applied in a process known as holographic renormalization where one can obtain renormalized QFT correlation functions by performing computations on the gravity side of the correspondence\cite{HR}. This has been used to compute such quantities as the expectation value of the boundary CFT stress-energy-momentum tensor associated with a gravitating system in asymptotically anti-de Sitter space\cite{Tmn}.

By relating thermodynamic properties on both sides of the correspondence, it was shown\cite{Witten} that $\N=4$ SYM theory in a thermal state is dual to a large mass, AdS-Schwarzschild black hole. This was developed by \cite{Maldacena1} to argue that an eternal black hole in AdS spacetime can be holographically described by two identical copies of the dual CFT associated with the spacetime and an initial entangled state. This allowed a whole swathe of research on trying to understand the basic non-perturbative aspects on both sides of the correspondence. Characteristic properties of confinement and asymptotic freedom in QCD have been investigated using the AdS/QCD correspondence\cite{AdS/QCD1,AdS/QCD2,AdS/QCD3,AdS/QCD4}, where one can perform calculations in a weakly coupled string theory. More recently, superfluidity and superconductivity in, for example, strongly coupled plasmas have been studied by applying AdS/CFT to condensed matter systems, known as AdS/CMT\cite{AdS/CMT1,AdS/CMT2,AdS/CMT3,AdS/CMT4}. On the gravity side, properties such as causal structure, event horizons and singularities could equally well be understood using strongly coupled gauge theories. In particular, one might hope to better understand the physics of black holes, including the description of physics behind the horizon. 

By identifying field theoretic observables, in particular boundary correlation functions, work was done on identifying signals of the bulk curvature singularity\cite{Fid,Bal,Louko,Kraus,Fest,Freiv}. It was observed that signals of black hole singularities in the bulk can be identified by considering nearly null spacelike probes. These provide a large contribution to the boundary field theory correlation function, giving rise to ``light-cone'' like singularities. These results, however, were limited to static geometries, and one would prefer to investigate properties of CFT correlators which correspond to manifestly time-dependent spacetimes, such as gravitational collapse scenarios. Fortunately, it was shown by\cite{hubeny} that this can be achieved by examining the structure of singularities for generic Lorentzian correlators. It is already known that CFT correlators will exhibit light-cone singularities when the operator insertion points are connected by strictly null geodesics. However, by considering boundary-to-boundary null geodesics which penetrate the bulk, the CFT correlator exhibits additional singularities inside the boundary light cone, called\textit{``bulk-cone singularities''}\cite{hubeny}. 

Through this relationship, one can probe the geometry of certain asymptotically AdS spacetimes given information of boundary data, namely the locus bulk-cone singularities. More specifically, one can attempt to reconstruct the metric of certain classes of spacetimes given such boundary data. This was achieved numerically in\cite{hammer1}, for a general class of static, spherically symmetric, asymptotically AdS $(d+2)$-dimensional spacetimes with metric
\be
\label{genads}
ds^2 = -f(r)\,dt^2+h(r)\,dr^2+r^2\,d\Om^{2}_d.
\ee
In the simplified case where $h(r)=\frac{1}{f(r)}$,\cite{hammer1} was able to extract the metric function $f(r)$ given information of the locus of endpoints of boundary-to-boundary null geodesics encapsulated by the function's worth of information $\D t(\D\p)$. $\D t$ and $\D\p$ are the time and angular separation of null geodesic boundary endpoints respectively. Complementing this work, we show \cite{Bilson} in chapter \ref{Null} that one can achieve the extraction analytically via the observation that the expression for $\D t(\D\p)$ in terms of $f(r)$ reduces to a standard integral equation with known solution. By applying this solution to various expressions for $\D t(\D\p)$, we were able to recover the same limitation on the extraction as illustrated in \cite{hammer1}, namely that spacetimes admitting null circular orbits with radius $r_m$ can only be recovered in the region $r\in(r_m,\infty)$. Unique to this analytical approach, we also show in \S\ref{pertmeth}, that given $\D t(\D\p)$ as a convergent series, one can always find a convergent series solution for the metric function $f(r)$. This is verified in the case of the AdS-Schwarzschild black hole solution.

Pursuing the theme of extracting metric data given information of the dual CFT, one can look for alternative bulk geodesic probes corresponding to some observable in the dual CFT which go behind the region $r\in(r_m,\infty)$. As identified in the review \cite{Hub1}, for fixed energy and angular momentum, boundary spacelike geodesics provide such a probe. Thus, by identifying boundary observables corresponding to such a probe, one may attempt to learn more about the bulk geometry. This was achieved in \cite{Ryu,Ryu1}, which related the area of co-dimension 2 static\footnote{In this thesis, we restrict our analysis to static spacetimes where $\pd_t$ is Killing. However one can extend this relationship to time-dependent cases such that the area/entropy formula is fully covariant\cite{Hub}. Extensions to these cases are discussed in chapter \ref{discussion}.} minimal surfaces anchored to the boundary, with the entanglement entropy of some subsystem in the boundary CFT. In the specific case of a $2+1$-dimensional bulk, the area of the static minimal surface is the proper length of a zero-energy boundary spacelike geodesic. 

It had been known for a while\cite{Suss} that black hole entropy, which can be calculated from the horizon area using \eqref{BHent}, is related to a quantity known as entanglement entropy (or geometric entropy) of a QFT. This is defined as the von Neumann entropy of a reduced density matrix by tracing out the degrees of freedom of a certain subsystem. Thus it measures how closely entangled a quantum system is. Since the eternal black hole in AdS is related to an entangled state in the CFT\cite{Bal1}, it is natural to assume a relationship between horizon area in the bulk, and entanglement entropy on the boundary. It was this proposal by Ryu \textit{et.al.} that lead to the existence of such a holographic description\footnote{For a good review of all aspects of holographic entanglement entropy, see \cite{Ryu2}.} exist in the form of minimal surface areas. More specifically, it was shown that 
\begin{Con}
\textit{The entanglement entropy $S_A$ of a static subsystem $A$ in a $(d+1)-$dimensional CFT can be determined from a $d$-dimensional static minimal surface $\gamma_A$, in the dual $(d+2)$-dimensional bulk, whose boundary is given by the $(d-1)$-dimensional manifold $\pd\g_A=\pd A$. The entropy is given by applying the usual Bekinstein-Hawking area/entropy relation}
\end{Con}
\be
\label{entent}
S_A=\dfrac{\Area(\g_A)}{4G_N^{(d+2)}}.
\ee
It is this proposal which was used in\cite{hammer2} to numerically recover the metric function $h(r)$ in a $(2+1)$-dimensional bulk described by the metric \eqref{genads}. The boundary observable in this case is the entanglement entropy $S_A(l)$ in an infinitely long one dimensional system where $A$ is an interval of length $l$. This was combined with bulk-cone singularities in null boundary probes to show the one can extract both metric functions in \eqref{genads} in $(2+1)$-dimensions. Following the same reasoning, in chapter \ref{nullspace} we show \cite{Bilson} that expressions for $\D t(\D\p)$ and $S_A(l)$ reduce to simple integral equations in terms of the metric functions $f(r)$ and $h(r)$. Thus providing an analytical method for determining metric \eqref{genads}. We show that for spacetimes with an event horizon, $h(r)$ can be extracted in the region $r\in(r_+,\infty)$, where $r_+$ is the horizon radius. Since for all physical spacetimes $r_+\leq r_m$, this method allows one to reconstruct deeper regions of the bulk, thus agreeing with the analysis of \cite{Hub1}.

A major limitation of the extraction methods laid out in chapter \ref{nullspace}, is the restriction to a $(2+1)-$dimensional bulk spacetime. As is pointed out in \cite{Hub1}, for sensible spacetimes the surfaces anchored to the boundary of higher dimensionality probe deeper into the bulk. So, by not restricting ourselves to 1-dimensional probes, and utilising the full form of \eqref{entent}, one would hope to recover a larger region of the bulk. This is the main motivation behind chapter \ref{strdsk}, where we consider the static, planar symmetric asymptotically AdS metric
\be
ds^2=R^2\left(\dfrac{-h(z)^2\,dt^2+f(z)^2\,dz^2+\sum_{i=1}^d dx_i^2}{z^2}\right),
\ee
and ask whether one can recover the real metric function\footnote{Since we are considering only static surfaces, expressions for the area do not depend on $h(r)$, and so cannot be determined using the method outlined.} $f(z)$ given the entanglement entropy of certain multi-dimensional subsystems of the boundary CFT using \eqref{entent}. This is discussed in the context of two particular simple subsystems defined in \cite{Ryu2}, namely the infinite strip $A_S$ and the circular disk $A_D$. We will show \cite{Bilson2} that in the case of the infinite strip, one can again reduce the problem to a known integral equation, thus solving for $f(z)$ in terms of the entanglement entropy $S_{A_S}(l)$. We find the solution consistent with the known pure AdS result and find a series solution for $f(z)$ in perturbed AdS spacetimes. We then turn our attention to the circular disk $A_D$. We find \cite{Bilson2} that the reduction in symmetry of the minimal surface equations means that we must resort to a perturbative analysis. We outline a method for determining the area of minimal surfaces anchored to $\pd A_D$ by perturbing around the pure AdS solution of a hemisphere and recover analytical expressions at first order. By doing so, we illustrate in the simple case of the planar black hole, the positive correlation between surface dimension and the depth probed in the bulk.

In the next chapter we make the relationships and concepts outlined here more quantitative by building up the string/field theoretical framework required to describe them.

\chapter{Background}
\label{background}
\textit{This chapter is a review of background material taken from other authors' published work.}\\
\numberwithin{equation}{section}

In this chapter, we introduce the basic framework required to describe the relationships and concepts in this Thesis. The aim of this chapter is to build up towards two main results, namely the holographic interpretation of entanglement entropy\cite{Ryu,Ryu1,Ryu2}, and bulk-cone singularities in CFT correlators\cite{hubeny}. Both results stem from the properties of a gauge/gravity duality known as the AdS/CFT correspondence introduced in \S\ref{adscftint}. Therefore, to fully understand these two results, one must begin by motivating and explaining AdS/CFT. This will be achieved in this chapter by looking at both sides of the correspondence: String theory and supergravity on the one side; gauge theory and D-branes on the other. Once AdS/CFT has been  fully motivated, the correspondence will be stated and its basic properties explained, most notably the operator/field correspondence and the correspondence at finite temperature. Once this is done, the two main relationships will have been given more context and will be explained in more detail.  

\section{String Theory}

As has already been outlined in the introduction, string theory\cite{string1,string2,string3} serves as the underlying framework for all the results used in this thesis. It is therefore prudent and illustrative to provide an outline of the main features and results that string theory has provided. We will begin by describing the bosonic string and the superstring, followed by certain objects which must exist within string theory, most notably D-branes\footnote{See, for example, \cite{Dbranesj} for a more detailed introduction to the subject.}.

\begin{figure}[h!t]
\centerline{\includegraphics[width=10cm]{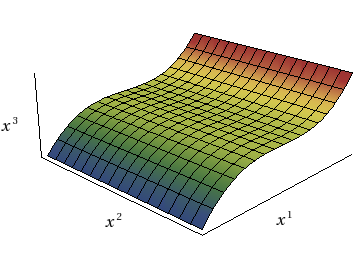}}
\caption{\label{worldsheet}This a diagram of the open string worldsheet parametrised by the coordinates $(\t,\s)$, embedded in a target spacetime with coordinates $x^\m$ and metric $g_{\m\n}(x)$. The grid lines are lines of constant $\s$ and $\t$.}
\end{figure}

The string is a relativistic one-dimensional object which sweeps out a worldsheet with coordinates $(\s^1,\s^2)\equiv (\tau,\s)$. The string's path in a $(d+1)-$dimensional spacetime is described by the $(d+1)-$dimensional vector field $x^\m(\tau,\s)$, embedding the worldsheet in a target (background) spacetime with metric $g_{\m\n}(x)$ where $\m,\n=0,1,\cdots,d$ (see figure \ref{worldsheet}). The induced metric on the worldsheet is given by the pull-back of the background metric
\be
h_{ab}=\pd_a x^{\m}\pd_bx^\n g_{\m\n}\quad(\pd_a\equiv\pd/\pd\s^a).
\ee

This is the natural two-dimensional extension of the one-dimensional relativistic point particle which traces out a worldline $x^\m(\tau)$ .

\subsection{The Bosonic String}

Bosonic string theory is the original and the most basic string theory as it contains no fermionic excitations. As such, it cannot be a fundamental theory, but it is useful to describe the framework and its basic properties, as these persist when one studies superstrings.
 
We can define the bosonic string action (Nambu-Goto action) in a similar way to the action of a point particle of mass $m$, given by
\be
\label{N-G}
S=-m\int ds=-m\int d\t\sqrt{-g_{\m\n}\dot{x}^\m\dot{x}^\n},\quad \dot{x}^\m=\pd_\tau x^\m(\tau),
\ee
where $s$ is the total length of the particle's trajectory defined by the worldline $x^\m(\t)$.

Since strings are massless, the natural physical quantity to parametrise the string is its tension $T$, with the action given as the total area $A$ of the string worldsheet,
\be
S=-T\int dA=-T\int d\tau d\s\sqrt{-h},\quad h=\det(h_{ab}),
\ee
where $T$ is related to the characteristic length scale of the string $l_s$, and the ``universal Regge slope'' $\a'$ by,
\be
T\equiv\dfrac{1}{2\pi l_s^2}=\dfrac{1}{2\pi\a'}.
\ee
The Nambu-Goto action can be rewritten without the awkward square root by including a rank two symmetric tensor field $\g_{ab}(\t,\s)$ such that
\be
\label{Polyakov}
S=-T\int d^2\s\sqrt{-\g}\g^{ab}h_{ab}
\ee
This is known as the Polyakov action and one can think of $\g_{ab}$ as the metric on the string worldsheet. The Nambu-Goto action is then recovered by applying the Virasoro constraints,
\be
T_{ab}\equiv\dfrac{-2}{\sqrt{\g}}\dfrac{\delta S}{\delta\g^{ab}}=0.
\ee
This constraint on the energy-momentum tensor $T_{ab}$ represents two symmetries of \eqref{Polyakov}, namely reparametrization invariance
\be
\label{reparam}
(\t,\s)\rightarrow (\t'(\t,\s),\s'(\t,\s))),
\ee
and Weyl invariance
\be
\g_{ab}\tend\g'_{ab}=e^{2f(\t,\s)}\g_{ab},
\ee
where $f$ is some arbitrary real valued function. Weyl invariance, combined with Poincar\'{e} invariance of the worldsheet metric requires the Polyakov action be conformally invariant.

The worldsheet metric $\g_{ab}$ is a symmetric $2\times 2$ matrix, thus it has three degrees of freedom. The reparametrisation and Weyl invariance of the metric reduces this number to one degree of freedom. Thus we can pick a particular gauge which makes this manifest. Some useful choices are the ``conformal gauge''
\be
\g_{ab}=e^{\p(\t,\s)}\eta_{ab},
\ee
and the ``static gauge'', which identifies the worldsheet time with the background time
\be
x^0(\t,\s)=\t,\quad\forall\,\s\in\R.
\ee
  
Extremising the Polyakov action $\frac{\delta S}{\delta x^\m}=0$ in the conformal gauge leave us with the bulk equation of motion
\be
\label{steom}
\eta^{ab}\pd_a\pd_b x^\m(\t,\s)=0,
\ee
and the boundary term
\be
x'^\m\delta x_\m |_{\s=0,\pi}=0,\quad x'^\m=\pd_\s x^\m.
\ee

The bulk equation is the two-dimensional wave equation, as would be expected from an oscillating string. The boundary term provides us with different boundary conditions for closed and open strings. For closed strings, we need the endpoints to join up in a smooth fashion, and so we have
\be
x^\m(\t,0)=x^\m(\t,\pi)\quad\text{and}\quad x'^\m(\t,0)=x'^\m(\t,\pi).
\ee

For open strings, two types of boundary conditions are allowed. ``Neumann'' boundary conditions are defined by
\be
x'^\m(\t,\s)|_{\s=0,\pi}=0.
\ee
These correspond to the open string having zero momenta at the endpoints. Alternatively, ``Dirichlet'' boundary conditions are defined as
\be
\delta x^\m(\t,\s)|_{\s=0,\pi}=0.
\ee
This condition fixes the endpoints of the string. We will see later in \S\ref{dbranes} that the spacetime points $x^\m(\t,0)$ and $x^\m(\t,\pi)$ on which the open string end define objects called ``D-branes''.

Quantisation of bosonic string theory leads to an infinite number of simple harmonic oscillators. Requiring that the states of the Fock space be physical, by satisfying the Virasoro constraints $T_{ab}=0$, the number of background dimensions is fixed to be $26$. The ground states of both the open and closed string spectra are tachyonic with $m^2<0$. This means bosonic string theory is not a consistent quantum theory. As well as an infinite tower of massive states ($m^2>0$), there are massless states ($m^2=0$), which are given as follows: 
\begin{description}
\item[Open Strings] $A_\m$
\begin{itemize}
\item $A_\m$ is the massless spin 1 gauge field with $U(1)$ gauge symmetry. It has 24 physical polarizations, which is the same number required by a background photon field. Thus the low-energy oscillations of bosonic open strings correspond to a $U(1)$ gauge theory.
\end{itemize}
\item[Closed Strings] $B_{\m\n},\,\,g_{\m\n},\,\,\Phi$
\begin{itemize}
\item $B_{\m\n}$ is the massless anti-symmetric spin 2 tensor called the ``Neveu-Schwarz B-field''. It couples to strings with charge $q$ through a term in the action of the form
\be
q\int d^2\s\,\e^{ab}\pd_a x^{\m}\pd_bx^\n B_{\m\n},
\ee
and is thus sourced by the fundamental string. This is analogous to a relativistic particle with worldline $x^\m(\t)$ and charge $q$ sourcing an electromagnetic potential $A_\m$ through the term $q\int d\t\dot{x}^\m(\t)A_\m$. 
\item $\Phi$ is the massless spin 0 scalar field called the dilaton. $\Phi$ determines the string coupling constant $g_s$ through
\be
g_s=e^{\langle\Phi\rangle},
\ee
where $\langle\Phi\rangle$ is the vacuum expectation value (vev) of the dilaton.
\item $g_{\m\n}$ is the symmetric, traceless spin 2 tensor known as the graviton. Thus the low-energy oscillations of closed bosonic strings correspond to classical gravity.
\end{itemize}
\end{description}
These are the only states that remain in the low-energy ($T\tend\infty, \a'\tend 0$) limit where strings look like point particles. These massless background fields can be consistently added to the string action such that the $\b$-functions vanish requiring the bosonic theory remain conformal beyond quantisation. This requirement implies that the worldsheet theory is a $(1+1)-$dimensional conformal field theory (CFT). Since CFT's are scale invariant, the $\b$-functions of the closed string massless fields must vanish. These equations turn out to be the equations of motion of the fields $B_{\m\n},g_{\m\n},\Phi$ for a particular Einstein-Hilbert action, with spacetime $g_{\m\n}$, containing background fields $B_{\m\n}$ and $\Phi$. This is known as the closed bosonic ``string frame'' action given by
\be
\label{bossf}
S_{\text{cl}}=\dfrac{1}{4\pi\a'}\int d^{26} x\sqrt{-g}e^{-2\Phi}\left(R_g+4\nabla_\m\Phi\nabla^\m\Phi-\dfrac{1}{12}H^2+\O(\a')\right),
\ee
where $H=dB$ and $R_g$ is the Ricci scalar for the metric $g_{\m\n}$.

One can do the same for the open string field $A_\m$ whose $\b$-function equation is the equation of motion of a particular gauge theory with open string frame action
\be
S_{\text{op}}=-C\int d^{26}x\,e^{-\Phi}\text{Tr}(F^2)+\O(\a'),
\ee
where $F=dA$ and $C$ is a dimensionful constant.

Worldsheet quantum string theory provides the structure to its background spacetime by fixing the metric, dimension and field content. In fact, since the string coupling constant is also fixed by the vev of the dilaton, the only free parameter in string theory is its tension $T$. This remarkable fact makes it a good candidate as a fundamental theory. 

\subsection{Superstrings}

Unfortunately, as was already noted in the previous section, quantisation of bosonic string theory leads to tachyonic ground states. Once we require that string theory be invariant under supersymmetry, the theory stabilizes, and we find that ground states no longer have negative mass squared. The requirement of supersymmetry also adds spacetime fermions to the theory, which is a step towards a theory of everything.

The Polyakov action \eqref{Polyakov} of the bosonic string can be extended to include fermions. The worldsheet action for superstring theory in the conformal gauge then takes the form
\be
S=-\dfrac{T}{2}\int d\tau d\s\,\left(\pd_ax^\m\pd^ax_\m-i\bar{\psi}^\m\rho^a\pd_a\psi_\m\right),
\ee
where $\psi$ is a two-component Majorana spinor and $\rho^a,\,a=0,1$ are $2\times 2$ Dirac matrices.

Due to the addition of fermions, additional boundary conditions can be imposed on the Dirac fields. These are the periodic ``Ramond'' (R) sector
\be
\label{R}
\psi^\m(\t,0)=\bar{\psi}^\m(\t,0)\quad\text{and}\quad\psi^\m(\t,\pi)=\bar{\psi}^\m(\t,\pi),
\ee
and the anti-periodic ``Neveu-Schwarz'' (NS) sector
\be
\psi^\m(\t,0)=-\bar{\psi}^\m(\t,0)\quad\text{and}\quad\psi^\m(\t,\pi)=-\bar{\psi}^\m(\t,\pi).
\ee
In a similar way to the bosonic string, quantisation fixes the number of spacetime dimensions. The ``superstring critical dimension'' in this case is
\be 
d=10. 
\ee
The ground states of the NS sector in the open and closed superstring spectra are again tachyonic. However, the requirement of supersymmetry allows the spectrum to admit a consistent truncation known as the Gliozzi-Scherk-Olive or ``GSO projection''\cite{GSO}, which removes half the number of fermionic excitations. In particular, the physical masses in the spectrum are only allowed to be integer multiples of $1/\a'$, removing the tachyonic $m^2=-\frac{1}{2\a'}$ ground state. The R sector vacuum energy vanishes via the convention of the boundary conditions \eqref{R}, and the action of GSO projection in this case introduces a chirality into the spacetime spinors $\psi^\m$. Thus GSO projection stabilises quantum superstring theory, and so it is worthwhile studying this truncated spectrum.

The GSO projected open string spectrum has two sectors, R and NS. The NS massless sector describes the eight physical polarizations of the spacetime photon field. The massless R sector gives rise to a sixteen component spacetime Dirac fermion which satisfies the massless Dirac wave equation. It decomposes into two possible Majorana-Weyl spinor ground states depending on the spacetime chirality. It was already noted that GSO projection of the R sector selects out a particular chirality. This reduces the number of physical polarizations of the massless R sector to eight, which is equal to the number of spacetime bosons in the NS sector. Thus the physical states of the massless open string sector is the same number required for spacetime supersymmetry, and in particular they form the vector supermultiplet of the $d=10$ supersymmetric Yang-Mills theory. This analysis can be extended to include massive states to show that fully interacting superstring theory has spacetime supersymmetry\cite{GS3}.

The GSO projected closed superstring spectrum is analogous to the open superstring sector except now there is a tensor product of left and right-moving modes. This leads to four distinct sectors: the bosonic sectors NS-NS and R-R, and the fermionic sectors NS-R and R-NS. The massless states of the NS-NS sector are, in analogy with that of the closed bosonic string massless spectrum, the dilaton $\Phi$, the symmetric two-form $B_{\m\n}$, and the graviton $g_{\m\n}$, except now in ten spacetime dimensions. The NS-R and R-NS massless sectors produce gravitino and dilatino fields. The massless R-R fields are anti-symmetric tensors of rank $n=0,1,\cdots,10$, or $n$-forms $F_n$. Whether the rank of $F_n$ is even or odd depends on the action of the GSO projection on the R-R states. If we take the opposite GSO projection on the two R sectors, $n$ is even, and the resulting theory is chiral. This is known as Type IIA superstring theory. If we take the same GSO projection on the two R sectors, $n$ is odd, and the resulting theory is non-chiral. This is known as Type IIB superstring theory. The NS-NS and R-R massless spectra together form the bosonic components of $d=10$ Type IIA or Type IIB supergravity respectively. This result can be generalised to show that the low-energy limit of a superstring theory is a supergravity theory, and is the natural supersymmetric extension of the low energy limit of the closed bosonic string spectrum corresponding to a gravitational theory (see equation \eqref{bossf}.

The fields $F_n$ satisfy the equivalent of Maxwell's equation of motion $dF_n=0$ and therefore one can define a R-R gauge potential $C_{n-1}$ such that $F_n=dC_{n-1}$. One finds that it is the R-R fields which couple to the string, and not the R-R potentials. As we have already seen in bosonic string theory, the fact that the NS-NS B-field couples to the fundamental string means that the string carries charge under the B-field. This means that perturbative string states cannot carry any charge with respect to the R-R potentials. This requires string theory to be complemented by non-perturbative objects which carry the R-R charges of the potentials $C_{p+1}$. These objects are known as ``Dirichlet $p$-branes'', or ``D$p$-branes''. In the next section we will show how D-branes arise naturally through an inherent symmetry in string theory known as ``T-duality''. It is through the study of D-branes that we will get our first look at a correspondence between gauge theory and gravity.

\subsection{D-Branes}
\label{dbranes}

D-Branes are most easily understood in the context of a symmetry which exist within string theory, called ``T-duality''. Consider the closed bosonic string with embedding function $x^\m(\t,\s)$. Let us compactify the coordinate $x^{25}$ on a circle of radius $R$. The string can wind around the spacetime circle $w$ number of times, where $w\in\Z$ is the ``winding number'' such that
\be
x^{25}\sim x^{25}+2\pi R,
\ee
where $2\pi w R$ is the length of the string.

For the embedding to be single valued, it is also required that the centre of mass momentum in the $x^{25}$ direction be quantised such that
\be
p_0^{25}=\dfrac{n}{R},\quad n\in\Z.
\ee
These conditions modify the mass spectrum of the closed string, giving
\be
\label{mcomp}
m^2=-p^\m p_\m=\dfrac{n^2}{R^2}+\dfrac{w^2R^2}{\a'^2}+(\text{original oscillator terms}).
\ee
The first term in \eqref{mcomp} contributes a tower of Kaluza-Klein momentum states, whereas the second term is a purely stringy phenomenon contributing a tower of winding states. T-duality arises from examining what happens to these two terms as one considers the extreme limits of $R$. If we consider the limit as $R\tend\infty$, the winding states become infinitely heavy, and the momentum states and the $w=0$ state becomes a continuum. This is now the usual mass spectrum where we have decompactified the $x^{25}$ direction. Now consider the limit as $R\tend 0$. The momentum states become energetically unfavourable, but the winding states form a continuum, and we have an effective uncompactified dimension. Thus, the two limits of $R$ provide the same mass spectra. This can be made more precise by noting that the mass formula \eqref{mcomp} is invariant under the transformations
\be
n\leftrightarrow w\quad\text{and}\quad R\leftrightarrow R'\equiv\a'/R.
\ee
which is equivalent to a spacetime parity transformation on the right movers defined by
\be
\label{Tdual}
x^{25}(\t,\s)=x_L^{25}(\t+\s)+x_R^{25}(\t-\s)\longleftrightarrow x'^{25}(\t,\s)=x_L^{25}(\t+\s)-x_R^{25}(\t-\s).
\ee
This transformation is known as a ``T-duality'' and the theory compactified on the circle of radius $R'$ is known as the ``T-dual'' theory. Since all physical quantities, even after quantisation, are invariant under this transformation, the $R$ and the T-dual $R'$ theories are physically equivalent. T-duality is an exact quantum symmetry of perturbative closed string theory.

Now let us consider T-duality in the case of the bosonic open string. The first thing to note is that there is no equivalent of a winding number for open strings and so no winding states in the mass spectrum of equation \eqref{mcomp}. This means, in the limit as $R\tend 0$, open strings look like a quantum field theory in one less dimension. More precisely, we can again compactify the $x^{25}$ coordinate on a circle of radius $R$ such that
\be
x^{25}\sim x^{25}+2\pi R,
\ee
and so the momentum in the $x^{25}$ direction is again quantised such that $p_0^{25}=n/R,\,\,n\in\Z$. The embedding coordinate is given as a solution of the bulk string equation of motion \eqref{steom},
\be
x^\m(\t,\s)=x^\m(\t+\s)+x^\m(\t-\s),
\ee
where
\be
x^\m(\t\pm\s)=1/2(x_0^\m\pm x'^\m_0+2\a'p_0^\m(\t\pm\s))+(\text{oscillator terms}\propto e^{in(\t\pm\s)}).
\ee
The T-dualised coordinate $x'^{25}(\t,\s)$, compactified on a circle of radius $R'=\a'/R$, is found by using the spacetime parity transformation in \eqref{Tdual} to give
\be
x'^{25}(\t,\s)=x^{25}(\t+\s)-x^{25}(\t-\s)=x'^{25}_0+2\a'p_0^{25}\s+(\text{oscillator terms}\propto\sin(n\s)).
\ee
The first thing to note is that
\be
p'^{25}_0=\int^{2\pi}_0 d\s\dfrac{\pd x'^{25}}{\pd\t}=0,
\ee
and so the T-dual open string has no momentum in the $x'^{25}$ direction. The second point is that the oscillator terms vanish at the endpoints. Thus the usual Neumann boundary conditions for the original coordinate $\pd_\s x^{25}\mid_{\s=0,\pi}=0$ have been replaced by Dirichlet boundary conditions for the T-dualised coordinate $\pd_\t x'^{25}\mid_{\s=0,\pi}=\delta x'^{25}\mid_{\s=0,\pi}=0$. This fixes the separation of the endpoints
\be
x'^{25}(\t,\pi)-x'^{25}(\t,0)=\dfrac{2\pi\a'n}{R}=2\pi n R',
\ee
where $n\in\Z$ is defined as the open string winding number. The T-dual string endpoints are free to move in the extra $24+1$ dimensions and so defines a hyperplane called a D-brane. More generally, if one T-dualises $m$ spatial directions, the T-dual open string endpoints are fixed to a hyperplane with $p=25-m$ spatial dimensions called a D$p$-brane.

T-duality not only exists in bosonic string theory, but in superstring theory as well. Thus D-branes are natural objects which exist in superstring theory as has already been argued in the previous section. For the specific case of Type II superstrings, T-duality flips the chirality of the GSO projection since it is essentially a parity transformation. Thus, T-duality interchanges Type IIA and IIB superstring theories and explains part of figure \ref{Mtheory}. Now we can begin to describe D-branes more formally within superstring theory and explain their relevance for the later sections.
 
\begin{figure}[h!t]
\centerline{\includegraphics[width=8cm]{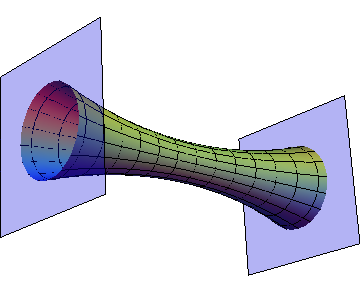}}
\caption{\label{wsduality}In this diagram we have a string worldsheet fixed between two D-branes. One can interpret this as an open string with endpoints fixed on each D-brane moving in a circle. Alternatively, one can see this as a closed string being produced on one brane and absorbed on the other brane. This alternative descriptions of the same object is due to the reparametrisation invariance of the worldsheet string action in equation \eqref{reparam}.} 
\end{figure}

A D$p$-brane is a ($p+1$)-dimensional hypersurface, where $p$ is the number of spatial directions, onto which end points of open strings attach. In analogy with the 2-dimensional worldsheet of string theory, D$p$-branes sweep out a ($p+1$)-dimensional worldvolume. It is specified by choosing Neumann boundary conditions parallel to the hypersurface and Dirichlet boundary conditions in the transverse directions,
\bea
\pd_\s x^\m\mid_{\s=0,\pi}&=&0\,,\,\m=0,1,\cdots,p\,,\nn
\delta x^\m\mid_{\s=0,\pi}&=&0\,,\,\m=p+1,\cdots,9\,.
\eea
In string perturbation theory, the position of the D$p$-brane is fixed by the boundary coordinates $x^{p+1},\cdots,x^9$ in spacetime, corresponding to a particular string theory background, while strings are free to move in the $x^0,\cdots,x^p$ directions. D$p$-branes couple to R-R potentials $C_{p+1}$ and are thus a source of closed strings. This can be seen via worldsheet duality (see fig \ref{wsduality}).
D-branes also have a non-perturbative description. This is because massless modes of the open string are associated with the fluctuation modes of the D-branes themselves, so non-perturbatively, D-branes become dynamical $p$-dimensional objects. We have seen that the massless open string spectrum contains the gauge field $A_\m(x)$. The gauge field decomposes into fields living on the brane, and those transverse to it. In particular, a 10-dimensional gauge field $A_\m(x),\,\m=0,1,\cdots,9$ splits into a $U(1)$ gauge field $A_a(x),\,a=0,1,\cdots,p$ living on the D$p$-brane, and scalar fields $\Phi^m(x),\,m=p+1,\cdots,9$ describing the fluctuations of the D$p$-brane in the $9-p$ transverse directions.

The low-energy dynamics of D-branes can be described in terms of a gauge theory on the worldvolume. One can write down an action (known as the ``Dirac-Born-Infeld'' (or DBI action) which describes the low-energy dynamics of a D$p$-brane in a flat background
\be
S_{\text{DBI}}=-\dfrac{T_p}{g_s}\int d^{p+1}\xi\ds\sqrt{-\det(\eta_{ab}+\pd_a x^m\pd_b x_m+2\pi\a'F_{ab})},
\ee
where $\xi^a=x^a,\,a=0,1,\cdots,p+1$ are the $p+1$-brane coordinates, $F_{ab}=\pd_aA_b-\pd_bA_a$ is the field strength of the gauge fields $A_a(\xi)$ living on the brane, and $T_p$ is the tension of the $p$-brane, given by,
\be
\label{ptension}
T_p=\dfrac{1}{\sqrt{\a'}}\dfrac{1}{(2\pi\sqrt{\a'})^p}
\ee
$S_{DBI}$ can be easily interpreted when one switches off gauge field fluctuations, in which case the DBI action becomes
\be
S_{\text{DBI}}\stackrel{A\tend 0}{=}-\dfrac{T_p}{g_s}\int d^{p+1}\xi\sqrt{-h}
\ee
where $h_{ab}=\eta_{\m\n}\pd_a x^\m\pd_b x^\n$ is the pull-back of the flat background metric to the D$p$-brane.

This is the Nambu-Goto action for strings extended to $p$-branes (see \eqref{N-G}) and so the DBI action can be geometrically interpreted as the Nambu-Goto action for $p$-branes with the addition of worldvolume gauge fields. It is a good low-energy approximation to the full D-brane dynamics in the static gauge and for slowly-varying field strengths $F_{ab}$. One can generalise the DBI action to include supergravity fields of the massless NS-NS closed string sector, thus representing D-branes living in a curved background as 
\be
S_{\text{DBI}}=-\dfrac{T_p}{g_s}\int d^{p+1}\xi e^{-\Phi}\sqrt{-\det(g_{ab}+B_{ab}+2\pi\a'F_{ab})},
\ee
where $g_{ab},\,B_{ab}$ are pull-backs of the spacetime supergravity fields to the D$p$-brane worldvolume.

The exact gauge theory that lives on the D$p$-brane can be found by expanding $S_{DBI}$ in the low-energy limit (in powers of $\a'$) and comparing the field content with a known gauge theory. In the case of a flat space background, the theory on the D$p$-brane is equivalent to the dimensional reduction to $p+1$ dimensions of $U(1)$ Yang-Mills theory in 10 spacetime dimensions, defined by the action
\be
\label{YM}
S_{YM}=-\dfrac{1}{4g_{YM}^2}\int d^{10}x\,F_{\m\n}F^{\m\n}.
\ee
One must take the Yang-Mills fields to depend on the $p+1$-brane coordinates $A_a=A_a(\xi)$, $A_m=\frac{1}{2\pi\a'}x^m(\xi)$, and identify the Yang-Mills coupling constant
\be
\label{pcoupling}
g_{YM}^2=g_sT_p^{-1}(2\pi\a')^2.
\ee
This result can be generalised to supersymmetric spacetimes with multiple D-branes, leading to the following result\cite{Witten2}:

\begin{Con}
\label{conj}
\it{The low-energy dynamics of $N$ parallel, coincident Dirichlet $p$-branes in flat space, is described in the static gauge by the dimensional reduction to $p+1$ dimensions, of $\N=1$ supersymmetric Yang-Mills theory (SYM), with gauge group $U(N)$ in ten spacetime dimensions.}
\end{Con}

This result makes the relationship between D-branes and gauge theory manifest. Since D-branes source closed strings (see fig \ref{wsduality}), and the low-energy dynamics of closed strings is described by a gravitational theory, this is the first indicator of a deeper relationship between gauge theory and gravity. We will see more evidence for this in the next sections which will eventually lead us to the AdS/CFT correspondence.

\section{Gauge Theories}

\subsection{Supersymmetric Yang-Mills Theories}

In this section, we describe the basic properties and symmetries of supersymmetric Yang-Mills (SYM) theories. As we have seen in \S\ref{dbranes}, the low-energy dynamics of D-branes are described by a gauge theory, in particular SYM theories. We will see in the later sections that a particular configuration of a stack of $N$ D3-branes gives rise to a particular SYM theory, namely $\N=4$ $SU(N)$ in $3+1$ dimensions. This will be the CFT side of AdS/CFT and so it will be useful to describe its basic properties. 

$\N=4$ SYM in four dimensions can be recovered from the ten dimensional action of $\N=1$ SYM with gauge group $U(N)$, given by
\be
\label{sym10}
S_{\text{SYM}}=\dfrac{1}{2g^2_{YM}}\int d^{10}x\,\text{Tr}\left(-F_{\m\n}F^{\m\n}+2i\bar{\psi}\Gamma^{\m}D_{\m}\psi\right),
\ee
where $\Gamma^\mu$ are $16\times 16$ Dirac matrices, 
\be F_{\m\n}=\pd_{\m}A_\n-\pd_{\n}A_\m-i[A_\m,A_\n] \ee is the field strength of the $U(N)$ gauge field $A_\m$, \be D_\m\psi=\pd_\m\psi-i[A_\m,\psi] \ee is the covariant derivative, and $\psi$ is a 16 component Majorana-Weyl spinor which transforms in the adjoint ($N\times N$) representation. This action preserves $16$ supersymmetries, which is the smallest algebra in $(9+1)-$dimensions.

Via dimensional reduction of \eqref{sym10} on a flat 6-dimensional torus $T^6$, we recover \cite{SYM1} a $(3+1)$-dimensional action. The ten-dimensional gauge field $A_\m$ reduces to a four-dimensional gauge field $A_a$ and six scalar fields $\phi_i$, and the Majorana spinor $\psi$ reduces to four 4-dimensional Weyl spinors $\lambda^a_\a$ where $a=1,2,3,4$. This is the field content of $\N=4$ SYM in $3+1$ dimensions with action given by
\bea
\label{N4}
S_{\N=4}=\int d^4x\,\text{Tr}\Big\{-\frac{1}{2g_{YM}^2}F^{ab}F_{ab}+\frac{\th_I}{8\pi^2}F^{ab}\tilde{F}_{ab}-i\bar{\lambda}\bar{\sigma}^a D_a\lambda\nn
-(D_a\phi_i)^2+\frac{g_{YM}^2}{2}\ds\sum_{i,j}\left[\phi_i,\phi_j\right]^2+\text{fermion-scalar terms}\Big\},
\eea

$\N=4$ SYM has a remarkable number of symmetries due to its maximal number of supersymmetries. Classically, it is Weyl (scale) invariant. It is also has $SO(1,3)$ Poincar\'e invariance. These two symmetries combine to form a larger \textit{conformal} symmetry with group $SO(2,4)\cong SU(2,2)$. The theory also has $\N=4$ Poincar\'e supersymmetry, which endows it with 16 Poincar\'e supercharges. Since the theory is also conformally invariant, it has an additional 16 conformal supersymmetries so the theory has 32 supercharges in total. The requirement of supersymmetry enlarges the algebra so that it is invariant under \textit{superconformal} transformations. Upon perturbative quantisation, $\N=4$ SYM exhibits no ultraviolet divergences. As a result, the renormalisation group $\beta$-function vanishes identically. Thus conformal symmetry is protected at the quantum level, which means that the theory is a conformal field theory (CFT).

The \textit{Montonen-Olive conjecture} states that the quantum theory of $\N=4$ SYM is invariant under a global discrete transformation. If we combine the Yang-Mills coupling $g_{YM}$ and the real instanton angle $\th_I$ into a single complex coupling
\be
\tau\equiv\dfrac{\th_I}{2\pi}+i\dfrac{4\pi}{g_{YM}^2},
\ee
then the quantum theory is invariant under the S-duality group $SL(2,\mathbb{Z})$, generated by
\be
\tau\tend\dfrac{a\tau+b}{c\tau+d},\quad ad-bc=1,\quad a,b,c,d\in\mathbb{Z}.
\ee
Finally, the theory exhibits an $R$-symmetry. This is a global symmetry which rotates the 6 scalars $\phi_i$, and thus has symmetry group $SO(6)_R\cong SU(4)_R$. This symmetry combines with the $SU(2,2)$ conformal symmetry group to form the supergroup $PSU(2,2|4)\supset SU(2,2)\times SU(4)_R$.

\subsection{Large $N$ Gauge Theories}
\label{largeN}

Gauge theories have been extremely useful in describing the physics of strong and electroweak interactions, as shown by the consistency of the Standard Model with experimental evidence. In particular, strong interactions are described in terms of an $SU(3)$ gauge theory (QCD). Gauge theories of this type have negative beta function, meaning that the theory suffers from confinement at low energies and asymptotic freedom at high energies. In the low energy regime one cannot do perturbation theory as the coupling is large, so it would be useful to search for another dimensionless parameter in which one can do perturbation theory at and below the QCD scale $\Lambda_{QCD}$ where the gauge theory may simplify. One such parameter is the integer number $N$ in $SU(N)$ gauge theories, where one can perform a perturbative expansion in powers of $1/N$ for large $N$.

Let us consider an $SU(N)$ pure Yang Mills theory with coupling $g_{YM}$. The beta function equation is given by
\be
\b(g_{YM})\equiv\m\dfrac{dg_{YM}}{d\m}=-\dfrac{11}{3}N\dfrac{g^3_{YM}}{16\pi^2}+\O(g^5_{YM}).
\ee
One can show that if we take the limit $N\tend\infty$, but keeping the quantity $\la\equiv g_{YM}^2N$ fixed, the order of the terms in the beta function does not change. This is known as the \textit{t' Hooft limit}, and $\la$ is known as the \textit{t'Hooft coupling}. We will now show that in this limit, gauge theories behave like string theories.

We have already seen from equation \eqref{YM} that the Yang Mills Lagrangian is of the form
\be
\L_{YM}\sim-\dfrac{1}{g_{YM}^2}\mathrm{Tr}(F^2)=-\dfrac{N}{\la}\mathrm{Tr}(F^2).
\ee
We can see from the Feynman rules for such a theory that each vertex has a coefficient proportional to $N/\la$, and each propagator has a $\la/N$ factor. The fermionic fields in this theory live in the adjoint  ($N\times N$ dimensional) representation of the gauge group, and so can be represented and a direct product of fields in the fundamental and anti-fundamental irreducible representations of $SU(N)$. Thus we draw the propagators in Feynman diagrams as two lines, carrying indexes in the fundamental and anti-fundamental irreps. Since each index has $N$ possible values, a closed loop will contribute a factor of $N$. Thus, a diagram with $E$ edges (propagators), $V$ vertices (interactions) and $F$ faces (closed loops), comes with a contribution
\be
N^{\chi}\la^{E-V},
\ee
where $\chi=F-E+V=2-2g$ is the Euler number of the closed oriented surface, and $g$ is the genus. So if we take the t'Hooft limit, the diagrams which dominate have genus zero with contribution of order $N^2$, and so lie on a sphere (plane). We call these \textit{planar diagrams}. World-sheet string perturbation theory tells us that all world-sheet diagrams for closed oriented strings are characterized by their Euler number, with contribution of order $g_s^{-\chi}$, where $g_s$ is the string coupling. Planar diagrams in perturbative string theory have a contribution $g_s^{-2}$. Thus we can identify perturbation theory of closed oriented strings and perturbation theory of $SU(N)$ gauge theories in the t'Hooft limit if we let $g_s\sim 1/N$. This relation is one of the strongest cases for believing that string theories and gauge theories are related.

In this section we have built up a lot of evidence for a connection between gauge theories and string theory, thus motivating the CFT side of AdS/CFT. In the next section we will be more specific about the AdS side of the correspondence, showing how a gravitational theory with AdS structure can arise from string theory. 

\section{Supergravity}

One important fact about superstring theory is that its low-energy dynamics can be described in terms of a local supersymmetric field theory, or ``supergravity'' theory. This can be motivated by looking at the massless spectra of superstring theories and comparing it with the field content of supergravity theories (SUGRA). Of particular importance to the AdS/CFT conjecture presented here is that the low energy limit ($\a'\tend 0$) of Type IIB superstring theory is Type IIB supergravity. The bosonic action for Type IIB superstring theory can be written in a similar way to the bosonic string frame action \eqref{bossf}
\bea
\label{2B}
S_{\mathrm{IIB}}&=&\dfrac{1}{(2\pi)^7l_s^8}\int d^{10}x\sqrt{-g}\Big(e^{-2\Phi}(R_g+4(\nabla\Phi)^2\nn
&&-\dfrac{1}{12}H\wedge H)-\dfrac{2}{(8-p)!}F^2_{p+2}-2(dK)^2\Big),
\eea
where $F_{p+2}$ is the field strength of the R-R $(p+1)$-form $C_{p+1}$ ($p=-1,1,3,5,7,9$), $H$ is the field strength of the NS-NS $B$ field, and $\Phi,\, K$ are the dilaton and the R-R scalar field (axion) respectively. The R-R four form $C_4$ is self-dual such that $F_5=*F_5$.

This action is invariant under the non-compact symmetry group $SL(2,\R)$. If we identify the axion-dilaton field as $\t\equiv K+ie^{-\Phi}$, then the action of the $SL(2,\R)$ group on $\t$ leads to a M\"{o}bius transformation of the form,
\be
\tau\tend\dfrac{a\tau+b}{c\tau+d},\quad ad-bc=1,\quad a,b,c,d\in\R
\ee
Referring to conjecture \ref{conj}, it is useful to consider geometric solutions of Type IIB SUGRA which describe the low-energy dynamics of $N$ coincident D$p$-branes in Type IIB superstring theory. We can then begin to relate a specific gauge theory (namely SYM) with a specific gravitational theory. One important fact about D-branes is that its Dirichlet boundary conditions are only invariant under half the background spacetime supersymmetries, meaning that D-branes are ``$1/2$-BPS'' states where the conserved charge is equal to the mass of the state. The solitonic geometries which correspond to such states are known as extremal black $p$-branes. They have conserved charge $N$ with respect to the R-R $(p+1)$-form $C_{p+1}$ equal to the mass $M$ of the black hole with horizon radius $r_+$. A spherically symmetric solution in $(10-p)$ dimensions with a R-R source at the origin, 
\be
\int_{S_{8-p}}\ast F_{p+2}=N,
\ee
is given by the metric
\be
\label{pbrane}
ds^2=H_p(r)^{-1/2}\eta_{\m\n}\,dx^\m dx^\n+H_p(r)^{1/2}(dr^2+r^2d\Om^2_{8-p}),
\ee
where $\m,\n=0,\cdots,p,$ and,
\bea
e^{2\Phi}&=&g_s^2H_p(r)^{\frac{3-p}{2}},\nn
H_p(r)&=&1+\left(\dfrac{r_+}{r}\right)^{7-p},\nn
r_+^{7-p}&=&d_pg_sNl_s^{7-p},\nn
C_{p+1}&=&g_s^{-1}(1-H_p(r)^{-1})\,dx^0\wedge\cdots\wedge dx^p.
\eea
$g_s$ is the asymptotic string coupling constant and $d_p$ is the numerical factor,
\be
d_p=2^{5-p}\pi^{\frac{5-p}{2}}\,\G\left(\frac{7-p}{2}\right).
\ee
For $p>6$, there is no curvature singularity. When $3<p\leq6$, the horizon and the singularity coincide at $r=0$, since the radius of the $S^{8-p}$ vanishes there, and so there is a null singularity. In the special case of $p=3$, the horizon has finite size at $r=0$, where $r_+^4=4\pi g_s\a'^2N$, since the $S^5$ does not vanish. In the limit where $r\ll r_+$, and so $H_3(r)\sim (r_+/r)^4$, the metric in equation \eqref{pbrane} reduces to
\be
ds^2=\dfrac{r^2}{r_+^2}\eta_{\m\n}dx^\m dx^\n+\dfrac{r_+^2}{r^2}dr^2+r_+^2d\Om^2_5,
\ee
where $\m,\n=1,\cdots,4$.

This metric looks singular at $r=0$, but via a change of coordinates $z=r^2_+/r$, the metric becomes
\be
\label{ads5s5}
ds^2=r^2_+\left(\dfrac{\eta_{\m\n}dx^\m dx^\n+dz^2}{z^2}+d\Om^2_5\right).
\ee
We see that this is the geometry of $AdS_5\times S^5$ in Poincar\'{e} coordinates if we identify $r_+$ with the AdS radius $R$. 

\begin{figure}[h!t]
\centering
\begin{overpic}[trim = 0mm 30mm 0mm 0mm, clip, width=10cm]{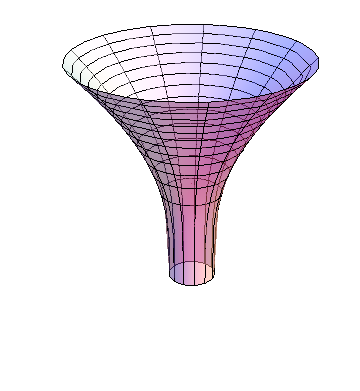}
\thicklines
\put(10,5){\color{black}\vector(0,1){75}}
\put(15,6){$r=0$}
\end{overpic}
\caption{\label{deepthroat}This is a graphical representation of how a stack of D3-branes back-reacts on the target spacetime, resulting in a ``deep throat'' geometry. Inside the throat , the horizon radius is kept constant by the finite size of the $S^5$ (represented here by an $S^1$), leading to the $AdS_5\times S^5$ geometry in equation \eqref{ads5s5}. Away from the throat at asymptotic infinity the space is flat, and so we have a Minkowski $\R^{1,9}$ geometry.}
\end{figure}

Alternatively, if we take the the asymptotic limit where $r\gg r_+$, and so $H_3(r)\sim 1$, we see that the extremal 3-brane solution reduces to Minkowski space in 10-dimensions. Thus, the stack of D3-branes back-reacts on the spacetime to provide us with a ``deep throat'' geometry which has finite size where the branes are located (see figure \ref{deepthroat}).

The classical supergravity description of the black $p$-brane is only valid when stringy effects are small, i.e. when $l_s\ll r_+$, since $r_+$ sets the size of the black hole. To suppress string loop corrections, we also need the effective string coupling $e^\Phi$ to be small. When $p=3$ the dilaton is constant everywhere and so the string coupling $g_s$ can be set less than one through the entire spacetime, and thus $l_P<l_s$. Thus, using the definition of $r_+$, the supergravity approximation for $p=3$ is valid when
\be
\label{largela}
1\ll\la<N,
\ee
where $\la=g_sN$ is the t'Hooft coupling.

\section{Anti-de Sitter Spacetimes}
\label{ads}
AdS space was mentioned in the previous section as a particular limit of the geometry of a stack of D3-branes. As different coordinate patches of AdS space will be used throughout this and the remaining chapters, it will be useful to explain its basic properties before we proceed to examining AdS/CFT in its full glory.

Anti-de Sitter spacetimes (or AdS spaces) are maximally symmetric solutions of Einstein's equations with negative cosmological constant $\Lambda$. This requires that the Riemann tensor $R_{\m\n\rho\s}$ satisfies
\be
R_{\m\n\rho\s}\propto g_{\m\rho}g_{\n\s}-g_{\m\s}g_{\n\rho},
\ee
where $g_{\m\n}$ is the AdS metric and is a solution to
\be
R_{\m\n}-\frac{1}{2}Rg_{\m\n}+\Lambda g_{\m\n}=0,\quad\text{where}\quad\Lambda=-\dfrac{d(d+1)}{2R^2}.
\ee
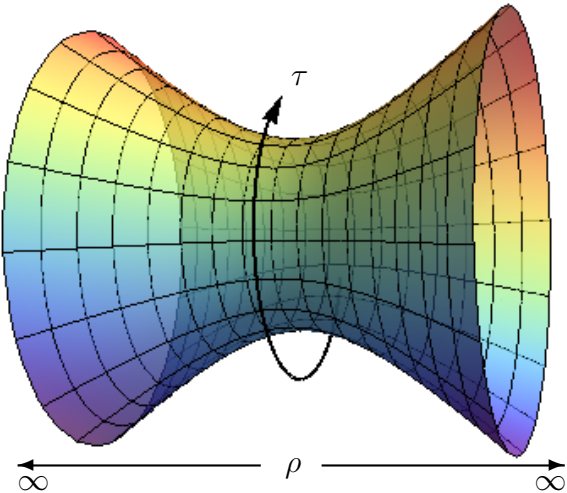
\begin{figure}[h!t]
\centering
\begin{overpic}[trim = 0mm 5mm 0mm 25mm, clip, width=8cm]{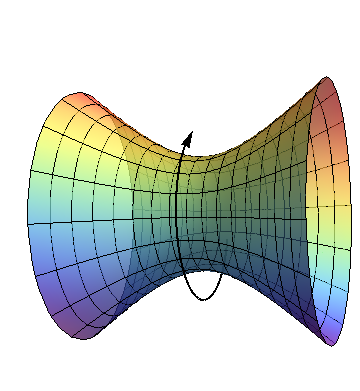}
\thicklines
\put(60,7){\color{black}\vector(1,0){40}}
\put(50,7){\color{black}\vector(-1,0){40}}
\put(54,6){$\rho$}
\put(95,3){$\infty$}
\put(10,3){$\infty$}
\put(55,70){$\tau$}
\end{overpic}
\caption{\label{AdS1}This is $AdS_2$ in global coordinates given in \eqref{sads}. The geometry is that of the hyperboloid $\mathbb{H}^{1,1}$ embedded in $\R^{2,1}$. Closed timelike curves can be removed by decompactifying the time direction $\t$ and working with the universal covering space $CAdS_2$.}
\end{figure}
Anti-de Sitter space in $d+2$ dimensions ($AdS_{d+2}$) has a natural geometric representation as a Lorentzian hyperboloid $\mathbb{H}^{1,d+1}$ embedded in the flat spacetime $\mathbb{R}^{2,d+1}$ in $d+3$ dimensions with signature $(-,-,+,+,+,\cdots)$.

The flat spacetime $\mathbb{R}^{2,d+1}$, with coordinates $\{X_{-1},X_0,X_1,\cdots,X_{d+1}\}$, has the metric
\be
\label{mads}
ds^2=-dX_{-1}^2-dX_0^2+\sum_{i=1}^{d+1}dX^2_i.
\ee
The spacetime is homogeneous and isotropic with isometry group $SO(2,d+1)$.

The hyperboloid $\mathbb{H}^{1,d+1}\subset\mathbb{R}^{2,d+1}$ is defined by the condition
\be
\label{cads}
-X_{-1}^2-X_0^2+\sum_{i=1}^{d+1}X^2_i=-R^2,
\ee
where $R$ is defined as the $AdS$ radius.

To find the induced metric on $\mathbb{H}^{1,d+1}$, we find a solution to \eqref{cads} as an embedding of \eqref{mads}. One such solution which covers the entire hyperboloid once is known as ``global'' coordinates on $AdS$. These are defined as
\bea
\label{sads}
X_{-1}&=&R\,\cosh\rho\,\cos\tau,\nn
X_0&=&R\,\cosh\rho\,\sin\tau,\nn
X_1&=&R\,\sinh\rho\,\cos\th_1,\nn
X_{1<N<d+1}&=&R\,\sinh\rho\,\cos\th_N\prod_{i=1}^{N-1}\sin\th_i,\nn
X_{d+1}&=&R\,\sinh\rho\,\prod_{i=1}^d\sin\th_i.
\eea
This set of equations satisfies equation \eqref{cads} by iterative use of the trigonometric and hyperbolic identities $\sin^2\th+\cos^2\th=1$ and $\cosh^2\rho-\sinh^2\rho=1$.

By substituting the set of equations \eqref{sads} into \eqref{mads}, we find the metric on $AdS_{d+2}$ in global coordinates is given by
\be
\label{gads1}
ds^2=R^2(-\cosh^2\rho\,d\tau^2+d\rho^2+\sinh^2\rho\,d\Om_d^2),
\ee
where\footnote{Since the choice of coordinates identifies $\tau=0$ and $\tau=2\pi$, $AdS_{d+2}$ has closed timelike curves. These can be removed by passing to the universal covering space $CAdS_{d+2}$ where $\tau$ is decompactified to cover the whole real line. This process does not change the metric, but gives it an $SO(d)\times\mathbb{R}_{\tau}$ symmetry} $0\leq\rho<\infty$, $0\leq\tau\leq 2\pi$, and $d\Om_d^2$ is the metric on the Euclidean $d$-sphere $S^d\subset\mathbb{R}^{d+1}$. This space is represented for $d=0$ in figure \ref{AdS1}.
Via a change of coordinates $r=R\,\sinh\rho$, the metric in \eqref{gads1} can be rewritten\footnote{From this point on we will consider $AdS$ space to be its universal covering space $CAdS$. As such we will only consider the decompactified time coordinate $t$.} as
\be
\label{gads2}
ds^2=-\left(1+\dfrac{r^2}{R^2}\right)d t^2+\left(1+\dfrac{r^2}{R^2}\right)^{-1}dr^2+r^2\,d\Om_d^2,
\ee
where the boundary of this spacetime is at $r=\infty$. It is this coordinate patch of the metric which will be used in chapters 3 \& 4.

Via a different change of coordinates, $\cosh\rho=\frac{1+y^2}{1-y^2}$, the metric in \eqref{gads1} is converted to
\be
\label{adsball}
ds^2=\dfrac{R^2}{(1-y^2)^2}\left\{-(1+y^2)^2dt^2+4dy^2+4y^2d\Om^2_d\right\},
\ee
where $0\leq y<1$ with boundary at $y=1$.

It is often convenient to make the coordinate transformation $r=R\tan\chi$ with $\chi\in [0,2\pi]$. The metric in \eqref{gads2} then becomes:
\be
\label{compads}
ds^2=\dfrac{R^2}{\cos^2\chi}\left(-dt^2+d\chi^2+\sin^2\chi d\Om_d^2\right)
\ee
\begin{figure}[h!t]
\centering
\begin{overpic}[width=7cm]{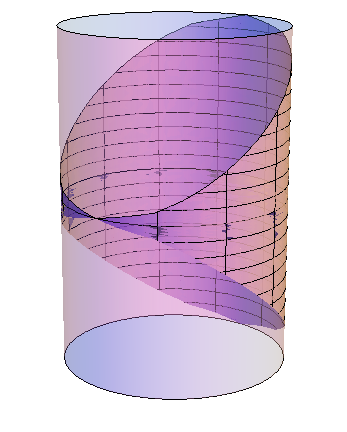}
\thicklines
\put(5,15){\color{black}\vector(0,1){80}}
\put(1,90){$t$}
\end{overpic}

\caption{\label{AdS2}$AdS_{d+2}$ can be viewed as a ``soup-can'' with conformal timelike boundary $S^d\times\R$. Poincar\'{e} coordinates only cover half of $AdS_{d+2}$, which can be seen here for $AdS_3$ as the space in-between the two slices, known as a Poincar\'{e} wedge. The boundary in these coordinates is Minkowski space, which can be viewed here for $\R^{1,1}$ as a handkerchief pinched at two corners, thus wrapping the Poincar\'{e} wedge. The second half of $AdS_{d+2}$ can be covered by labelling the hyperboloid as in \eqref{pads} but now letting $-\infty<z\leq 0$. These two patches together cover $AdS_{d+2}$, while to cover $CAdS_{d+2}$ one assembles a vertical tower of such patches.}
\end{figure}
Thus $AdS_{d+2}$ has the topology of an $d+1$-dimensional sphere times the real line or $S^{d+1}\times\R$. The boundary of the spacetime at $\chi=\pi/2$ is timelike and has topology $\mathbb{R}\times S^d$, which is otherwise known as the Einstein static universe in $d+1$ dimensions ($ESU_{d+1}$). Thus $AdS_{d+2}$ maps to half of $ESU_{d+2}$. In general, if a spacetime can be conformally compactified into a region which has the same boundary structure as one half of the Einstein static universe, the spacetime is called \textit{asymptotically AdS}.

Another solution of \eqref{cads} is given by the set of equations,
\bea
\label{pads}
X_{-1}&=&\dfrac{1}{2z}(z^2+\vec{x}^2-t^2+R^2),\quad X_0=R\dfrac{t}{z}\nn
X_{d+1}&=&\dfrac{1}{2z}(z^2+\vec{x}^2-t^2-R^2),\quad X_{0<i<d+1}=R\dfrac{x_i}{z},
\eea
which gives a metric of the form,
\be
ds^2=R^2\left(\dfrac{-dt^2+dz^2+d\vec{x}^2}{z^2}\right),
\ee
where $\vec{x}$ is a vector in $\mathbb{R}^d$, $t,x_i\in\mathbb{R}$ and $z\in [0,\infty)$.

These are known as Poincar\'e coordinates or local coordinates. Due to the range of the $z$ coordinate, they only cover one half of the hyperboloid and thus one half of $AdS_{d+2}$ (see fig. \ref{AdS2}). The boundary at $z=0$ is conformal to flat Minkowski space $\mathbb{R}^{1,d}$. However, the boundary only covers half of the cylindrical boundary of global $AdS_{d+2}$. It is this coordinate patch of the metric which will be considered in chapter 5.

\section{The AdS/CFT Correspondence}

We have now built up all the machinery to describe the main example of the AdS/CFT correspondence as proposed by Maldacena\cite{Maldacena}, relating a five dimensional bulk gravity theory to a four dimensional boundary CFT. Other such examples in different dimensions include: $AdS_3/CFT_2$ (see \cite{Maldacena,ads3cft2}); $AdS_4/CFT_3$, where one considers a stack of M2-branes (the main examples of which are ABJM theories \cite{ABJM}); and $AdS_7/CFT_6$, where one considers a stack of M5-branes \cite{M2M5}.  

Consider $N$ parallel coincident D3 branes extended in ($3+1$) dimensions in a ($9+1$) dimensional spacetime. The string theory consists of open strings ending on the branes (they can end on the same brane, on different branes, or can be open-ended) describing excitations of the D-branes, and closed strings describing excitations in the bulk. In the low energy limit ($\a'\tend 0$), only massless string modes exist and so we can write an effective action given by
\be
S=S_{\text{bulk}}+S_{\text{brane}}+S_{\text{int}}.
\ee
In the low energy limit, open strings and closed strings decouple and interactions between the brane and the bulk disappear. This means that the interaction term in the effective action $S_{\text{int}}$ vanishes. $S_{\text{brane}}$ is the effective action on the the stack of D3 branes and so, in a similar way to conjecture \ref{conj}, is just the action of $\N=4\,SU(N)$ SYM theory in $3+1$-dimensions given in equation \eqref{N4}. $S_{\text{bulk}}$ is the effective action of the bulk, and so we only have massless NS-NS field excitations, meaning the action is given by SUGRA in a $(9+1)$-dimensional Minkowski background. This is equivalent to \eqref{2B} with the massless R-R fields turned off.

We can now compare this to the extremal $3$-brane description in the low energy limit. We see that the theory decouples to an $AdS_5\times S^5$ geometry close to $r=0$ (the location of the D3 branes) and SUGRA in $9+1$-dimensional Minkowski space for large $r$. Since the two decoupled theories in the asymptotic region are equivalent, one can make the conjecture that
\begin{Con}
\label{AdS/CFT}
\it{Type IIB superstring theory with $\mathrm{AdS}_5\times \mathrm{S}^5$ boundary conditions and string coupling $g_s$, where both $AdS_5$ and $S^5$ have the same radius $R$, and where the 5-form $F_5$ has integer flux $N=\int_{S^5}F_5$, is equivalent to $\mathcal{N}=4\,SU(N)$ super-Yang-Mills theory in $3+1$ dimensions with Yang-Mills coupling $g_{YM}$.}
\end{Con}

Using equations \eqref{ptension} and \eqref{pcoupling}, and taking $p=3$, we see that the relationship between the couplings of the two theories is given by
\be
g^2_{YM}=2\pi g_s.
\ee
We also have the following relationships between the physical parameters
\be
R^4=4\pi\a'^2g_sN,\quad\langle K\rangle=\th_I,
\ee
where $\langle K\rangle$ is the axion vev given in equation \eqref{2B}, and $\th_I$ is the real instanton angle given in equation \eqref{N4}.

The theories are ``equivalent'' in the sense that there is a precise relationship between the physical observables of both theories, not in the sense that it has been proven (despite concentrated effort). The fact that we have taken the low-energy limit does not mean this conjecture would not hold for massive stringy excitations, since the limit is taken for an observer at infinity, and so high energy excitations would be heavily red-shifted. This equivalence provides unexpected connections between seemingly different structures, and the absence of a proof suggests that there are still new insights to be found. However, we can still look for necessary (but not sufficient) conditions of the validity of the correspondence. The first of these is outlined in \S\ref{mapsym}. 

\subsection{Mapping the Symmetries}
\label{mapsym}

A necessary check of the correspondence is that the global symmetries of the two theories must match. We have already identified certain symmetries of $\N=4$ SYM. Firstly the theory is invariant under the conformal symmetry $SO(2,4)$. This can be identified with the $SO(2,4)$ isometry group of $AdS_5$. Since conjecture \ref{AdS/CFT} refers to string theory (gravity) with $AdS_5\times S^5$ boundary conditions, one can have any number of geometric or topological deformations in the bulk spacetime. As such, the only possible $SO(2,4)$-invariant relation between $AdS_5\times S^5$ and the background $3+1$ Minkowski space ($M_4$) of $\N=4$ SYM is to make the identification $\pd AdS_5\equiv M_4$. Thus one can think of the field theory as ``living of the boundary'' of the bulk spacetime containing gravity.

$\N=4$ SYM is also invariant under the R-symmetry $SO(6)$. This can immediately be identified with the rotational symmetry group of $S^5$. We also note that $\N=4$ SYM has 32 supersymmetries, which is exactly the number of Killing spinors of the $AdS_5$ supergroup. Finally, $\N=4$ SYM has a discrete $SL(2,\Z)$ global symmetry from the Montonen-Olive conjecture. We recall that type IIB SUGRA has an $SL(2,\R)$ symmetry of the axion-dilaton field. However, the correspondence pertains to the fully quantised type IIB superstring theory. In this scenario, the axion expectation value is quantised such that $\langle K\rangle\in\Z$, and so $\t\in\Z$. Therefore, the allowed M\"{o}bius transformations must be elements of $SL(2,\Z)\subset SL(2,\R)$ which matches the symmetry group of the Montonen-Olive duality.

Apart from matching the global symmetries, another non-trivial check is matching the spectra of the two theories. It was shown (see \cite{MAGOO} for a review) that there is a one-to-one correspondence between the spectrum of chiral primary operators in $\N=4$ SYM and the spectrum of fields in Type IIB superstring theory on $AdS_5\times S^5$.

\subsection{Limits of the Correspondence}
\label{limit}

Conjecture \ref{AdS/CFT} is referred to as the ``strong form'', as it holds for all values of $N$ and $g_s$ and so is non-perturbative. However, quantising string theory is notoriously hard on curved backgrounds such as $AdS_5\times S^5$. So it is useful to look at limits of the correspondence where one can perform calculations. We do this by looking at different limits of the 't Hooft coupling $\la\equiv g_{YM}^2N=2\pi g_sN$ which was identified as a natural parameter of gauge theories in the previous section.\\

\emph{$\la$ fixed, $N\tend\infty$:}

This is known as the 't Hooft limit. We have seen (see \S\ref{largeN}) that this limit allows a $1/N$ expansion of $\N=4$ SYM. Since $1/N\sim g_s\tend 0$ , but $\la\sim R/l_s$ is kept fixed, the other side of the correspondence is classical Type IIB string theory.\\

\emph{$\la$ large, $N\tend\infty$:}

This is the large $\la$ limit which gives a $1/\sqrt{\la}$ expansion of $\N=4$ SYM. This corresponds to large $R/l_s$ and so stringy effects become negligible. Thus in this limit, which was given in \eqref{largela}, we have Type IIB supergravity. It is in this limit that the following chapters will be most concerned with as one can use the technology of classical GR to provide insight into strongly coupled gauge theories.

The next sections will be concerned with the properties and applications of the correspondence in the various limits outlined above. This will eventually lead us to our two main results outlined at the start of this chapter.

\subsection{The Holographic Principle}

One of the most important aspects of the AdS/CFT correspondence is that it is ``holographic'', in the sense that it obeys the holographic principle. In this section we explain what the holographic principle is, why it is important and what it has to do with AdS/CFT.

The holographic principle\cite{Hol1,Hol2,Hol3,Hol4} arose by asking questions of what happens when one studies the physics of black holes using quantum mechanics. Hawking and Bekenstein raised deep questions about the nature of black holes by considering what happens to information that falls into a black hole. Hawking argued that information is irretrievably lost when matter falls behind the horizon of the black hole\cite{Haw1,Haw2}. This led to the idea that black holes are thermodynamic objects with specific temperature $T_H$ and entropy $S_{BH}$ given by\cite{Bek1,Bek2,Bek3}
\be
\label{TH}
T_H=\dfrac{\kappa}{2\pi}\quad\text{and}\quad S_{BH}=\dfrac{A}{4},
\ee
where $\kappa$ is the surface gravity of the black hole and $A$ is its horizon area.

Consider a gravitating system in a region of space $\G$ with boundary $\pd\G=A$. Consider a thermodynamic system within $\G$. The maximum amount of mass that can be contained within $\G$ is the mass of a black hole of horizon area $A$. Thus from the second law of thermodynamics it must be the case that 
\be
\label{seb}
S\leq S_{BH}=\dfrac{A}{4},
\ee
and so the maximum entropy of $\G$ is proportional to the area of $\pd\G$ measured in Planck units. This is called the spherical entropy bound. Bousso\cite{Hol4,Hol9} generalised this bound by making it covariant as follows:

\textit{Let $A(B)$ be the area of an arbitrary $d-1$ dimensional spatial surface $B$ (which need not be closed). A $d$ dimensional hypersurface $L(B)$ is called a light-sheet of $B$ if $L$ is generated by light rays which begin at $B$, extend orthogonally away from $B$ to a caustic (where the light rays intersect). Let $S$ be the entropy on any light-sheet of $B$. Then}
\be
\label{bbound}
S[L(B)]\leq\dfrac{A(B)}{4}.
\ee
Thermodynamic entropy has a statistical interpretation in terms of measuring the lack of knowledge we have about the details of the microscopic theory given a macroscopic theory. Therefore the covariant entropy bound expresses the fact that information contained within $L(B)$ is somehow related to information on the boundary $B$. Using Boltzmann's formula for entropy, $S=\ln\,\N$, to find the number of states $\N$, the covariant entropy bound can be recast (using \cite{Hol9}) as:\\

\textbf{The Holographic principle}

\textit{The covariant entropy bound is a law of physics which must be manifest in an underlying theory. This theory must be a unified quantum theory of matter and spacetime. From it, Lorentzian geometries and their matter content must emerge in such a way that the number of independent quantum states describing the light-sheets of any surface $B$ is manifestly bounded by the exponential of the surface area}:
\be
\label{holpr}
\N[L(B)]\leq e^{A(B)/4}.
\ee
This contradicts the intuition that degrees of freedom are local in space, and so complexity grows with volume. The holographic principle implies a radical reduction in the number of degrees of freedom we use to describe nature. It exposes quantum field theory, which has degrees of freedom at every point in space, as a highly redundant effective description, in which the true number of degrees of freedom is obscured.

The most concrete example of a holographic theory to date (within a limited set of spacetimes) is the AdS/CFT correspondence. Anti-de Sitter spacetimes have a natural boundary on which holographic data can be stored and evolved forward using a conformal field theory.

To describe the holographic nature of this correspondence, it will be useful to consider the following form of the metric on $AdS_5\times S^5$ (see \eqref{adsball}),
\be
ds^2=\dfrac{R^2}{(1-r^2)^2}\left\{-(1+r^2)^2dt^2+4dr^2+4r^2d\Om^2_3\right\}+R^2d\Om^2_5,
\ee
The radius of curvature $R$, is related to the flux $N$ (defined in \ref{AdS/CFT}), by the formula
\be
\label{RtoN}
R=N^{1/4},
\ee
in Planck units.

If we apply a conformal rescaling by a factor of $R^2/(1-r^2)^2$ on our metric, the new metric becomes
\be
ds^2=\left\{-(1+r^2)^2dt^2+4dr^2+4r^2d\Om^2_3\right\}+(1-r^2)^2d\Om_5^2.
\ee
This has the topology of a $4$-dimensional open ball times the real time axis times the five-sphere. As we approach the boundary at $r=1$, the 5-sphere shrinks to zero, leaving us with a $3+1$ dimensional boundary with topology $\mathbb{R}\times S^3$. This is the conformal boundary of $AdS_5\times S^5$. We observe that this is the same dimensionality of the spacetime in which the dual CFT resides. Since the CFT is conformal, and complementing the arguments in \S\ref{adscftint} and \S\ref{mapsym}, this is another way of saying that the dual CFT ``lives'' on the boundary of $AdS_5\times S^5$.

This property of AdS/CFT, namely that boundary data completely describes all physics in the interior, is suggestive of the holographic principle. But to check if this is indeed the case, one need to compute the CFT's degrees of freedom $N_{dof}$, which must not exceed the boundary area $A$.

The proper area of the boundary of $AdS_5$ is divergent, as is the degrees of freedom of a conformal field theory due to scale invariance. Thus to compare these two quantities, one need to find a natural regulator. This was achieved in \cite{Hol5} by making use of the following fact about AdS/CFT: Infra-red effects in AdS space correspond to ultraviolet effects in the boundary theory. This relation is known as the \textit{UV/IR connection}, and is key to the information bound that is an important part the holographic principle.

In AdS space the boundary is at an infinite proper distance from the bulk, but light can reach the boundary in finite time. It can be shown that a null geodesic going form $r=0$ to $r=1$ and back again does it in time $t=\pi R$. We can then make the approximation that AdS space behaves like a finite cavity of size $\O(R)$. With this assumption we can now count the degrees of freedom of such a space. In order to do this we need to regulate the boundary of AdS by introducing a cut-off surface $L$ with topology $S^3\times S^5$ at $r=1-\delta$. Then the area $A$ of $L$ is approximately
\be
A\sim\dfrac{R^3}{\delta^3}R^5
\ee
To count the degrees of freedom of the CFT on the boundary, we note from the UV/IR connection that the infrared cut-off $\delta$ in the bulk corresponds to an ultraviolet cut-off in the CFT. We have argued that the CFT lives on a spacial 3-sphere, and so $\delta$ partitions the sphere into $\delta^{-3}$ cells. A $U(N)$ gauge theory has roughly $N^2$ independent quantum fields, and assuming one stores one bit of information per cell, per quantum field, the total number of degrees of freedom is given by
\be
N_{dof}\sim\dfrac{N^2}{\delta^3}
\ee
Using \eqref{RtoN}, we find that the CFT number of degrees of freedom saturates the holographic bound,
\be
N_{dof}\sim A
\ee

\subsection{The Field $\leftrightarrow$ Operator Correspondence}

Witten\cite{Witten} gave a precise description of the AdS/CFT correspondence by relating the physical observables of the two theories.

Consider a scalar field $\phi$ propagating in 10-dimensional $AdS_5\times S^5$ space. We can decompose this field via Kaluza-Klein reduction on $S^5$, so effectively all fields $\phi_\D(\vec{z},z)$, where $\vec{z}=(t,\vec{x})$, are on $AdS_5$ (in Poincar\'{e} coordinates) 
\be
\p(\vec{z},z,y)=\sum_\D\phi_\D(\vec{z},z)\,Y_\D(y),
\ee
where $Y_\D(y)$ are the spherical harmonics on $S^5$ labelled by the rank $\D$ of the totally symmetric traceless representations of $SO(6)$.

The action of the massive scalar field is given by
\be
S_{\p_\D}=\int dz\,d^4\vec{z}\sqrt{g}\left(\dfrac{1}{2}g^{\m\n}\pd_\m\p_\D\pd_\n\p_\D+\dfrac{1}{2}m_\D^2\p^2_\D\right).
\ee
Close to the boundary, bulk interaction terms disappear, and so the dynamics as $z\tend 0$ are given by the Klein-Gordon equation
\be
\label{KG}
(\nabla^2+m_\D^2)\phi_\D=0\quad\text{where}\,\,m^2_\D=\D(\D-4).
\ee
The boundary fields $\phi_\D^0$ in the full interacting theory are associated with the non-normalizable solutions to \eqref{KG} (see \cite{Witten} for more detail) and are thus defined by
\be
\label{fBC}
\phi_\D^0(\vec{z})=\lim_{z\tend 0}z^{4-\D}\phi_\D(\vec{z},z).
\ee
To evolve data such as $\phi_\D$ on a spacetime, the Cauchy problem must be well defined. I.e. given $\p_\D$ on some spacelike hypersurface at $t=t_0$, can one determine $\p_\D$ on some spacelike hypersurface at $t>t_0$? This is not the case for AdS space due to its timelike boundary. Thus one must add boundary conditions for $\phi_\D\forall t$. This can be achieved by adding a term to the Lagrangian of $\N=4$ SYM which includes the boundary fields $\phi_\D^0$. These fields should couple to the dynamical observables of the theory, which are the local gauge invariant CFT operators $\O_\D$ giving a term $\int d^4\vec{z}\,\phi^0_\D(\vec{z})\O_\D(\vec{z})$.

Bulk fields sourced by such a boundary term are given by
\be
\p_\D(\vec{z},z)=\int d^4\vec{z}K_\D(\vec{z},z,\vec{u})\p_\D^0(\vec{u}),
\ee
where the bulk-to-boundary propagator is given by
\be
K_\D(z,\vec{z},\vec{u})=C_\D\left(\dfrac{z}{z^2+|\vec{z}-\vec{u}|^2}\right)^\D,
\ee
with
\be
C_\D=\dfrac{\G(\D)}{\pi^2\G(\D-2)}.
\ee
One can define a generating functional $W[\phi^0_\D]$ for all the correlators of $\O_\D$ in terms of the source fields $\phi_\D^0$,
\be
\exp(-W[\phi^0_\D])\equiv\left\langle\exp\left(\int d^4\vec{z}\phi^0_\D(\vec{z})\O_\D(\vec{z})\right)\right\rangle,
\ee
then one can define a map between observables for any operators/fields via the Witten prescription,
\be
\label{f-O}
\exp(-W[\phi^0_\D])=Z_{\text{string}}[\phi_\D\rvert_{z=0}=\phi_\D^0],
\ee
where the right hand side is the full string theory partition function whose boundary fields are defined by \eqref{fBC}.

In the large $\la$ limit, stringy effects are ignored, and one can approximate the string partition function by $\exp(-S_{\text{SUGRA}})$, where $S_{\text{SUGRA}}$ is the supergravity action, and so \eqref{f-O} becomes,
\be
W[\phi^0_\D]\simeq\text{extr}\,S_{\text{SUGRA}}[\phi_\D].
\ee
One can apply this prescription to obtain the 2-point correlation function of operators $\O_\D(\vec{x})$ in $\N=4$ SYM. More specifically we use the AdS bulk-to-boundary propagator and extract the $z^\D$ behaviour as $z\tend 0$, giving
\be
\label{2pfn}
\langle\O(\vec{z})\O(\vec{u})\rangle=\lim_{z\tend 0}z^{-\D}K_\D(\vec{z},z,\vec{u})=C_\D(2\D-2)\dfrac{1}{|\vec{z}-\vec{u}|^{2\D}}.
\ee
This result is what one would expect from the conformal invariance of the CFT with operators having conformal dimension $\D$.

\subsection{The Correspondence at Finite Temperature}
\label{finiteT}

We arrived at the correspondence between $\N=4$ SYM and gravity in $AdS_5\times S^5$ by taking the near horizon limit of the extremal D3-brane solution. If one applies the same procedure to the non-extremal case we find the solution is at finite temperature.

If we consider the action in equation \eqref{2B}, find a non-extremal spherically symmetric solution, and take the near-horizon limit $r\ll r_\pm$ at $p=3$. Then the resulting metric, written in AdS Poincar\'{e} coordinates, is given by
\be
\label{pBH}
ds^2=R^2\left(\dfrac{-h(z)\,dt^2+h(z)^{-1}dz^2+d\vec{x}^2}{z^2}+d\Om_5^2\right),
\ee
where
\be
h(z)=1-(z/z_+)^4,\quad z_+=(\pi T)^{-1}.
\ee
This is the planar black hole in $AdS_5$ times a five-sphere. It is a particular limit of the AdS-Schwarzschild black hole (AdSBH), which we shall now describe. 

The AdSBH is a $d\geq 3$ dimensional, two-parameter family of solutions, given by the size of the black hole (the horizon radius $r_+$) and the AdS radius $R$. The metric for AdSBH in $d+2$ dimensions, with topology $\R^2\times S^{d-1}$, is written in global AdS coordinates by,
\be
\label{AdSBH}
ds^2=-f(r)\,dt^2+\dfrac{dr^2}{f(r)}+r^2d\Om^2_d,\quad\text{where}\quad f(r)=1+\dfrac{r^2}{R^2}-\dfrac{r_+^{d-1}}{r^{d-1}}\left(1+\dfrac{r_+^2}{R^2}\right),
\ee
which solves the vacuum Einstein equations with a negative cosmological constant
\be
R_{\m\n}=\La g_{\m\n},\quad \La=-\dfrac{d(d+1)}{2R^2},
\ee
and has mass (in Planck units)
\be
M=\text{Vol}(S^d)\dfrac{r_+^d(d-1)}{16\pi}\left(1+\dfrac{r_+^2}{R^2}\right).
\ee
Black holes are thermodynamic objects, and the AdSBH is no different. Via the usual procedure of Wick rotating to a Euclidean time $t_{E}=it$, and demanding that the geometry in the near-horizon limit be smooth, the temperature of the black hole is given by the period $\b$ of $t_E$ as
\be
T=\dfrac{1}{\b}=\dfrac{r^2_+d+R^2(d-1)}{4\pi R^2r_+}.
\ee
$T(r_+)$ or equivalently $T(M)$ has a minimum at $r_+=\sqrt{(d-1)/(d+1)}R$ given by,
\be
T_{min}=\dfrac{\sqrt{(d+1)(d-1)}}{2\pi R}.
\ee
For $r_+<\sqrt{(d-1)/(d+1)}R$, the specific heat $C=\pd M/\pd T$ of the black hole is negative, which means the black hole is thermodynamically unstable. This is known as the small AdSBH. For $r_+>\sqrt{(d-1)/(d+1)}R$, where $M$ is large, has $C>0$, and so is thermodynamically stable. This is known as the large AdS black hole. If we take $r_+\gg R$, and make the coordinate transformation $z=R^2/r$, we reproduce the planar black hole given in equation \eqref{pBH}.

It was proposed by Maldacena\cite{Maldacena1} that there is a dual non-perturbative description of the maximally extended version of the AdSBH spacetime. From AdS/CFT, when we are given a particular asymptotically AdS spacetime, we know there is a holographic dual CFT in a particular state. If we let $\H=\H_1\times\H_2$ be the Hilbert space consisting of two copies of the associated CFT labelled by $1,2$. The wave function $|\Psi\rangle\in\H$ is given by
\be
\label{entst}
|\Psi\rangle=\dfrac{1}{\sqrt{Z(\b)}}\sum_ne^{-\b E_n/2}|E_n\rangle_1\otimes|E_n\rangle_2,
\ee
where $|E_n\rangle$ are the energy eigenstates of the two copies of the CFT and $Z(\b)$ is the partition function of one copy of the CFT at temperature $\b^{-1}$, then it was argued that
\begin{Con}
\textit{Two copies of the CFT in the particular pure (entangled) state given by \eqref{entst}, is approximately described by gravity on the extended AdSBH spacetime}\cite{Maldacena1}.
\end{Con}
\section{Probing the Bulk}
\label{probulk}

Examining the 2-point function given in \eqref{2pfn}, we see that it is singular when $|\vec{z}-\vec{u}|^2=0$, i.e. when the boundary points are null separated. From properties of pure AdS spacetimes, we know that all boundary-to-boundary null geodesics end at the same point, including those coinciding with the light cone of the boundary manifold. Thus, we can assert that in pure AdS spacetimes, a boundary correlation function $G(\vec{x},\vec{x}')=\langle\O(\vec{x})\O(\vec{x}')\rangle$ is singular if there is a null geodesic connecting $\vec{x}$ and $\vec{y}$. It is then natural to investigate properties of 2-point functions in the case of more general asymptotically AdS spacetimes, more specifically those containing a singularity. This is useful in the context of quantum gravity where one would like to understand the physics inside the horizon of black holes. Much work has been done in this fashion by considering boundary-to-boundary geodesic probes with endpoints connected by boundary correlators\cite{Bal,Bal1,Louko,Kraus,Levi}. In particular, by considering nearly-null spacelike boundary-to-boundary geodesics in eternal AdSBH spacetimes, it was shown\cite{Fid} that the boundary CFT correlators contain ``light-cone'' like singularities. Motivated by this, one can consider whether correlators have similar properties for null geodesics in perturbed asymptotically AdS spacetimes. Such analysis leads to additional singularities called ``bulk-cone singularities'', as the following argument taken from \cite{hubeny} will imply.

In the large $\lambda$, $N\tend\infty$ limit of AdS/CFT, where the bulk theory reduces to a classical Einstein-Hilbert action (see \S\ref{limit}), one can define the boundary correlation function $G(x,x')$ in terms of the bulk propagator $\mathcal{G}(x,r;x',r')$ in global AdS coordinates (see \S\ref{ads}), in a similar way to \eqref{2pfn} as
\be
\label{gg}
G(x,x')=2\n\lim_{r,r'\tend\infty}(rr')^\D\mathcal{G}(x,r;x',r'),
\ee
where the gauge invariant operators $\O(x)$ have conformal dimension $\D$, and are dual to free scalar field of mass $m$ living in a $d+1$ dimensional bulk where
\be
\D=\dfrac{d}{2}+\n,\quad\n=\sqrt{m^2+\dfrac{d^2}{4}}.
\ee
We now want to determine the behaviour of $G(x,x')$ from $\mathcal{G}(x,r;x',r')$ given $x$ and $x'$ are null separated in the bulk. 

In the limit of large $m$ so that $\D\sim m$ (from \eqref{KG}), one can regularise the bulk propagator by introducing a cut-off surface $r=r'=\La\tend\infty$. Then, if one considers two spacelike separated points $A=(x,\La)$ and $B=(x',\La)$ on the cut-off surface, the behaviour of $\mathcal{G}(A,B)$ in the semi classical WKB approximation (see \cite{Bal}) is given by
\be
\mathcal{G}(A,B)\propto e^{-m\,d(A,B)},
\ee
where $d(A,B)$ is the proper distance between $A$ and $B$ in the bulk.

If one considers some point $C(x'',\La)$ where $x''$ is in the neighbourhood of $x'$ on the boundary and using \eqref{gg}, then
\be
G(x,x')=\lim_{x''\tend x'}G(x,x'')\sim 2m\dfrac{1}{(\delta x^2)^m},
\ee
where $\delta x$ is the proper distance between $x$ and $x'$ on the boundary. 

And so $G(\vec{x},\vec{x}')=\langle\O(\vec{x})\O(\vec{x}')\rangle$ is singular if there exists null geodesics connecting $x$ and $x'$. Thus, boundary-to-boundary null geodesics with endpoints $\vec{x}$ and $\vec{x}'$ in asymptotically AdS spacetimes can be divided into two types, ones that lie on the boundary (Type $\A$) and those that enter the bulk (Type $\B$). Type $\A$ null geodesics give rise to the standard light-cone singularities of $G(\vec{x},\vec{x}')$. Type $\B$ null geodesics for perturbed AdS spacetimes give rise to additional singularities of $G(\vec{x},\vec{x}')$, such singularities are called ``\textit{bulk-cone singularities}''. It is precisely this relationship that will be used in the following chapters to determine the geometry of certain perturbed asymptotically AdS spacetimes.

\section{Entanglement Entropy}
\label{ent2}

In this final section of chapter 2, we motivate the second important result which will be used in the following chapters. In contrast to \S\ref{probulk}, where one was focused on the properties of correlation functions of local operators in the CFT, here we find that properties of non-local quantities are equally useful in probing the bulk. We have already seen in \S\ref{finiteT} that the entropy of the AdSBH has a holographic dual in the entanglement entropy of two copies of the boundary CFT. So entanglement entropy might provide such a quantity from which one can can find a certain holographic dual. This is what will be shown below after first introducing entanglement entropy more formally.

Consider a quantum mechanical system at zero temperature. The total quantum system is described by the pure ground state $|\Psi\rangle$. The density matrix is then given by the pure state
\be
\rho=|\Psi\rangle\langle\Psi|.
\ee
The von Neumann entropy is the quantum extension of the classical statistical entropy and is given in terms of the density matrix of a system as
\be
S(\rho)=-\text{tr}(\rho\log\rho).
\ee
In the case of a pure state, the entropy is zero.

The concept of entanglement entropy provides a way of expressing how closely entangled a subsystem is with the other subsystems. More specifically, consider dividing the total system into two subsystems $A$ and $B$. The total Hilbert space can be written as a direct product of two spaces $\H=\H_A\otimes\H_B$ corresponding to the subsystems $A$ and $B$ respectively. The reduced density matrix for the subsystem $A$ is defined by tracing of the degrees of freedom in subsystem $B$ giving
\be
\rho_A=\tr_B\rho.
\ee
The entanglement entropy of the subsystem $A$ is defined as the von Neumann entropy of the reduced density matrix $\rho_A$
\be
S_A=-\tr_A(\rho_A\log\rho_A).
\ee
One can immediately apply such a concept to a QFT defined on a $d+1$ dimensional spacetime manifold $\pd\M$ living on the boundary of a bulk gravitational theory with $d+2$ dimensional asymptotically AdS spacetime $M$ with conformal boundary $\pd M$ a la AdS/CFT. Assume that $\pd M$ allows the foliation by $d$ dimensional spacelike time-slices $\pd N_t$ as $\pd M=\prod_t\pd N_t\times\R_t$. Consider a subsystem $A_t\subset\pd N_t$ at a fixed time $t$. Denote the complement of $A_t$ with respect to $\pd N_t$ by $B_t$ such that $A_t\cup B_t=\pd N_t$. Then the entanglement entropy of $A_t$ is given in a similar way as
\be
S_{A_t}(t)=-\tr(\rho_{A_t}(t)\log\rho_{A_t}(t)).
\ee
Consider the case of a static system such that $A_t=A$. One can then calculate the entanglement entropy associated with the region $A$. For an ordinary QFT, there are infinitely many degrees of freedom, thus the entanglement entropy suffers from ultraviolet divergences. These can be regulated by introducing a lattice spacing $a$. Then $S_A$ is proportional to the area of the boundary $\pd A$ of the subsystem $A$ as follows
\be
S_A=\a\dfrac{\text{Area}(\pd A)}{a^{d-1}}+\text{sub-leading terms}.
\ee
Since the AdS/CFT correspondence relates physical quantities in the bulk to physical quantities on the boundary, it is natural to consider if there is a geometrical representation in the bulk of the boundary entanglement entropy. This was found to be the case in \cite{Ryu,Ryu1}. We note that the foliation of the boundary manifold $\pd M$ by spacelike time-slices $\pd N_t$ can be extended into the bulk, due to the existence of a timelike Killing field, such that the bulk manifold $M$ is foliated by spacelike time-slices $N_t$ giving $M=\prod_tN_t\times\R_t$. If we consider the case of a static system again, we can construct a minimal surface $\g_A$ in the bulk which ends on $\pd A\subset\pd N$. This is then related to the entanglement entropy associated with region $A$ via the formula
\be
\label{holent}
S_A=\dfrac{\text{Area}(\g_A)}{4G_N^{d+2}},
\ee
where $G_N^{d+2}$ is the $d+2$ dimensional Newton constant.

Motivated by the Bousso's bound \eqref{bbound}, the authors of \cite{Hub} made the above statement fully covariant
\be
S_A(t)=\dfrac{\Area(\g_A(t))}{4G_N^{d+2}},
\ee
where $\g_\A(t)$ is the extremal surface in the entire Lorentzian spacetime $M$ with boundary condition $\pd\g_A(t)=\pd A(t)$.

We will see in chapters four and five how the holographic entanglement entropy proposal can be used to extract information of bulk geometries.

\chapter{Recovering the Bulk I}
\label{Null}
\textit{This chapter is the author's original work based on \cite{Bilson}.}\\

In chapter two, we motivated two main results using AdS/CFT, namely the existence of bulk-cone singularities, and the holographic interpretation of entanglement entropy. Using such results, and motivated by the discussion in chapter one, it is natural to consider how much information of the bulk spacetime can be determined given boundary CFT data. More specifically, this and the remaining chapters will be concerned with the following question:

\textit{Given a $d$-dimensional boundary CFT where one has knowledge of the location of its singular boundary correlators and knowledge of its entanglement entropy for certain subsystems of the boundary CFT, can one determine the bulk geometry of the $d+1$-dimensional perturbed asymptotically AdS spacetime dual to this CFT?}

In this chapter, we consider a limiting version of the above question, where one has knowledge of the location of the bulk-cone singularities only. To answer such a question, one must first build up the framework upon which the question is to be analysed. As was stated in \S\ref{probulk}, bulk-cone singularities arise in boundary correlators when they are connected by Type $\B$ boundary-to-boundary null geodesics, and so it is necessary to describe the properties of such objects.

\section{Boundary-to-Boundary Null Geodesics in AdS}
\label{btbngeo}

As with most problems in GR, we begin with a metric. We consider a minimal generalised version of the metric of AdS space in global coordinates (see equation \eqref{gads2}), which corresponds to a class of static, asymptotically AdS metrics with spherical symmetry, given in $(d+2)-$dimensions as
\be
\label{sm}
ds^2 = -f(r)\,dt^2 + \dfrac{dr^2}{f(r)} +r^2\,d\Om^{2}_d.
\ee
$f(r)$ is a well behaved real function of one variable and we recover AdS space when we set $f(r)=1+(r/R)^2$, where $R$ is the AdS radius. In general, this spacetime is asymptotically AdS, which requires the condition for $r\tend\infty$,
\be
\label{asymads}
f(r)\tend\dfrac{r^2}{R^2}+1+\O\left(\dfrac{1}{r^{d-1}}\right).
\ee
The main example of a solution to Einstein's equations which satisfy such a condition is the AdS black hole in $d+2$-dimensions, given in equation \eqref{AdSBH}.

To simplify the problem of solving for the null geodesic equations of motion, we use spherical symmetry to pick a particular spatial plane on which the null geodesics live. A helpful choice is the equatorial plane, where $\{\th_1,\th_2,\cdots,\th_d\}=\{\phi,\pi/2,\cdots,\pi/2\}$ (using the coordinate set given in \eqref{sads}). Then the equations of motion for $t$ and $\p$ yield 2 constants of motion
\be
\dot{t}\,f(r) = E \quad\text{and}\quad r^{2}\dot{\p} = J,
\ee
where $\dot{x}=\frac{\d x}{\d\lambda}$ and $\lambda$ is the affine parameter.

For null trajectories, $\d s = 0$, thus the equation of motion for $r$ is given by
\be
\dot{r}^2 = E^2 - \dfrac{f(r)\,J^2}{r^2}.
\ee
By redefining the affine parameter $J\la\tend\la$, we can just consider the ratio $\a = \frac{E}{J}$ to parametrise the null trajectories. Thus, the geodesic equations of motion become
\be
\label{ge}
\dot{t} = \dfrac{\a}{f(r)},\quad\dot{\p} = \dfrac{1}{r^2},\quad\dot{r}^2=\a^2-V(r),
\ee
where $V(r)=f(r)/r^2$ is the effective potential in the radial direction.

This gives us two equations of motion
\be
\dfrac{dt}{dr}=\dfrac{\a}{f(r)\sqrt{{\a}^2-V(r)}}\quad\text{and}\quad\dfrac{d\p}{dr}=\dfrac{1}{r^2\sqrt{{\a}^2-V(r)}}.
\ee

Now we can quantify the location of the singularities of the correlation functions in the CFT in terms of the boundary coordinates of the bulk metric. Considering the metric given by equation \eqref{sm}, we see that these quantities can be described in terms of a function $\D\p(\D t)$, where if $G(x,x')$ is some singular correlator on the equatorial plane between the points $x=(t,\p)$ and $x'=(t',\p')$, then $\D t=t'-t$ and $\D\p=\p'-\p$. Given that the quantities $\D\p$ and $\D t$ are also the location of the endpoints of boundary-to-boundary null geodesics of the bulk metric, one can also express them in terms of the geometry of the bulk. Thus using the null geodesic equations of motion, we have
\be
\label{dt}
\D t(\a) = 2\int^{\infty}_{r_{min}(\a)}\dfrac{{\a}\,dr}{f(r)\sqrt{{\a}^2-V(r)}},
\ee
and
\be
\label{dph}
\D\p(\a) = 2\int^{\infty}_{r_{min}(\a)}\dfrac{dr}{r^2\sqrt{\a^2-V(r)}},
\ee
where $\a=E/J$ parametrises the null geodesics, and $r_{min}(\a)$ is the minimum radius that the null geodesic, with angular momentum $J=E/\a$, reaches into the bulk. The factor of 2 comes from the fact that the null trajectory goes from the boundary, down to the minimum value $r_{min}(\a)$, then back to the boundary at $r=\infty$. 

For AdS space where $f(r)=1+r^2/R^2$, integrals \eqref{dt} and \eqref{dph} can be solved to find that $\D t(\a)=R\pi$ and $\D\p(\a)=\pi$ with $r_{min}(\a)=1/\sqrt{\a^2-1/R^2}$. Thus all boundary-to-boundary null geodesics in AdS space with radius $R$ which start at the same point, end at the same point. The same will be shown not to be true for perturbed asymptotically AdS spacetimes.

\begin{figure}[h!t]
\centering
\begin{tabular}{cc}
\begin{minipage}{200pt}
\centerline{\includegraphics[width=200pt]{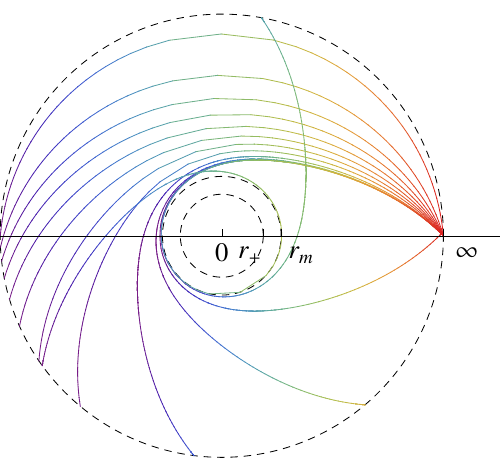}}
\end{minipage}
&
\begin{minipage}{200pt}
\centerline{\includegraphics[width=215pt]{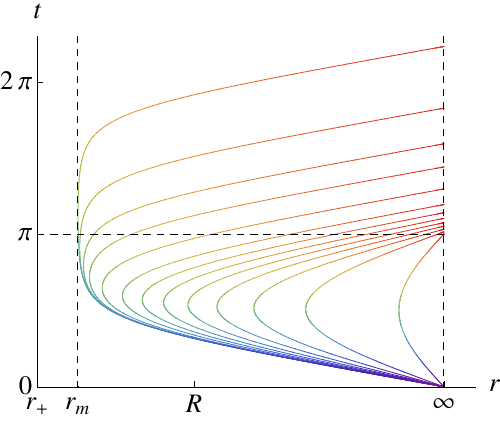}}
\end{minipage}
\end{tabular}
\caption{\label{BHplot}These are $2$-dimensional plots of boundary-to-boundary null geodesics in a small AdS-black hole geometry with metric given by \eqref{sm} with $f(r)=1+r^2-0.1/r^2$, where $d-1$ angular directions have been projected out. To plot the null trajectories, the spacetime has been compactified using the coordinate transformation $r=\mathrm{tan}(\tilde{r})$, thus the boundary is a timelike cylinder with radius $\tilde{r}=\frac{\pi}{2}$. The left figure illustrates the quantity $\D\p(\a)$ via projection of the null geodesics on the $r$-$\p$ plane at constant $t$. The right figure illustrates the quantity $\D t(\a)$ via projection of the null geodesics on the $r$-$t$ plane, where $\a=\frac{E}{J}$ parametrises the null curves.}
\end{figure}

Now we can quantify the distinction between Type $\A$ and Type $\B$ boundary-to-boundary null geodesics referred to in \S\ref{probulk}. Type $\A$ geodesics sit on the boundary at $r=\infty$, and so from \eqref{ge} and condition \eqref{asymads}, we have $\a_\A=1/R$. For Type $\B$ geodesics, we have $\a_\B>1/R$ since $\dot{r}^2>0$. The upper bound on $\a_\B$ is set by the radius of null circular orbits, $r_m$, for which null geodesics going inside such a radius will not return to the boundary. The quantity $r_m$ is defined by the global maximum of the potential, i.e. where $r_m$ is the smallest real root of $V'(r)=0$. Then using \eqref{ge}, the upper bound is given by $\sqrt{V(r_m)}$, and so $\a_\B\in(1/R,\sqrt{V(r_m)})$. See figure \ref{BHplot} to see the trajectories of boundary-to-boundary null geodesics in a spacetime containing null circular orbits. Of course for spacetimes where the global maximum is infinity, such as AdS space, there is no upper bound on $\a_\B$.
\begin{figure}
\centering
\includegraphics[width=4in]{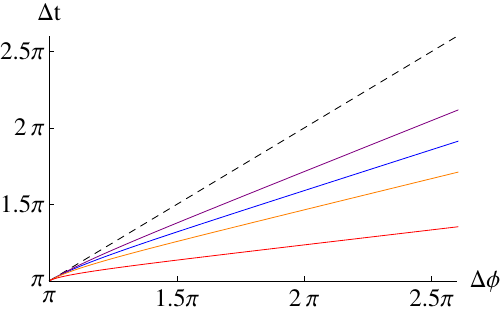}
\caption{\label{BHplotdtdp}This is a plot of $\D t(\D\p)$ for the AdSBH given in \eqref{sm} with $d=3$ and $f(r)=1+r^2-\m/r^2$. The dashed curve $\D t=\D\p$ is the Type $\A$ boundary null geodesic. The blue curve represents the spacetime depicted in figure \ref{BHplot} with $\m=0.1$. The purple, green and red curves correspond to $\m=0.2,0.05,\text{and}\,0.01$ respectively. We see that all curves converge to the AdS solution $\D t=\D\p=\pi$ as $\a\tend 1$, and that $\D t<\D\p$ for $\a>1$.} 
\end{figure}
Since $\a$ is purely a bulk quantity, expressions \eqref{dt} and \eqref{dph} do not yet provide a relationship between the data in the CFT and the geometry of the bulk. But it can be shown \cite{hammer1,Bilson} that one can relate the bulk quantity $\a$ with the boundary quantities $\D\p$ and $\D t$.

Starting with equations \eqref{dt} and \eqref{dph}, we introduce a regulator at $r=\La\tend\infty$, such that one can compare divergent peices of the derivatives. Multiplying each term top and bottom by $V'(r)$, we have
\be
\label{dadt}
\D t(\a)=2\lim_{\La\tend\infty}\left(\int^\La_{r_{min}}\dfrac{\a\,V'(r)}{V(r)\sqrt{\a^2-V(r)}}\dfrac{dr}{r^2V'(r)}\right),
\ee
and
\be
\D\p(\a)=2\lim_{\La\tend\infty}\left(\int^\La_{r_{min}(\a)}\dfrac{V'(r)}{\sqrt{\a^2-V(r)}}\dfrac{dr}{r^2V'(r)}\right).
\ee
Integrating both expressions by parts, we have
\bea
\D t(\a)=&-&4\lim_{\La\tend\infty}\left[\mathrm{arctanh}\left(\dfrac{\sqrt{\a^2-V(r)}}{\a}\right)\dfrac{1}{r^2V'(r)}\right]^\La_{r_{min}(\a)}\nn
&+&4\lim_{\La\tend\infty}\int^\La_{r_{min}(\a)}\mathrm{arctanh}\left(\dfrac{\sqrt{\a^2-V(r)}}{\a}\right)\left(\dfrac{1}{r^2V'(r)}\right)'\,dr,
\eea
and
\bea
\D\p(\a)=&-&4\lim_{\La\tend\infty}\left[\sqrt{\a^2-V(r)}\dfrac{1}{r^2V'(r)}\right]^\La_{r_{min}(\a)}\nn
&+&4\lim_{\La\tend\infty}\int^\La_{r_{min}(\a)}\sqrt{\a^2-V(r)}\left(\dfrac{1}{r^2V'(r)}\right)'\,dr.
\eea
Using the fact that $V(r_{min}(\a))=\a^2$, we see that the boundary terms at $r_{min}(\a)$ vanish. Now taking derivatives with respect to $\a$ yields
\bea
\label{dadt}
\D t'(\a)=&-&4\lim_{\La\tend\infty}\left(\dfrac{1}{\sqrt{\a^2-V(\La)}}\dfrac{1}{\La^2V'(\La)}\right)\nn
&+&4\lim_{\La\tend\infty}\int^\La_{r_{min}(\a)}\dfrac{1}{\sqrt{\a^2-V(r)}}\left(\dfrac{1}{r^2V'(r)}\right)'\,dr,
\eea
and
\bea
\label{dadph}
\D\p'(\a)=&-&4\a\lim_{\La\tend\infty}\left(\dfrac{1}{\sqrt{\a^2-V(\La)}}\dfrac{1}{\La^2V'(\La)}\right)\nn
&+&4\a\lim_{\La\tend\infty}\int^\La_{r_{min}(\a)}\dfrac{1}{\sqrt{\a^2-V(r)}}\left(\dfrac{1}{r^2V'(r)}\right)'\,dr,
\eea
where primed functions are partial derivatives with respect to $\a$. We have utilised the Leibniz rule where if $a(\a), b(\a), c(\a,r)$ are continuous, well behaved functions, then
\bea
a(\a)&=&\int^{\infty}_{b(\a)}c(\a,r)\,dr\nn
\thus a'(\a)&=&-b'(\a)\,c(\a,b(\a))+\int^{\infty}_{b(\a)}c'(\a,r)\,dr.
\eea
By identifying the limits of \eqref{dadt} and \eqref{dadph}, we can write
\be
\D\p'(\a)=\a\,\D t'(\a),
\ee
which can be rewritten in terms of the function $\D t(\D\p)$ as
\be
\label{atodt}
\dfrac{d\D t(\D\p)}{d\D\p}=\dfrac{1}{\a}.
\ee
For Type $\A$ geodesics, we know that $\a_\A=1/R$, and so using \eqref{atodt} we have $\D t_\A(\D\p_\A)=R\D\p_\A$. Note that $\D t_\A(\pi)=R\pi$, which is consistent with the pure AdS result. For Type $\B$ geodesics, $\a_\B>1/R$, and so $\D t_\B<R\D\p_\B\,\forall\,\a_\B$ (see figure \ref{BHplotdtdp}). 

Through equation \eqref{atodt}, we have found a way to relate the bulk parameter $\a$ to the boundary parameters $\D\p$ and $\D t$, and so equations \eqref{dt} and \eqref{dph} indeed become relationships between purely bulk data and purely boundary data. Thus these are the equations we need to use to find an expression for $f(r)$ and thus the metric itself.

\section{Solving for the Metric}
\label{solfr}

Given that we know $\D t(\a)$ and $\D\p(\a)$ from boundary data, we can now proceed to solve for $f(r)$ given equation \eqref{dph}\footnote{one can equally well use equation \eqref{dt} for this purpose but \eqref{dph} is a slightly simpler expression so we will choose this one}.

The extraction of $f(r)$ is complicated by the fact that equation \eqref{dph} is non-linear in $f(r)$. Non-linearity in any type of equation makes finding an analytical solution much harder, so it would be useful to find a quantity which is linear in the integrand and invert that instead. This can be done by identifying a linear equation which control the trajectory of the null geodesics. Such an equation is the ``energy equation'' taken from \eqref{ge}
\be
\label{EC}
\dot{r}^2+V(r)=\a^2.
\ee
Instead of integrating over the variable $r$, let us perform a substitution to integrate over the variable $V=V(r)$ instead. If we let $r_{min}(\a) = x$, we see an immediate simplification for the expression of $\D\p(\a)$,
\be
\D\p(x)=2\int^{\infty}_x\dfrac{\d r}{r^2\sqrt{V(x)-V(r)}}.
\ee
Now if we perform the substitution $V(r)=V$, and integrate over $V$, we have
\be
\D\p(x) = \int^{V(x)}_{V(\infty)}\dfrac{p(V)\,dV}{\sqrt{V(x)-V}},\quad\text{where}\quad p(V)=-2\,\dfrac{dr(V)}{dV}\dfrac{1}{r(V)^2}.
\ee
We can now see that the integrand is linear in $p(V)$. To make the integral easier to digest, we let $V(x)=V_{max}$, and applying the asymptotically AdS condition \eqref{asymads}, we have
\be
\label{pV}
\D\p(V_{max}) = \int^{V_{max}}_{1/R^2}\dfrac{p(V)\,\d V}{\displaystyle\sqrt{V_{max} - V}}.
\ee
It is first noted that equation \eqref{pV} is an integral equation of the first kind. The simplest equation of this type is $a(x)=\int^x_0b(r)\,dr$, which has the trivial solution $b(r)=a^\prime(r)$. More precisely, equation \eqref{pV} is a Volterra equation of the First Kind. It is a particular form of {\it Abel's equation} \cite{Abel} (see Appendix A), which is of the form
\be
\label{abelint}
\int^x_a\dfrac{y(t)}{\sqrt{x-t}}\,dt=f(x),
\ee
with solution,
\be
y(t)=\dfrac{1}{\pi}\dfrac{d}{dt}\int_a^t\dfrac{f(x)\,dx}{\sqrt{t-x}}.
\ee
Applying this to equation \eqref{pV}, we have,
\be
p(V)=\dfrac{1}{\pi}\dfrac{d}{dV}\int^{V}_{1/R^2}\dfrac{\D\p(V_{max})\,dV_{max}}{\ds\sqrt{V-V_{max}}}.
\ee
Now substituting back in for the original variables,
\be
-2\,\dfrac{\d r}{\d V}\,r^{-2}=\dfrac{2}{\pi}\dfrac{d}{dV}\int^{\sqrt{V}}_{1/R}\dfrac{\a\,\D\p(\a)}{\ds\sqrt{V-\a^2}}\,\d\a.
\ee
Integrating up, and applying the boundary conditions $V(r)=V$ and $V(\infty)=1/R^2$, we finally arrive with,  
\be
\label{exfr}
\int^{\sqrt{V(r)}}_{1/R}\dfrac{\a\,\D\p(\a)}{\ds\sqrt{V(r)-\a^2}}\,\d\a-\dfrac{\pi}{r}=0.
\ee
Thus, given the location of the endpoints of null geodesics by the function $\D t(\D\p)$, one can find $\D\p(\a)$ using equation \eqref{atodt}, and in principle, solve for the metric function\footnote{One can use the same method to recover $f(r)$ using $\D t(\a)$ instead via equation \eqref{dt}. In this case, using the same inversion techniques, we find
\be
\label{exfr2}
V(r)\int^{\sqrt{V(r)}}_{1/R}\dfrac{\D t(\a)}{\sqrt{V(r)-\a^2}}\,d\a-2\int^{\sqrt{V(r)}}_{1/R}\D t(\a)\sqrt{V(r)-\a^2}\,d\a-\dfrac{\pi}{r}=0
\ee} $f(r)=r^2V(r)$ using \eqref{exfr}. In the next section, we will define what we mean by ``in principle'' by analysing the method of extraction using test functions.

\section{Analysing the Extraction} 

Before considering the limits of the extraction method, it is useful to check the validity of \eqref{exfr} in simple cases where $\D\p(\a)$ is known, namely AdS space.

Given $f(r)=1+r^2/R^2$, we know that $\D\p(\a)=\pi$. So substituting this result into \eqref{exfr} we have,
\bea
\int^{\sqrt{V(r)}}_{1/R}\dfrac{\a}{\ds\sqrt{V(r)-\a^2}}\,d\a&=&\dfrac{1}{r}\nn
\Rightarrow \ds\sqrt{V(r)-\dfrac{1}{R^2}}&=&\dfrac{1}{r}\nn
\Rightarrow f(r)&=&r^2\,V(r)=1+\dfrac{r^2}{R^2}.
\eea
Thus \eqref{exfr} has reproduced the the known result for pure AdS space.

\begin{figure}
\centering
\includegraphics[width=4in]{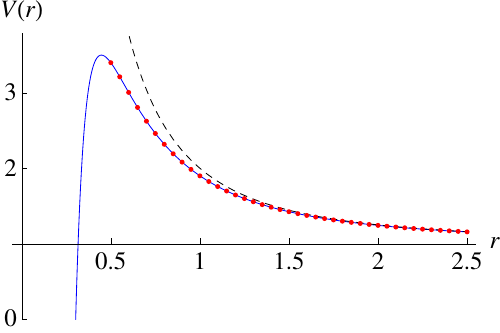}
\caption{\label{BHplotextr}This is a plot of $V(r)=1+1/r^2-0.1/r^4$ (blue), compared with the data extracted using \eqref{exfr} (red). The dashed line is pure AdS. Notice the spacetime can only be extracted up to the maximum of the potential at $r_m=\sqrt{0.2}$.} 
\end{figure}

We can also test the extraction for a known perturbed AdS solution, namely the AdSBH. In this case we take the metric in \eqref{sm} with $f(r)=1+r^2-0.1/r^2$ (see figure \ref{BHplot}), use this to find a numerical solution for $\D\p(\a)$, and then numerically solve for $V(r)=f(r)/r^2$ using \eqref{exfr}. The result is shown in figure \ref{BHplotextr}. Here we find the first limitation of the extraction method outlined in the previous section. We see that the spacetime can only be probed down to the radius of null geodesics $r_m$. This makes sense as Type $\B$ geodesics do not exist for $\a<\a_m$, where\footnote{see the argument in \S\ref{btbngeo}} $\a_m=\sqrt{V(r_m)}$.

\begin{figure}
\centering
\includegraphics[width=4in]{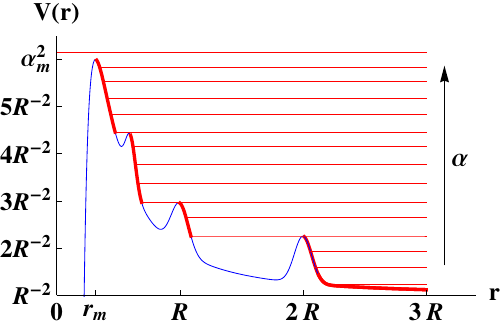}
\caption{\label{nullprobbumpy}This figure illustrates how much of the domain of the metric $f(r)$ can be recovered by firing null geodesic probes from the boundary (shown here by the parallel red lines), for a potential $V(r)=r^2f(r)$ (plotted in blue) with multiple maxima. The turning points of the null geodesics are shown by the thick red lines. We see that the presence of the maxima causes the null probes to ``miss'' parts of the spacetime, thus restricting the domain of $f$ which can be recovered.} 
\end{figure}

Now we can analyse the extraction for a more general type of function $V(r)$. We do this by adding some Gaussian ``bumps'' to an AdSBH solution so that
\be
\label{bumpyfr}
V(r)=\dfrac{f(r)}{r^2}=1+\dfrac{1}{r^2}-\dfrac{1}{5r^4}+\displaystyle\sum_{i=1}^3\dfrac{\mathrm{A_i}}{\sqrt{2\pi}\sigma_i}\,e^{-\frac{(r-\mu_i)^2}{2{\sigma_i}^2}},
\ee
where $\mathrm{A_i}=(0.05,0.03,0.1),\,\mu_i=(1,1.3,2)\,\,\text{and}\,\,\sigma_i=(0.1,0.1,0.3)$.
\begin{figure}
\centering
\includegraphics[width=4in]{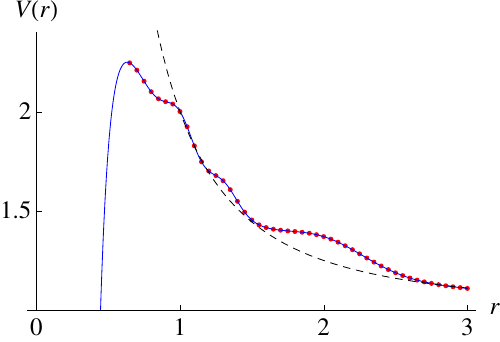}
\caption{\label{BHbumpyextr}This is a plot of $V(r)$ given in \eqref{bumpyfr} (blue), compared with the data extracted using \eqref{exfr} (red). The dashed line is pure AdS. Notice the spacetime can only be extracted up to the maximum of the potential at $r_m=\sqrt{0.4}$.} 
\end{figure}
The extracted data is given in figure \ref{BHbumpyextr}. Again, we have the same limitation that the metric can only be determined down to $r_m$. This is only true if $V(r)$ is monotonically decreasing for $r>r_m$, if there are additional local maxima, then the range of $r$ for which $f$ can be extracted is more complicated. This is illustrated in figure \ref{nullprobbumpy}. 
 
\section{The Series Solution}
\label{sersol}

Now that we have the expression \eqref{exfr2} for the metric function $f(r)$ in terms of $\D\p(\a)$, we can now start to look for general examples where the integral in equation \eqref{exfr} can be solved analytically in the form of a series solution. In this section we explore the limitations of such a process. 

\subsection{The Method}
\label{pertmeth}

Given a generic function $\D\p(\a)$ which describes a perturbed AdS spacetime, we can always expand about the pure AdS solution $\D\p(\a)=\pi$ in a power series by introducing a parameter $\e>0$ such that
\be
\label{pertdp}
\D\p(\a)=\pi+\pi\sum_{n=1}^{\infty}a_n\e^n(\a^2-1/R^2)^{n/2}.
\ee
The parameter $\e$ is a bookkeeping tool to count the order of the perturbation, but its value is unimportant and will eventually be set to $1$.

From the energy equation of null geodesics,
\be
\dot{r}^2+V(r)=\a^2,
\ee
we have the boundary values $\a\tend\frac{1}{R}$ as $r\tend\infty$, and so equation \eqref{pertdp} has the correct boundary behaviour $\D\p(1/R)=\pi$, since all boundary null geodesics end at the antipodal point.

We can then ansatz a power series solution
\be
\label{pertV}
V(r)=\dfrac{1}{r^2}+\dfrac{1}{R^2}+\sum_{n=1}^{\infty}\e^nV_n(r),
\ee
where we have expanded about the pure AdS potential $\dfrac{1}{r^2}+\dfrac{1}{R^2}$ in powers of $\e$. To have the correct AdS asymptotics (see \eqref{asymads}), $V_n(r)\sim\O(1/r^{d+1+n})$ for large $r$, where $d+2$ is the dimension of the spacetime.

One can now solve for the the functions $V_n(r)$ by substituting \eqref{pertdp} and \eqref{pertV} into \eqref{exfr}, and solving order by order in powers of $\e$.

Substituting \eqref{pertdp} into \eqref{exfr} and solving the integral, we have
\be
\sqrt{V(r)-\dfrac{1}{R^2}}+\dfrac{\sqrt{\pi}}{2}\sum_{n=1}^{\infty}a_n\e^n\dfrac{\G(\frac{2+n}{2})}{\G(\frac{3+n}{2})}\left(V(r)-\frac{1}{R^2}\right)^{\frac{n+1}{2}}=\dfrac{1}{r}.
\ee
Now substituting in the series \eqref{pertV}, and performing a Taylor series expansion about $\e=0$, we can solve for the functions $V_n(r)$ order by order in powers of $\e$. The first terms in the series are given by,
\bea
V_1(r)&=&-\dfrac{\pi}{2}a_1\dfrac{1}{r^3},\nn
V_2(r)&=&\left(\dfrac{5\pi^2}{16}a_1^2-\dfrac{4}{3}a_2\right)\dfrac{1}{r^4},\nn
V_3(r)&=&-\pi\left(\dfrac{7\pi^2}{32}a_1^3-2a_1a_2+\dfrac{3}{8}a_3\right)\dfrac{1}{r^5},\nn
&\vdots&\nn
V_n(r)&=&\dfrac{b_n(a_1,a_2,\cdots,a_n)}{r^{n+2}}.
\eea
By setting $\e=1$, we have the metric given as a power series solution,
\be
\label{fpert}
f(r)=\dfrac{r^2}{R^2}+1+\sum_{n=1}^\infty\dfrac{b_n}{r^n}.
\ee
The range of values of $r$ for which \eqref{fpert} is a good approximation of the actual geometry is determined by the convergence of the power series. Using the ratio test for convergence, we find the range of validity
\be
\label{convr}
r_c<r<\infty,
\ee
where $r_c=\lim_{n\tend\infty}|\frac{b_{n+1}}{b_n}|$ is the radius of convergence.

Thus, given any boundary data for which the function $\D\p(\a)$ which can be written as the series of the form \eqref{pertdp}, one can determine the geometry of the perturbed AdS bulk with metric \eqref{sm}, over the region \eqref{convr} to good approximation.

\subsection{An Example: the ``star'' geometry}

In \S\ref{pertmeth}, we described the general method for determining $f(r)$ as a series solution. In this section we will check the results with a known solution that behaves like the interior of a star\footnote{The geometry described here is a simplified version of the ``star'' geometry considered in \cite{hubeny}, where we have let the metric depend on one function only. See the referenced paper for an analysis of the full geometry}.

\begin{figure}[h!t]
\centering
\begin{tabular}{cc}
\begin{minipage}{200pt}
\centerline{\includegraphics[width=200pt]{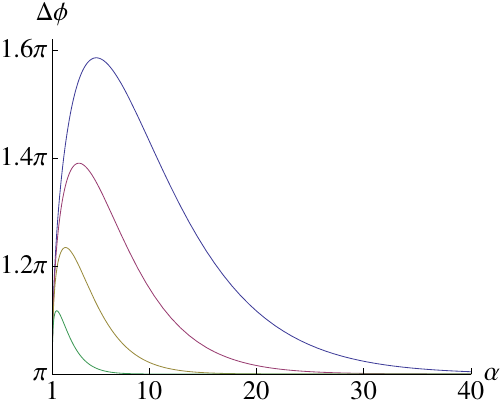}}
\end{minipage}
&
\begin{minipage}{200pt}
\centerline{\includegraphics[width=200pt]{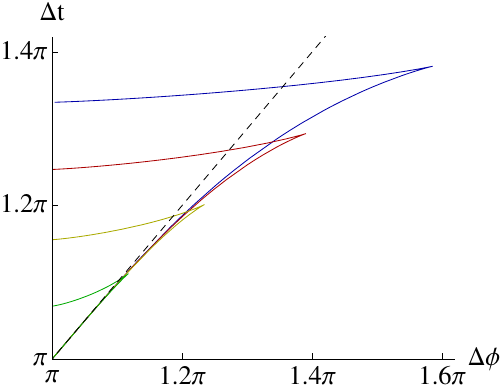}}
\end{minipage}
\end{tabular}
\caption{\label{starplot}To the left is a plot of $\D\p$ given in \eqref{stardp} for $A=R=1$ and $B=\{0.2,0.3,0.5,1\}$. We see the correct boundary behaviour where $\D\p(1)=\pi$. We also notice that as $\a\tend\infty,\,\,\D\p\tend\pi$. This represents the behaviour of radial null geodesics where $J\tend 0$. To the right is a plot of $\D t(\D\p)$ by using the relation \eqref{atodt} on the $\D\p(\a)$ curves in the left plot. The dashed curve is the Type $\A$ null boundary curve $\D t=\D\p$. We see as $\D t,\D\p\tend\pi$ we recover the behaviour of Type $\A$ boundary null geodesics. For radial null geodesics we see a shift in $\D t$ due to the presence of matter in the bulk. These plots are very similar to the full star geometry, and so we are justified in using the expression for $\D\p(\a)$ given in \eqref{stardp}}
\end{figure}

A class of boundary functions which model the angular separation of the endpoints of Type $\B$ null geodesics in star geometries is given by (see fig \ref{starplot})
\be
\label{stardp}
\D\p(\a)=\pi+A\ds\sqrt{\a^2-\frac{1}{R^2}}\exp\left(-B\ds\sqrt{\a^2-\frac{1}{R^2}}\right),
\ee
where $R$ is the AdS radius, and $A,B$ are arbitrary positive constants. 

\begin{figure}
\centering
\includegraphics[width=4in]{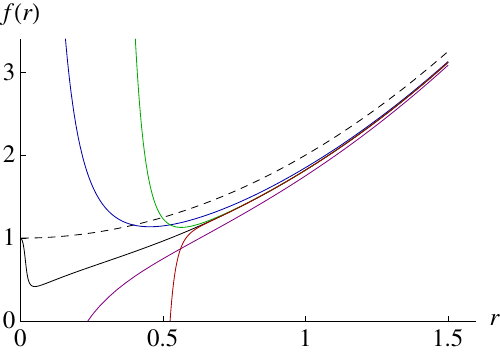}
\caption{\label{starpertplot}Here we compare the perturbative solution of \eqref{fpert} given the boundary function $\D\p$ in \eqref{pertdp} with $A=0.5$ and $B=0.1$, to the actual geometry solved numerically (shown in black). The order $n$ of the perturbation is given by $(n=1)=\,$Purple, $(n=2)=\,$Blue, $(n=10)=\,$Green, $(n=25)=\,$Red. The pure AdS metric with radius $R=1$ is given by the dashed line. We see that for $r\gg R$, all solutions tend to the AdS solution, which must be true by construction. For increasing order $n$, the series solution converges to the numerical solution for $r\gtrsim 0.6$.} 
\end{figure}

\begin{figure}
\centering
\includegraphics[width=4in]{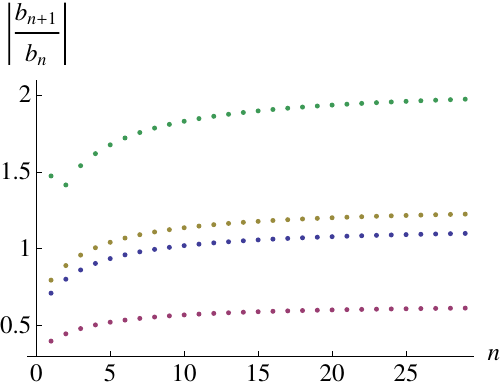}
\caption{\label{starpertconv}This plot shows the absolute value of the ratio of coefficients $b_{n+1}/b_n$ of the series \eqref{fpert} for increasing number of terms $n$. The series shown in figure \ref{starpertplot} is represented by the red points, with similar series solutions given by $(A=1,B=0.1)=\,$Blue, $(A=1,B=0.2)=\,$Yellow, $(A=1,B=1)=\,$Green. We see that the quantity $\lim_{n\tend\infty}|\frac{b_{n+1}}{b_n}|$ exists for all values of $A$ and $B$, where $\lim_{n\tend\infty}|\frac{b_{n+1}}{b_n}|\sim A(1+B)$.} 
\end{figure}

By performing a Taylor series expansion of \eqref{stardp} about $\sqrt{\a^2-1/R^2}=0$, we can reproduce the form of $\D\p$ given in \eqref{pertdp} with $\e=1$ and $a_n=\frac{A}{\pi}\frac{(-B)^{n-1}}{(n-1)!}$. One can then immediately use the method in the previous section to determine the constants $b_n$, such that the star metric $f(r)$ can be written as a series solution. The series solution is compared with the numerical solution for the specific case of \eqref{stardp} in figure \ref{starpertplot}. It illustrates that one can extract the metric function $f(r)$ perturbatively to high enough orders as long as the series is convergent. Given that the range of convergence is given by \eqref{convr}, and from the analysis of figure \ref{starpertconv}, we can make the approximation that $\lim_{n\tend\infty}|\frac{b_{n+1}}{b_n}|\sim A(1+B)$, the perturbation extraction using \eqref{stardp} is valid for
\be
A(1+B)\lesssim r<\infty.
\ee 

\subsection{Recovering the AdSBH Solution}

An alternative approach to check the consistency of the perturbation is to start with a known metric function and recover the boundary function which reproduces it as a series. This can then be compared with the numerical result.

The AdSBH in $d+1$ dimensions ($d>2$) has metric \eqref{sm}, where
\be
f(r)=\dfrac{r^2}{R^2}+1-\dfrac{r_+^{d-2}}{r^{d-2}}\left(1+\dfrac{r_+^2}{R^2}\right).
\ee
Comparing this with \eqref{fpert}, we can match the terms by setting $b_{n\neq d-2}=0$. This is achieved perturbatively by setting $a_{1<n<d-2}=0$ and picking $a_{n>d-2}$ such that $b_{n>d-2}=0$. Using the results in section \ref{pertmeth}, and setting $\e=1$ as before, we have
\be
f(r)=\dfrac{r^2}{R^2}+1+\dfrac{b_{d-2}}{r^{d-2}},
\ee
where
\be
b_{d-2}=-a_{d-2}\sqrt{\pi}\dfrac{\G(\frac{d}{2})}{\G(\frac{d+1}{2})}.
\ee
Comparing with the AdSBH metric, we have
\be
\label{mcoef}
a_{d-2}=\dfrac{r_+^{d-2}}{\sqrt{\pi}}\dfrac{\G(\frac{d+1}{2})}{\G(\frac{d}{2})}\left(1+\dfrac{r_+^2}{R^2}\right).
\ee
The other non-zero coefficients such that $b_{n\neq d-2}=0$ are $a_{p(d-2)}$ where $p\in\mathbb{N}$. The first few are given by,
\bea
a_{2(d-2)}&=&\dfrac{(2d-1)}{4}\dfrac{r_+^{2(d-2)}}{\sqrt{\pi}}\dfrac{\G(\frac{2d-1}{2})}{\G(d-1)}\left(1+\dfrac{r_+^2}{R^2}\right)^2,\nn
a_{3(d-2)}&=&\dfrac{(1-4d+3d^2)}{8}\dfrac{r_+^{3(d-2)}}{\sqrt{\pi}}\dfrac{\G(\frac{3d-3}{2})}{\G(\frac{3d-4}{2})}\left(1+\dfrac{r_+^2}{R^2}\right)^3,\nn
\vdots\nn
a_{p(d-2)}&=&P_{p-1}(d)\dfrac{r_+^{p(d-2)}}{\sqrt{\pi}}\dfrac{\G(\frac{3+p(d-2)}{2})}{\G(\frac{2+p(d-2)}{2})}\left(1+\dfrac{r_+^2}{R^2}\right)^p,
\eea
where $P_{p-1}(d)$ is an order ${p-1}$ polynomial in $d$.

\begin{figure}[h!t]
\centering
\begin{tabular}{cc}
\begin{minipage}{200pt}
\centerline{\includegraphics[width=200pt]{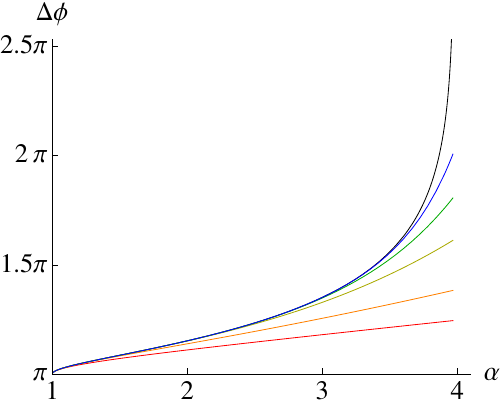}}
\end{minipage}
&
\begin{minipage}{200pt}
\centerline{\includegraphics[width=200pt]{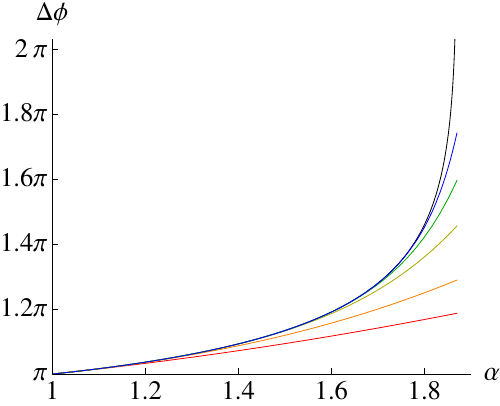}}
\end{minipage}
\end{tabular}
\caption{\label{BHpert}Here are graphs of $\D\p(\a)$ given an AdSBH spacetime with metric $f(r)=1+r^2-0.1/r^{d-2}$ with $d=3$ (left) and $d=4$ (right). The numerical result is shown in black and is compared with the perturbative solutions with increasing order $n=\{1,2,5,10,20\}$. We see that for large enough orders, the series given in \eqref{pertdp} can recover the AdSBH metric up to the radius of null circular orbits $r_m=\frac{3\m}{2}$ for $d=3$ and $r_m=\sqrt{2\m}$ for $d=4$.}
\end{figure}

Now we can determine the range of values of $\a$ for which the expansion of $\D\p(\a)$ given in \eqref{pertdp} is convergent. Using the ratio test, we have
\be
\dfrac{1}{R^2}<\a^2<\dfrac{1}{R^2}+\lim_{p\tend\infty}\left|\dfrac{a_{p(d-2)}}{a_{(p+1)(d-2)}}\right|^\frac{2}{d-2}.
\ee
For large $p$, we can make the approximation $P_{p-1}(d)\sim (2d/3)^{p-1}$, so substituting for the coefficients $a_{p(d-2)}$, the convergence condition reduces to
\be
\dfrac{1}{R}<\a\lesssim\dfrac{1}{R}\ds\sqrt{1+\dfrac{R^2}{r_+^2}\left[\dfrac{2d}{3}\left(1+\dfrac{r_+^2}{R^2}\right)\right]^{-\frac{2}{d-2}}}.
\ee
Thus for a small black hole where $r_+<R$, the range of $\a$ for which the series \eqref{pertdp} is a good approximation, is larger than for a large black hole where $r_+>R$. This reflects the fact that for a large black hole, the geometry is more perturbed, so the expansion around AdS will probe smaller distances away from the boundary. The range of $\a$ is also limited by a physical upper bound set by the radius of null circular orbits $r_m$. Type $\B$ null geodesics entering the region $r<r_m$ do not return to the boundary due to the presence of the singularity. In the specific case of the AdSBH, $r_m$ is determined by the the global maximum of the potential $V(r)=f(r)/r^2$, and so
\be
\dfrac{dV(r)}{dr}\Big|_{r=r_m}=0\quad\thus r_m=r_+\left[\dfrac{d}{2}\left(1+\dfrac{r_+^2}{R^2}\right)\right]^\frac{1}{d-2}.
\ee
The value of $\a$ at $r=r_m$ is given by the energy condition,
\bea
\a_m^2=V(r_m)&=&\dfrac{1}{R^2}+\left(\dfrac{d-2}{d}\right)\dfrac{1}{r_m^2}\nn
&=&\dfrac{1}{R^2}+\left(\dfrac{d-2}{d}\right)\dfrac{1}{r_+^2}\left[\dfrac{d}{2}\left(1+\dfrac{r_+^2}{R^2}\right)\right]^\frac{2}{2-d},
\eea
and so the range of $\a$ which parametrises Type $\B$ null geodesics in a $d+1$ dimensional AdSBH is given by
\be
\dfrac{1}{R}<\a<\dfrac{1}{R}\sqrt{1+\left(\dfrac{d-2}{d}\right)\dfrac{R^2}{r_+^2}\left[\dfrac{d}{2}\left(1+\dfrac{r_+^2}{R^2}\right)\right]^\frac{2}{2-d}}.
\ee
And so, we see that the power series \eqref{pertdp} is convergent for the entire physical range of $\a$ when
\bea
\dfrac{1}{R}\sqrt{1+\left(\dfrac{d-2}{d}\right)\dfrac{R^2}{r_+^2}\left[\dfrac{d}{2}\left(1+\dfrac{r_+^2}{R^2}\right)\right]^\frac{2}{2-d}}&\lesssim&\dfrac{1}{R}\ds\sqrt{1+\dfrac{R^2}{r_+^2}\left[\dfrac{2d}{3}\left(1+\dfrac{r_+^2}{R^2}\right)\right]^{-\frac{2}{d-2}}}\nn
\thus d&\gtrsim&2.
\eea
This is illustrated in figure \ref{BHpert} for $d=3,4$.

\section{Summary}
In this chapter, we have shown how one can extract a special class of asymptotically AdS metrics \eqref{sm}, given the location of singular correlation functions in the dual boundary CFT. This was achieved via an integral inversion and was verified using known cases. The limitations of such an extraction were discussed, and it was shown that for spacetimes admitting a singularity at the origin, the metric function could only be determined down to the radius of null circular orbits. Extracting the metric as a series solution was also investigated, and a general method was discovered. This was again tested using some known examples.

In the next chapter, we will investigate a more general class of spacetimes where there are two unknown metric functions. In this case the knowledge the locus of bulk-cone singularities expressed in terms of the location of the endpoints of Type $\B$ null geodesics will not be enough to determine the metric. One will require the knowledge of the entanglement entropy in special subsystems of the boundary CFT as mentioned at the beginning of this chapter.  

\newpage

\chapter{Recovering the Bulk II}
\label{nullspace}

\textit{This chapter is the author's original work based on \cite{Bilson}.}\\

\numberwithin{equation}{section}

In the previous chapter, we showed how to solve for a simple bulk metric \eqref{sm}, given information of the location of singular correlators in the dual boundary CFT. We then used this solution to probe the bulk spacetime given different functions of the endpoints of null geodesics $\D t(\D\p)$. Building on the previous chapter, we can now turn our attention to a more general class of spacetimes. As we will explain, knowledge of the location of the endpoints of null geodesics alone will not be enough to fully determine the metric. More information, in the form of the entanglement entropy between two points on the boundary, will be required to recover the metric. As with the example in the previous chapter, it will be modified forms of Abel's equation which are used for the inversion.

We begin with a class of metrics for static, spherically symmetric, asymptotically AdS spacetimes, which is given in $d+2$ dimensions by
\be
\label{gm}
ds^2=-f(r)\,dt^2+h(r)\,dr^2+r^2\,d\Om^2_d.
\ee
This is a generalisation of \eqref{sm} by including an additional real function $h(r)$. The main physical example of such a spacetime described by \eqref{gm} is the ``star'' geometry in AdS\footnote{see \cite{hubeny} for more details}.

The inclusion of another function's worth of information requires a modification to a number of results given in \S\ref{btbngeo}. Firstly, the requirement that \eqref{gm} be asymptotically AdS puts a condition on both metric functions, such that as $r\tend\infty$,
\be
\label{masymads}
\dfrac{f(r)}{r^2}\,,\,\dfrac{r^2}{h(r)}\tend\dfrac{1}{R^2}+\dfrac{1}{r^2}+\O\left(\dfrac{1}{r^{d+1}}\right).
\ee
The modified equations of motion become,
\be
\dot{t} = \dfrac{\a}{f(r)},\quad\dot{\p} = \dfrac{1}{r^2},\,\,\text{and}\quad\dot{r}^2 = \dfrac{1}{f(r)\,h(r)}\left[{\a}^2 - \dfrac{f(r)}{r^2}\right].
\ee
The time and angular separation of the endpoints of Type $\B$ geodesics in metric \eqref{gm} become
\be
\label{dt1}
\D t(\a) = 2\int^{\infty}_{r_{min}(\a)}\frac{\a\ds\sqrt{h(r)}}{\ds\sqrt{f(r)\,(\a^2 - V(r))}}\,dr,
\ee
and
\be
\label{dphi1}
\D\p(\a) = 2\int^{\infty}_{r_{min}(\a)}\frac{\ds\sqrt{f(r)\,h(r)}}{r^2\ds\sqrt{\a^2 - V(r)}}\,dr,
\ee
where $V(r)=f(r)/r^2$ as in \S\ref{btbngeo}.

In addition, one finds by an analogous calculation to \S\ref{btbngeo} that relation \eqref{atodt} indeed holds in this case also. This means we can follow a similar procedure for extracting $f(r)$, though complicated by the fact that there is another undetermined function $h(r)$. We will provide an analytical method in this chapter for how one can extract $h(r)$ in the specific case of $2+1$ dimensions. To show this, we need to first understand the nature of boundary-to-boundary spacelike probes.

\section{Boundary-to-Boundary Spacelike Geodesics in AdS}

In \S\ref{ent2}, it was shown that there is a relationship between the area of a minimal surface $\g$ and the entanglement entropy $S_A$ of static subsystem $A$ in the dual boundary CFT such that $\pd\g=\pd A$ (see \eqref{holent}). In the specific case of $(2+1)-$dimensional bulk static spacetimes, the dual CFT is $1+1$ dimensional, and so $A$ reduces to a closed subset $[\p_1,\p_2]\in S^1$, where $\p$ is the angular coordinate in the bulk. In this case, $\g$ is a boundary-to-boundary zero energy spacelike geodesic connecting $\p_1$ and $\p_2$. It then follows that the area of this minimal surface is just the proper length $\L$ of the spacelike geodesic, where $\L$ depends on the angular separation $\D\p=\p_2-\p_1$. The $(2+1)-$dimensional Area/Entropy law is then given by,
\be
\label{area3d}
S_A=\dfrac{\L(\D\p)}{4G_N^{(3)}}
\ee
We will find that an expression for $\L$ for the bulk metric \eqref{gm} can be obtained using the function $h(r)$ alone, and so given the entanglement entropy $S_A$, one can, in principle, solve for the metric function $h(r)$.

To calculate an expression for $\L$, we use the fact that the trajectory of spacelike geodesics in spacetime \eqref{gm} is described by the Lagrangian
\be
L=-f(r)\,{\dot{t}}^2 +h(r)\,\dot{r}^2 + r^2{\dot{\phi}}^2 = \epsilon,\,\,\dfrac{dx}{d\la}=\dot{x},
\ee
where $\epsilon > 0$ and can be normalised to 1 by redefining the affine parameter $\la$.

In the specific case of zero energy boundary-to-boundary spacelike geodesics, we have $\dot{t}=0$, and so the spacelike geodesic equations become
\be
h(r)\,\dot{r}^2 + r^2{\dot{\p}}^2 = 1\quad\text{and}\quad r^2\dot{\p}=J.
\ee
Combining the two equations gives us
\be
\label{EChr}
\dot{r}=\sqrt{\dfrac{r^2-J^2}{h(r)\,r^2}},
\ee
where $J$ is the angular momentum of the spacelike geodesic, and thus parametrises their trajectory.

The definition of the proper length $\L$ for any path through a metric $g_{\m\n}$ is given by
\be
\L=\int\sqrt{g_{\m\n}\dot{x}^\m\dot{x}^\n}\,d\la.
\ee
From the definition of spacelike geodesics we know that $L=1$, and so
\be
\L=\int d\la=\int\dfrac{dr}{\dot{r}}.
\ee
Since we are considering boundary-to-boundary spacelike probes, we can define a minimum radius $r_{min}(J)$, which the maximum radial distance the geodesic probes the bulk before returning to the boundary. Since the minimum radius is defined by the condition $\dot{r}\mid_{r_{min}(J)} = 0$, \eqref{EChr} implies
\bea
&& \dfrac{r_{min}(J)^2-J^2}{h(r_{min}(J))\,r_{min}(J)^2}=0\nn
&\thus& r_{min}(J)=J\quad\text{or}\quad h(r_{min}(J))=\infty.
\eea
$h(r)=\infty$ defines the event horizon for which zero energy spacelike geodesics cannot cross, and so we have the relation $r_{min}(J)=J$.

Since the spacetime is asymptotically AdS, and the boundary of AdS space is infinite, any proper length of geodesics ending on the boundary will be divergent. To regulate $\L$, we introduce a cut-off boundary at $r=r_b$, thus defining the region of integration $r_b\geq r\geq J$. This procedure is valid due to the UV/IR connection, where $r_b$ corresponds to a UV cut-off (or lattice spacing) $a$, in the dual CFT. The relationship for large $r_b/R$ is given by
\be
\label{uvir}
\dfrac{r_b}{R}=\dfrac{L}{2\pi a},
\ee
where $R$ is the AdS radius and $L$ is length of the compactified space in which the CFT lives. 

Putting this all together, the final expression for $\L$ is given by,
\be
\label{L}
\L(J) = 2\int^{r_b}_Jr\sqrt{\dfrac{h(r)}{r^2 - J^2}}\,dr.
\ee
Again, we find the same situation as in \S\ref{btbngeo}, namely that $J$ is a bulk parameter and so equation \eqref{L} is not yet a relationship between the boundary function $\L(\D\p)$ and the metric function $h(r)$. To achieve this, one can proceed in a similar manner to \S\ref{btbngeo}, and so taking a derivative with respect to $J$ and using the Leibniz rule, we have
\be
\label{lj}
\dfrac{d}{dJ}\L(J)=2J\lim_{r\tend J}\left(-\sqrt{\dfrac{h(r)}{r^2-J^2}}+\int_{r}^{r_b}\tilde{r}\dfrac{h(\tilde{r})}{(\tilde{r}^2-r^2)^{\frac{3}{2}}}\,d\tilde{r}\right).
\ee
A definition for the angular separation of boundary-to-boundary spacelike geodesics can be found from the geodesic equations,
\bea
\dfrac{d\p}{dr}&=&\dfrac{\dot{\p}}{\dot{r}}=\dfrac{J}{r}\sqrt{\dfrac{h(r)}{r^2 - J^2}},\nn
\thus \D\p(J)&=& 2\int^{r_b}_J\dfrac{J}{r}\sqrt{\dfrac{h(r)}{r^2 - J^2}}.
\eea
Taking a derivative, we have,
\be
\label{pj}
\dfrac{d}{dJ}\D\p(J)=2\lim_{r\tend J}\left(-\sqrt{\dfrac{h(r)}{r^2-J^2}}+\int_{r}^{r_b}\tilde{r}\dfrac{h(\tilde{r})}{(\tilde{r}^2-r^2)^{\frac{3}{2}}}\,d\tilde{r}\right).
\ee
Now comparing equations \eqref{lj} and \eqref{pj}, we see that
\be
\dfrac{d}{dJ}\L(J)=J\dfrac{d}{dJ}\D\p(J),
\ee
which gives
\be
\label{lpj}
\dfrac{d}{d\D\p}\L(\D\p)=J.
\ee
Now from equation \eqref{lpj}, we have a relationship between the CFT quantities $\L$ (from the area/entropy law \eqref{holent}) and $\D\p$, and the bulk quantity $J$. Thus $J$ can be fully expressed in terms of boundary variables, which in turn implies that the function $\L(J)$ can be recovered from boundary data alone. This means that equation \eqref{L} does indeed relate boundary data to bulk data, and so can be used to extract the bulk geometry given boundary information.

\newpage

\begin{figure}[h!t]
\centering
\begin{tabular}{cc}
\begin{minipage}{200pt}
\centerline{\includegraphics[width=200pt]{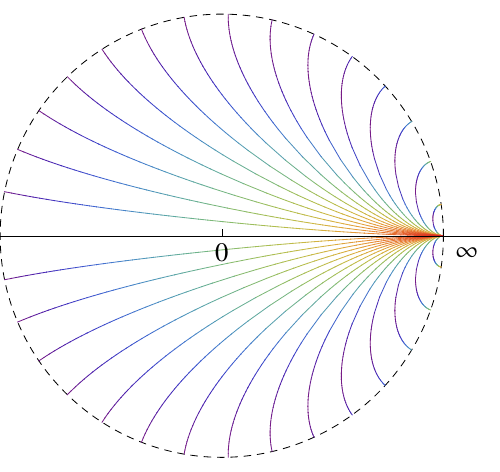}}
\end{minipage}
&
\begin{minipage}{200pt}
\centerline{\includegraphics[width=200pt]{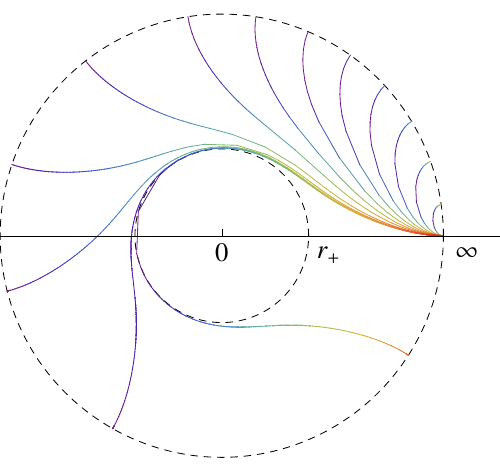}}
\end{minipage}
\end{tabular}
\caption{\label{BHspace}These 2 dimensional polar plots illustrate the behaviour of zero energy boundary-to-boundary spacelike geodesics in asymptotically AdS spacetimes given by \eqref{gm}. Both boundaries are compactified onto a circle of radius $\frac{\pi}{2}$ via the substitution $r\tend\arctan(r)$(see equation \eqref{compads}), and $d-1$ spacial directions are projected out. The left diagram is AdS space where $h(r)^{-1}=1+r^2$. The geodesics are parametrised by their angular momentum $J$ and the set of geodesics $J\in[0,\infty)$ cover the entire spacetime. Notice that the geodesics always meet the boundary at a right-angles. The diagram on the right represents a black hole located at $r=0$ in an AdS background with metric function $h(r)^{-1}=1+r^2-\frac{3}{4r^2}$. The spacetime has an event horizon at $r_+=1/\sqrt{2}$ which is shown here as the inner circle. Comparing the two diagrams, we see that for large values of $J$, the geodesics are more or less the same. But, as $J\tend r_+$, the geodesics are deformed by the presence of the singularity and begin to wrap around the horizon. Notice zero energy spacelike geodesics cannot enter the black hole as they have no timelike component, and so have no spacial extension inside the horizon. This means the set of zero energy boundary-to-boundary spacelike geodesics $J\in(r_+,\infty)$ only cover the same range of $r$, since $J$ is the minimum value of $r$.}
\end{figure}

\newpage

\section{Extracting the Metric using Spacelike Probes}
\label{extrhr}

We can now proceed with extracting $h(r)$ given $\L(\D\p)$ via the inversion of equation \eqref{L}.

Using the substitution $y(r) = -2r\sqrt{h(r)}$, one finds that
\be
\L(J)=\int^J_{r_b}\dfrac{y(r)}{\sqrt{r^2 - J^2}}\,dr.
\ee
Notice this is a linear integral equation and is in fact a modified version of Abel's equation (see Appendix A), which has solution,
\be
\label{modAbel}
y(r) = -\dfrac{2}{\pi}\dfrac{\d}{\d r}\int^r_{r_b}\dfrac{J\L(J)}{\sqrt{J^2 - r^2}}\,\d J.
\ee
Substituting back in for $h(r)$, we have
\be
\label{exhr}
h(r) = \left(\dfrac{1}{r\pi}\dfrac{d}{dr}\int^r_{r_b}\dfrac{J\L(J)}{\sqrt{J^2 - r^2}}\,dJ\right)^2.
\ee
Now we can see that given a function's worth of information of the proper length of boundary-to-boundary spacelike geodesics, namely $\L(\D\p)$, one can recover a function's worth of information about the metric, namely $h(r)$.

We can check whether equation \eqref{exhr} is indeed valid by considering some known cases where we have an explicit expression for $S_A$.\footnote{The first and simplest case to consider is Euclidean space in 2 dimensions. A boundary-to-boundary spacelike geodesic in this case is just a chord in a circle of radius $r_b$, at an angle $\D\p$. From Pythagoras' theorem we can deduce that
\be
\L(\D\p)=2r_b\sin\frac{\D\p}{2},
\ee
Using \eqref{lpj} this becomes
\be
\L(J)=2\sqrt{r_b^2-J^2},
\ee
Now solving for $h(r)$ using \eqref{exhr},
\bea
h(r)&=&\left(\dfrac{2}{r\pi}\dfrac{d}{dr}\int^r_{r_b}\dfrac{J\sqrt{r_b^2-J^2}}{\sqrt{J^2-r_b^2}}\,dJ\right)^2\nn
&=&\left(\dfrac{2}{r\pi}\dfrac{d}{dr}\int_0^{\sqrt{r_b^2-r^2}}\sqrt{r_b^2-r^2-x^2}\,dx\right)^2\nn
&=&\left(\dfrac{2}{r\pi}\dfrac{d}{dr}(r^2-r_b^2)\dfrac{\pi}{4}\right)^2\nn
\nn
h(r)&=&1
\eea
This is the Euclidean metric in 2 dimensions, thus we have recovered the metric given the proper length.}

\begin{figure}
\centering
\includegraphics[height=3in,width=4.5in]{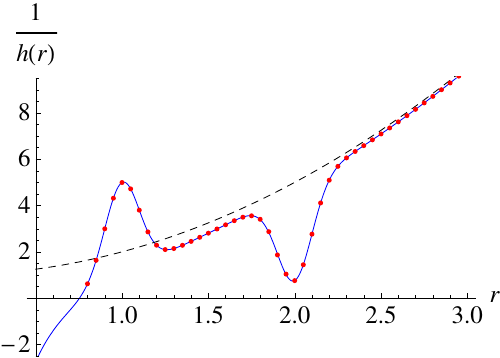}
\caption{\label{hextr}This plot illustrates the extraction, using \eqref{exhr} of a complicated metric function given by $\frac{1}{h(r)}=1+r^2-\frac{1}{r^2}+\frac{10}{\sqrt{2\pi}}(e^{-50(r-1)^2
}-e^{-50(r-2)^2})$. Notice one can only recover the metric down to the event horizon of the spacetime, defined by $h(r_+)^{-1}=0$ as zero energy spacelike geodesics cannot probe past this point (see figure \ref{BHspace}). In this particular case, $r_+\sim0.753$. The dashed line of pure AdS is given for comparison.} 
\end{figure}

Consider a subsystem $A$ defined by an arc of length $l$ on a circle of circumference $L$. The entanglement entropy of $A$ for a general 2$D$ CFT with conformal charge $c$ on this circle is given by,
\be
S_A(l)=\dfrac{c}{3}\,\ln\left(\dfrac{L}{\pi a}\sin\left(\dfrac{\pi l}{L}\right)\right).
\ee
We can translate this expression into bulk parameters using AdS/CFT. The central charge is given by\cite{3dcharge} $c=3R/2G_N^{(3)}$. Using the UV/IR connection we have from \eqref{uvir} that $L/a=2\pi r_b/R$, and the fraction $l/L$ is given in terms of the bulk angular separation $\D\p/2\pi$. Thus,
\be
S_A(\D\p)=\dfrac{R}{2G_N^{(3)}}\,\ln\left(\dfrac{2r_b}{R}\sin\left(\dfrac{\D\p}{2}\right)\right).
\ee
Using \eqref{area3d}, the proper length is given by
\be
\L(\D\p)=2R\,\ln\left(\dfrac{2r_b}{R}\sin\left(\dfrac{\D\p}{2}\right)\right),
\ee
Rewriting $\L$ as a function of $J$ using \eqref{lpj}, we have
\be
\L(J)=2R\,\ln\left(\dfrac{2r_b}{\sqrt{R^2+J^2}}\right).
\ee
Substituting this into \eqref{exhr},and integrating by parts, we have
\bea
\sqrt{h(r)}&=&\dfrac{2R}{r\pi}\dfrac{d}{dr}\Big\{-\sqrt{r_b^2-r^2}+\sqrt{R^2+r^2}\arctan\sqrt{\dfrac{r_b^2-r^2}{R^2+r^2}}\nn
&&-\sqrt{r_b^2-r^2}\,\ln\left(\dfrac{2r_b}{\sqrt{R^2+r_b^2}}\right)\Big\}
\eea
Taking the limit $r_b\tend\infty$ and differentiating, we find
\be
\sqrt{h(r)}=\dfrac{R}{\sqrt{R^2+r^2}}
\ee
This is the metric function for pure AdS space consistent with the AdS/CFT correspondence.

In a similar way to \S\ref{btbngeo}, one can also check \eqref{exhr} using various cooked up metric functions. This is verified in figure \ref{hextr} for a bumpy metric function which is singular at $r=0$ and models a perturbed AdSBH with horizon located at $r_+$. Here we see a restriction on the range of $r$ due to the existence of an event horizon, such that we can recover values for $h(r)$ over the range $r_+<r<\infty$.

\section{Extracting the Metric using Null Probes}

In \S\ref{extrhr} we have shown how one can recover the metric function $h(r)$ of \eqref{gm} in $2+1$ dimensions, given the entanglement entropy of the $(1+1)-$dimensional CFT on the boundary of the spacetime. Given $h(r)$, one can now proceed to recover the other metric function $f(r)$, with knowledge of the location of the endpoints of null geodesics through the bulk, in analogy with the method used in the previous chapter.

The location of the endpoints of null geodesics are given by the functions $\D t(\a)$, which gives the time separation, and $\D\p(\a)$, which expresses the angular separation, defined in \eqref{dt1} and \eqref{dphi1} respectively. The location of singular correlators in the CFT is again given by the boundary function $\D\p(\D t)$, so we must find a relationship between $\D\p(\D t)$ and $\a$ in order that we may use $\D\p(\a)$ as a boundary quantity. By inspection of the previous definitions of $\D\p(\a)$ and $\D t(\a)$ given in equations \eqref{dph} and \eqref{dt}, and the modified versions given in \eqref{dphi1} and \eqref{dt1}, we see that
\be
\dfrac{d\D\p(\D t)}{d\D t}=\a
\ee
must still hold in this case since factors of $h(r)$ and $f(r)$ just come along for the ride. So we can consider $\D\p(\a)$ (or alternatively $\D t(\a)$) for the remainder of the extraction procedure.

\subsection{The Method}

Since the definitions of $\D\p(\a)$ from the metric given in \eqref{sm} and \eqref{gm} are of similar form, we can proceed with the inversion in a similar way to \S\ref{solfr}.

Thus, via the substitutions $r_{min}(\a)=x,\,V(r)=V,\,V(x)=V_{max}$ and applying the condition the spacetime be asymptotically AdS, we have
\bea
\D\p\left(\sqrt{V_{max}}\right)&=&\int^{V_{max}}_{1/R^2}\dfrac{q(V)\,\d V}{\ds\sqrt{V_{max} - V}},\nn
\text{with}\quad q(V)&=&-2\dfrac{dr}{dV}\dfrac{\sqrt{Vh[r(V)]}}{r(V)}.
\eea
Now we see this is exactly the same integral equation, i.e. Abel's equation (see Appendix A), and so one can invert to find
\be
q(V)=\dfrac{1}{\pi}\dfrac{d}{dV}\int^{V}_{1/R^2}\dfrac{\D\p\left(\ds\sqrt{V_{max}}\right)\,dV_{max}}{\ds\sqrt{V-V_{max}}}.
\ee
Substituting for the original variables, we have
\bea
\label{star}
\int^{V(r)}_{1/R^2}\dfrac{I_1(V)}{\sqrt{V}}\,dV&=&\int^\infty_r\dfrac{\sqrt{h(r')}}{r'}\,dr',\nn
\text{where}\quad I_1(V)&=&\dfrac{1}{\pi}\dfrac{d}{dV}\int^{\sqrt{V}}_{1/R}\dfrac{\a\D\p(\a)\,d\a}{\sqrt{V-\a^2}}.
\eea
Thus we have an expression for $V(r)$ and therefore $f(r)$ given $\D\p(\a)$.

We can check that \eqref{star} is consistent with pure AdS space. By substituting $\D\p(\a)=\pi$ and $h(r)=(1+r^2/R^2)^{-1}$ into \eqref{star}, we have
\be
I_1[V]=\dfrac{d}{dV}\int^{\sqrt{V}}_{1/R}\dfrac{\a\,d\a}{\sqrt{V-\a^2}}=\dfrac{1}{2\sqrt{V-1/R^2}},
\ee
and so,
\be
\mathrm{ln}\left[R(\sqrt{V(r)}+\sqrt{V(r)-1/R^2})\right]=\mathrm{ln}\left[\dfrac{R(1+\sqrt{1+r^2/R^2})}{r}\right],
\ee
\be
\thus V(r)= \dfrac{1+r^2/R^2}{r^2}\thus f(r)=1+\dfrac{r^2}{R^2}\quad\text{as expected.}
\ee

\newpage

\begin{figure}
\centering
\includegraphics[width=4in]{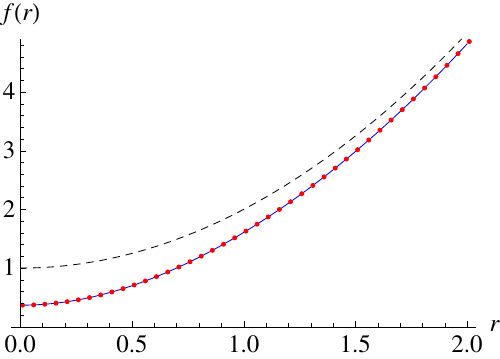}
\caption{\label{fhextr2}The figure compares the extracted data using \eqref{star} with the metric functions $f(r)=r^2+e^{-\frac{1}{1+r^2}}$ and $h(r)^{-1}=1+r^2$. The pure AdS case is given by the dashed line. As the spacetime does not admit null circular orbits, one can recover the entire metric.} 
\end{figure}

\begin{figure}[h!t]
\centering
\begin{tabular}{cc}
\begin{minipage}{200pt}
\centerline{\includegraphics[width=200pt]{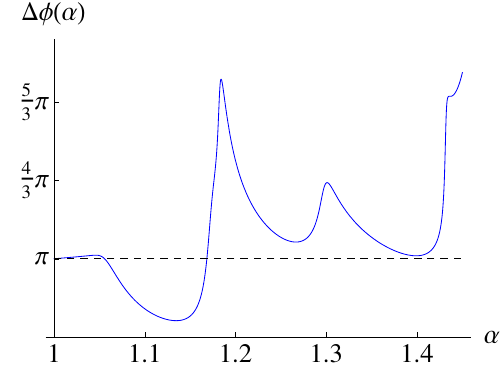}}
\end{minipage}
&
\begin{minipage}{200pt}
\centerline{\includegraphics[width=200pt]{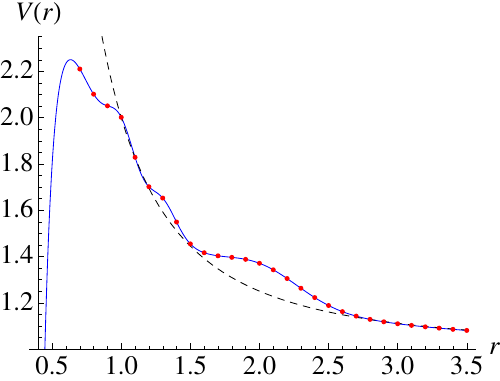}}
\end{minipage}
\end{tabular}
\caption{\label{fhextr}This figure illustrates the extraction of the metric function given in \eqref{bumpyfr} with $h(r)$ given in figure \ref{hextr}. To the left is a plot of $\D\p(\a)$ using \eqref{dphi1}. The plot to the right shows the data recovered by using \eqref{star} on $\D\p(\a)$. The pure AdS results are shown with dashed lines for comparison. We see the same limitations on the extraction as shown in figure \ref{BHbumpyextr}.}
\end{figure}

\newpage

Applying the same analysis as the previous chapter, one can also test \eqref{star} using various perturbed AdS cases. These are illustrated in figures \ref{fhextr2} and \ref{fhextr}. Again, we find the same limitation on the extracted domain of $f(r)$. For spacetimes admitting null circular orbits of radius $r_m$, the domain of $f(r)$ for which $f(r)$ can be recovered is given by $r_m<r<\infty$.

\section{Summary}

In this chapter we generalise metric \eqref{sm} by including an additional function $h(r)$. As a result, in the specific case of $2+1$ dimensions\footnote{The method in this chapter is restricted to $2+1$ dimensions via \eqref{area3d}. However, it is conceivable that there is a boundary quantity which corresponds to the proper length $\L$ in any dimension via AdS/CFT. If that is found to be the case, then this method presented in this chapter can be extended to any number of dimensions.} considered here, we require additional boundary information in the form of entanglement entropy of a CFT on a circle $S_A(\D\p)$. This, combined with information of bulk-cone singularities in the form of $\D t(\D\p)$, allows the extraction procedure to form a map $(\D t(\D\p),S_A(\D\p))\tend(f(r),h(r))$, with limits on the domains of $f(r)$ and $h(r)$ given by $(r_m,\infty)$ and $(r_+,\infty)$ respectively.

In the next chapter we will not restrict ourselves to $2+1$ dimensions. This will be achieved by calculating the area of various minimal surfaces, thus making full use of \eqref{holent}. We will find that AdS Poincar\'{e} coordinates is a simple set to analyse such an extraction.

\chapter{Recovering the Bulk III}
\label{strdsk}
\textit{This chapter is the author's original work taken from \cite{Bilson2}.} \\

\numberwithin{equation}{section}

In this chapter, we extend the work of the preceding chapter by attempting to extract metric functions in any dimension using the holographic entanglement entropy proposal \eqref{holent}, for certain static subsystems $A$ of the boundary theory. In particular, we will consider two specific forms of $A$, the straight belt of width $l$
\be
\label{infstr}
A_S=\{x_i|x_1=x\in[-l/2,l/2],x_{2,3,}\dots_{,m}\in\R\}, 
\ee
and the circular disk of radius $l$
\be
\label{cdisk}
\A_D=\{x_i|r\leq l\},\quad\text{where}\quad r=\sqrt{\sum_{i=1}^{m}x_i^2}.
\ee
Both subsystems are defined in terms of boundary planar spatial coordinates $\{x_i\}\in\R^m$, thus we should look for a class of metrics which respects this planar symmetry on the boundary. A simple class of asymptotically AdS metrics which preserves planar symmetry in the bulk and the boundary is given by
\be
\label{metric}
ds^2=R^2\left(\dfrac{-h(z)^2\,dt^2+f(z)^2\,dz^2+\sum_{i=1}^d dx_i^2}{z^2}\right).
\ee
The metric is written in AdS Poincar\'e coordinates (see \eqref{pads}), where $f(z)$ and $h(z)$ are arbitrary real functions which have the asymptotic behaviour
\be
h(z)^2,\,\dfrac{1}{f(z)^2}\tend 1+\O(z^{d+1}),\,\,\text{as}\,\,z\tend 0. 
\ee
The most well known example of such a spacetime given by \eqref{metric} is the AdS planar black hole with horizon at $z=z_+$. The metric functions are given by
\be
\label{pBH1}
h(z)^2=\dfrac{1}{f(z)^2}=1-\left(\dfrac{z}{z_+}\right)^{d+1}.
\ee
It can be recovered from the AdSBH in global coordinates (see \eqref{AdSBH}) by taking the large horizon radius limit $r_+/R\gg1$ and performing the coordinate transformation $z=R^2/r$.

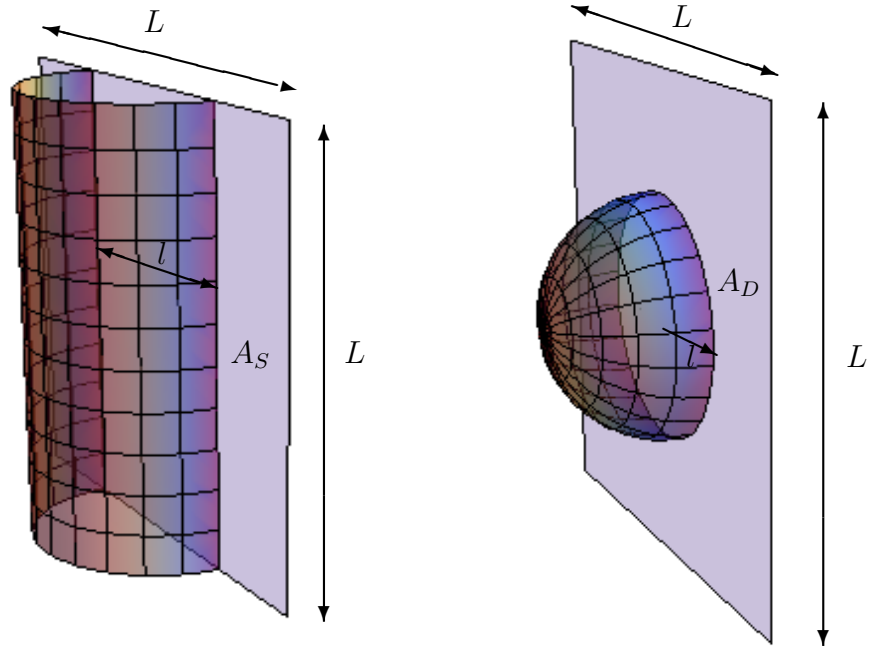
\begin{figure}
\centering
\begin{tabular}{cc}
\begin{minipage}{175pt}
\begin{overpic}[trim = 30mm 0mm 0mm 0mm, clip, width=250pt,height=250pt]{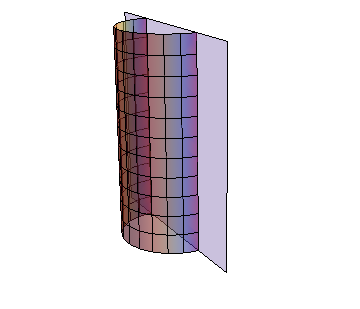}
\thicklines
\put(57,30){\color{black}\vector(0,1){56}}
\put(57,30){\color{black}\vector(0,-1){19}} 
\put(60,50){$L$}
\put(50,92){\color{black}\vector(-4,1){35}}
\put(50,92){\color{black}\vector(4,-1){3}}
\put(30,100){$L$}
\put(30,65){\color{black}\vector(-3,1){7}}
\put(30,65){\color{black}\vector(3,-1){11}}
\put(32,65){$l$}
\put(43,50){$A_S$}
\end{overpic}
\end{minipage}
&
\begin{minipage}{175pt}
\begin{overpic}[trim = 20mm 0mm 0mm 0mm, clip, width=225pt]{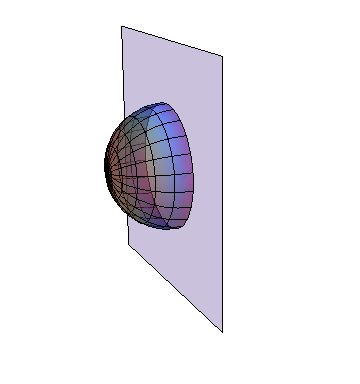}
\thicklines
\put(50,30){\color{black}\vector(0,1){55}}
\put(50,30){\color{black}\vector(0,-1){17}} 
\put(53,50){$L$}
\put(40,90){\color{black}\vector(-3,1){23}}
\put(40,90){\color{black}\vector(3,-1){4}}
\put(30,95){$L$}
\put(29,55){\color{black}\vector(2,-1){7}}
\put(32,50){$l$}
\put(36,60){$A_D$}
\end{overpic}
\end{minipage}
\end{tabular}
\caption{\label{surfplots}These plots depict the shape of static minimal surfaces $\g_A$ anchored to the straight belt $A_S$ of width $l$ and the circular disk $A_D$ of radius $l$. The subsystems $A_S$ and $A_D$ sit on the boundary, where we have regulated the boundary directions to be of length $L$.}
\end{figure}

\section{Static Minimal Surfaces in AdS}

The holographic interpretation of entanglement entropy relates the entanglement entropy $S_A$ in a $d+1$ dimensional boundary CFT to the area of an $d-$dimensional static minimal surface $\g_A$ (see figure \ref{surfplots}) whose boundary is the boundary of the subsystem $A$, i.e. $\pd\g_A=\pd A$. Since the purpose of this chapter is to extract the metric function\footnote{Since we will calculate the area of \textit{static} minimal surfaces, area expressions will not include $h(z)$. Thus one cannot recover $h(z)$ using the method outlined. Although one maybe able to extend such a method by requiring the entanglement entropy be time-dependent \cite{Hub}} $f(z)$ given $S_A$, one must first find an expression for the minimal surface $\g_A$ and thus $\Area(\g_A)$ in terms of $f(z)$. In this section we consider the case of the straight belt $A_S$ defined by \eqref{infstr}.

To calculate $\Area(\g_{A_S})$, we need to find the area of a generic surface $N$ such that $\pd N=\pd A_S$. In general, the area of an $m$ dimensional submanifold $N\subset M$ is given by
\be
\label{subarea}
\Area(N)=\int_Nd^mx\sqrt{|g_N|},
\ee
where $g_N=\det(g_{ab})$ and $g_{ab}$ is the pull-back of the metric $\tilde{g}_{\m\n}$ on manifold $M$ to $N$ such that $g_{ab}=\pd_ax^\m\pd_bx^\n\tilde{g}_{\m\n}$.

Since $\tilde{g}_{\m\n}$ is given by \eqref{metric}, one can define an embedding function $z=z(x)$ which respects the $\R^{m-1}$ planar symmetry so that $g_{ab}$ is given by
\be
ds_N^2=R^2\left(\dfrac{[z'f(z)]^2dx^2+dx^2+\sum_{i=2}^mdx_i^2}{z^2}\right),\quad\text{where}\,\,z'=\frac{dz}{dx}. 
\ee
Then using \eqref{subarea},
\be
\label{area}
\Area(N_{A_S})=\A_N(l)=R^mL^{m-1}\int_{-\frac{l}{2}}^{\frac{l}{2}}\!dx\,\dfrac{\sqrt{[z'f(z)]^2+1}}{z^m}=R^mL^{m-1}\int_{-\frac{l}{2}}^{\frac{l}{2}}\!\L(z',z,x)\,dx,
\ee
where we have regularized the directions $\{x_2,\cdots,x_m\}$ to be length $L$ (following the same procedure as \cite{Ryu}).

To find the minimal surface $\g_{A_S}$, we solve $\delta\A_N(l)=0$. Noting the the Hamiltonian
\be
\H(z,z')=z'\dfrac{d\L}{dz'}-\L=\dfrac{[z'f(z)]^2}{z^m\sqrt{[z'f(z)]^2+1}}-\dfrac{\sqrt{[z'f(z)]^2+1}}{z^m},
\ee
is independent of the variable $x$, we can write $\H$ as a constant such that
\bea
\label{ltz}
\H&=&-z_\ast^{-m},\quad\text{where}\,\,z'|_{z=z_\ast}=0\nn
\thus\dfrac{dz}{dx}&=&\dfrac{\sqrt{z_\ast^{2m}-z^{2m}}}{z^m f(z)}.
\eea
Along with the boundary conditions $z(\pm l/2)=0$, \eqref{ltz} defines the profile function $z(x)$ for the minimal surface $\g_{A_S}$.

\begin{figure}[h!t]
\centering
\begin{tabular}{cc}
\begin{minipage}{200pt}
\centerline{\includegraphics[width=200pt]{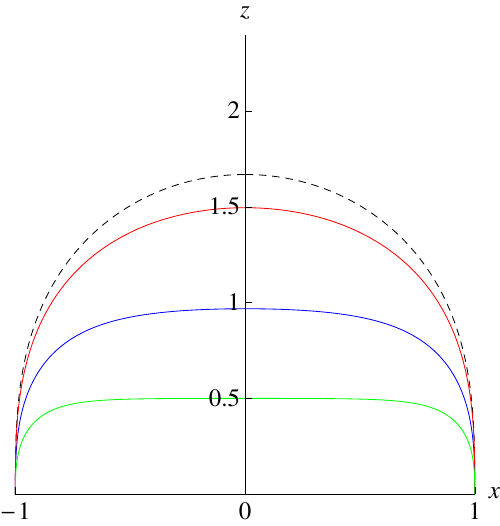}}
\end{minipage}
&
\begin{minipage}{200pt}
\centerline{\includegraphics[width=200pt]{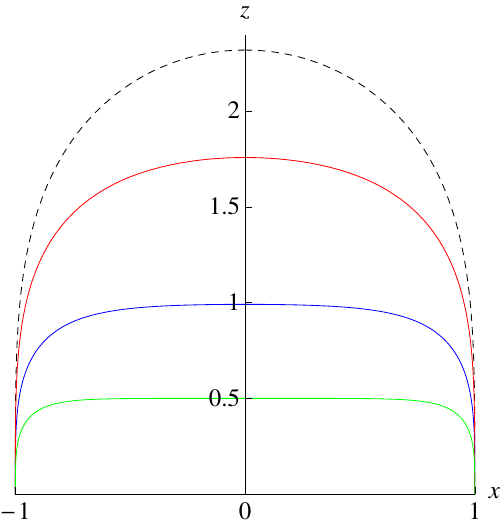}}
\end{minipage}
\end{tabular}
\caption{\label{minsurf}In this figure we plot the profile of static minimal surfaces bounded to a straight belt $A_S$ of width $l=2$ for planar black holes of dimension $4+1$. The left and right figures show minimal surfaces of spacial dimension $2$ and $3$ respectively for different values of horizon depth $z_+=\{0.5(\text{green}),1(\text{blue}),2(\text{red})\}$. The dashed lines correspond to the static minimal surfaces in pure AdS space (i.e. $z_+\tend\infty$).}
\end{figure}

\begin{figure}[!ht]
\centering
\includegraphics[height=3in,width=3in]{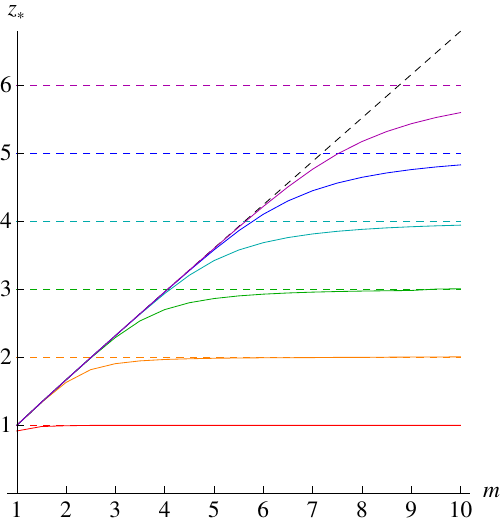}
\caption{\label{mvzplot}This is a plot of $z_\ast$ against the dimension of the minimal surface $m$ in a planar black hole spacetime of dimension $11+1$ for different depths of the horizon plane $z_+$ (shown here by the coloured dashed lines). The pure AdS plot (i.e. $z_+\tend\infty$) given in \eqref{mvz} is shown here by the black dashed line.}
\end{figure}

We illustrate the shapes of $z(x)$ for a $(4+1)-$dimensional planar black hole (see \eqref{pBH1}), with different horizon depths $z_+$, in figure \ref{minsurf}. It is noted that minimal surfaces in pure AdS probe furthest into the bulk. This is because $\g_{A_S}$ has no timelike component, so the presence of the null surface $z=z_+$ flattens the shape of $\g_{A_S}$ such that $z_\ast<z_+$. It is also observed that as one increases the dimension of $A_S$, $z_\ast$ increases. One can show this\footnote{Using $\int_0^1\!dx\,x^{\m-1}(1-x^\la)^{\n-1}=\frac{1}{\la}\frac{\G(\frac{\m}{\la})\G(\n)}{\G(\frac{\m}{\la}+\n)}$} for pure AdS by integrating \eqref{ltz} over the boundary values
\be
\label{lz}
l=2\ds\int_0^{z_\ast}\!dz\,\dfrac{z^m\,f(z)}{\sqrt{z_\ast^{2m}-z^{2m}}},
\ee
and then setting $f(z)=1$,
\bea
\label{mvz}
\za(m)&=&\dfrac{l}{2}\dfrac{\G(\frac{1}{2m})}{\sqrt{\pi}\,\G(\frac{1+m}{2m})}\nn
&\approx&\dfrac{l}{\pi}\left(m-2\g_E-2\psi^{(0)}(1/2)\right).
\eea
where $\g_E$ is the Euler–Mascheroni constant and $\psi^{(0)}$ is the Polygamma function of order zero, and so for pure AdS $z_\ast(m+1)>z_\ast(m)$. The same result is also observed for the planar black hole (see figure \ref{mvzplot}), where the lines \eqref{mvz} and $z_\ast=z_+$ asymptote the curves $z_\ast(m)$ for small and large $m$ respectively. Thus increasing the number of dimensions probes more of the bulk geometry\footnote{Intuitively, this makes sense, since probing extra dimensions costs energy, the minimal surface compensates by moving more into the bulk where distances are smaller.}, and so compared with the previous chapter where we look at a $2+1$ dimensional bulk, we will be able to recover more of the bulk geometry.

\newpage
 
\newpage

\section{The Method}

We can immediately recover an expression for $\Area(\g_{A_S})$ by substituting \eqref{ltz} into \eqref{area}
\be
\label{marea}
\Area(\g_{A_S})=\A_\g(z_\ast)=2R^mL^{m-1}\int_a^{z_\ast}\!dz\,\dfrac{z_{\ast}^m\,f(z)}{z^m\sqrt{z_\ast^{2m}-z^{2m}}},
\ee
where we have introduced a cut-off surface at $z=a$ close to the boundary to regulate the area functional in a similar way to the regularisation of the proper length in the chapter four.

Again we faced with the same problem as the previous two chapters where $\A_\g$ is a function of the bulk quantity $z_\ast$, but must be expressed in term of boundary variables only. The entanglement entropy $S_A$ is given as a function of the belt width $l$, and so one must find a map $S_A(l)\mapsto\A_\g(z_\ast)$ which is independent of the metric function $f(z)$. This can be achieved via the chain rule
\be
\label{cr}
\dfrac{dS_A(l)}{dl}=\dfrac{1}{4G_N^{(d+2)}}\dfrac{dz_\ast}{dl}\dfrac{d\A_\g(z_\ast)}{dz_\ast}\Big|_{m=d}.
\ee
Using \eqref{marea} and the Leibniz rule we have,
\be
\label{da}
\dfrac{d\A_\g}{dz_\ast}=2R^mL^{m-1}\lim_{z\rightarrow z_\ast}\left(\dfrac{f(z_\ast)}{\sqrt{z_\ast^{2m}-z^{2m}}}-m\int_0^{z}\!d\tilde{z}\,\dfrac{z^{m-1}\tilde{z}^m\,f(\tilde{z})}{(z^{2m}-\tilde{z}^{2m})^{\frac{3}{2}}}\right),
\ee
and from \eqref{lz} we have,
\be
\label{zl}
\dfrac{dl}{dz_\ast}=z_\ast^m\lim_{z\rightarrow z_\ast}\left(\dfrac{f(z_\ast)}{\sqrt{z_\ast^{2m}-z^{2m}}}-m\int_0^{z}\!d\tilde{z}\,\dfrac{z^{m-1}\tilde{z}^m\,f(\tilde{z})}{(z^{2m}-\tilde{z}^{2m})^{\frac{3}{2}}}\right).
\ee
Combining these equations we find that
\be
\label{stoa}
\dfrac{dS_A}{dl}=\dfrac{R^dL^{d-1}}{2z^d_\ast G_N^{(d+2)}}.
\ee
Thus, one can map $l\mapsto z_\ast$ and so $S_A(l)\mapsto\A_\g(z_\ast)$ independently of the metric and we are justified in using \eqref{marea} for the extraction of $f(z)$.

By rewriting \eqref{marea} in the more digestible form, where $\tilde{\A}_\g(z_\ast)=\dfrac{\A_\g(z_\ast)}{2R^dL^{d-1}z_\ast^d}$, $p(z)=\dfrac{f(z)}{z^d}$ and $g(z)=z^{2d}$, we have
\be
\label{genAbel}
\tilde{\A}_\g(z_\ast)=\int_a^{z_\ast}\!dz\,\dfrac{p(z)}{\sqrt{g(z_\ast)-g(z)}}.
\ee
The solution to the integral equation of the form $f(x)=\int_a^x\!\frac{y(t)dt}{\sqrt{g(x)-g(t)}}$, where $\frac{dg}{dx}=g'(x)>0$, is given by \cite{Abel} (see Appendix A),
\be
y(x)=\dfrac{1}{\pi}\dfrac{d}{dx}\int_a^x\!dt\,\dfrac{f(t)g'(t)}{\sqrt{g(x)-g(t)}}.
\ee
One can apply this solution to our particular case, and substituting back in for our original variables, we find,
\be
\label{invstr}
f(z)=\dfrac{d\,z^d}{\pi R^dL^{d-1}}\dfrac{d}{dz}\int_a^z\!dz_\ast\,\dfrac{\A_\g(z_\ast)\,z_\ast^{d-1}}{\sqrt{z^{2d}-z_\ast^{2d}}}.
\ee
Thus, using \eqref{stoa} and \eqref{invstr}, we have found a map $S_A(l)\mapsto f(z)$ which is independent of $f(z)$ and so determines the metric function uniquely in the limit $a/z\tend0$.

\section{Checking the Inversion}

We can check the validity of \eqref{invstr} by looking at a simple case where both the entanglement entropy and metric are known, i.e. pure AdS.

Using the result from \cite{Ryu}, the entanglement entropy of an $d$ dimensional straight belt of width $l$ defined by \eqref{infstr} is given by\footnote{The $d+2$ dimensional bulk Newton constant $G_N^{(d+2)}$ can be related to boundary CFT parameters via $AdS_{d+2}/CFT_{d+1}$ for a particular choice of boundary theory. For example, when $d=2$, $AdS_4\times S^7$ space in eleven dimensional supergravity is considered to be dual to $2+1$ dimensional $\N=8$ $SU(N)$ SCFT. In this case $\frac{G_N^{(4)}}{R^2}=\frac{3}{2\sqrt{2}}N^{-3/2}$.} 
\be
\label{Ssb}
S_A(l)=\dfrac{2R^d}{4G_N^{(d+2)}(d-1)}\left(\dfrac{L}{a}\right)^{d-1}-\dfrac{2^{d+1}\pi^{d/2}R^d}{4G_N^{(d+2)}(d-1)}\left(\dfrac{\G(\frac{d+1}{2d})}{\G(\frac{1}{2d})}\right)^d\left(\dfrac{L}{l}\right)^{d-1}.
\ee
Using \eqref{stoa}, we find that
\be
\label{lsb}
\dfrac{l}{2}=\za\sqrt{\pi}\dfrac{\G(\frac{d+1}{2d})}{\G(\frac{1}{2d})}.
\ee
Note that this is exactly the result given in \eqref{mvz}, only in this case we have not used the fact that $f(z)=1$.
By combining \eqref{Ssb} and \eqref{lsb} with \eqref{holent}, we see that
\be
\label{asb}
\A_S(z_\ast)=\dfrac{2R^d}{d-1}\left(\dfrac{L}{a}\right)^{d-1}-\dfrac{2\sqrt{\pi}R^d}{d-1}\dfrac{\G(\frac{d+1}{2d})}{\G(\frac{1}{2d})}\left(\dfrac{L}{z_\ast}\right)^{d-1}.
\ee
Substituting \eqref{asb} into \eqref{invstr}, we have 
\be
f(z)=\dfrac{2d}{\pi(d-1)}\left(\dfrac{z}{a}\right)^{d-1}I_1(z)-\dfrac{2d\sqrt{\pi}}{\pi(d-1)}\dfrac{\G(\frac{d+1}{2d})}{\G(\frac{1}{2d})}I_2(z),
\ee
where
\bea
I_1(z)&=&z\dfrac{d}{dz}\int^z_adz\dfrac{z^{d-1}_{\ast}}{\sqrt{z^{2d}-z^{2d}_\ast}}\nn
&=&z\dfrac{d}{dz}\int^1_{a/z}dx\dfrac{x^{d-1}}{\sqrt{1-x^{2d}}}\nn
&=&\left(\dfrac{a}{z}\right)^d\left(1+\dfrac{1}{2}\left(\dfrac{a}{z}\right)^{2d}+\O\left(\dfrac{a}{z}\right)^{4d}\right),
\eea
and
\bea
I_2(z)&=&z^d\dfrac{d}{dz}\int^z_a\dfrac{dz_\ast}{\sqrt{z^{2d}-z^{2d}_\ast}}\nn
&=&z^d\dfrac{d}{dz}\dfrac{1}{z^{d-1}}\int^1_{a/z}\dfrac{dx}{\sqrt{1-x^{2d}}}\nn
&=&z^d\dfrac{d}{dz}\dfrac{1}{z^{d-1}}\int^1_0\dfrac{dx}{\sqrt{1-x^{2d}}}-z^m\dfrac{d}{dz}\dfrac{1}{z^{d-1}}\int^{a/z}_0\dfrac{dx}{\sqrt{1-x^{2d}}}\nn
&=&-(d-1)\sqrt{\pi}\dfrac{\G(\frac{2d+1}{2d})}{\G(\frac{d+1}{2d})}+\left(\dfrac{a}{z}\right)\dfrac{1}{\sqrt{1-(a/z)^{2d}}}\nn
&&+(d-1)\int^{a/z}_0\dfrac{dx}{\sqrt{1-x^{2d}}}\nn
&=&-(d-1)\sqrt{\pi}\dfrac{\G(\frac{2d+1}{2d})}{\G(\frac{d+1}{2d})}+d\left(\dfrac{a}{z}\right)\left(1+\dfrac{3}{4d+2}\left(\dfrac{a}{z}\right)^{2d}+\O\left(\dfrac{a}{z}\right)^{4d}\right).
\eea
Combining the results, we have
\be
f(z)=1+\dfrac{2d}{\pi(d-1)}\left(\dfrac{a}{z}\right)\left(1-d\sqrt{\pi}\dfrac{\G(\frac{d+1}{2d})}{\G(\frac{1}{2d})}+\O\left(\dfrac{a}{z}\right)^{2d}\right).
\ee
Taking the limit $a/z\rightarrow 0$, we get the required result for pure AdS.

\section{The Series Solution}

One can consider examples of area functions for which one can recover analytic expressions for the metric function $f(z)$ using \eqref{invstr}. Take the infinite series describing a $d-$dimensional surface about the known AdS result \eqref{asb},
\be
\label{pertA}
\A_S(z_\ast)=2R^d\left(\dfrac{L}{z_\ast}\right)^{d-1}\sum_{n=0}^\infty b_n(d)z^n_\ast,
\ee
where $b_0(d)=-\frac{\sqrt{\pi}}{d-1}\frac{\G(\frac{d+1}{2d})}{\G(\frac{1}{2d})}$ and $b_{d-1}(d)=\frac{1}{d-1}\frac{1}{a^{d-1}}$. 

By applying \eqref{invstr}, we have
\bea
f(z)&=&\dfrac{2dz^d}{\pi}\dfrac{d}{dz}\int^z_adz_\ast\dfrac{\sum_{n=0}^\infty b_n(d)z^n_\ast}{\sqrt{z^{2d}-z^{2d}_\ast}}\nn
&=&\dfrac{2dz^d}{\pi}\sum_{n=0}^\infty b_n(d)\dfrac{d}{dz}z^{n+1-d}\int^1_0dx\dfrac{x^n}{\sqrt{1-x^{2d}}}+\O(a/z)\nn
&=&1+\dfrac{1}{\sqrt{\pi}}\sum^\infty_{n=1}b_n(d)(n+1-d)\dfrac{\G(\frac{n+1}{2d})}{\G(\frac{n+d+1}{2d})}z^n\nn
&=&1+\sum^\infty_{n=1}\tilde{b}_nz^n,
\eea
where $\tilde{b}_n=(n+1-d)\frac{\G(\frac{n+1}{2d})}{\G(\frac{n+d+1}{2d})}b_n(d)$.

Using this result, one can recover the planar black hole in $d+2$ dimensions with horizon radius $z_+$. The metric function in this case is given by
\bea
f(z)&=&\dfrac{1}{\sqrt{1-(z/z_+)^{d+1}}}\nn
&=&1+\sum^\infty_{n=1}\dfrac{\G(n+1/2)}{2n!\sqrt{\pi}}\left(\dfrac{z}{z_+}\right)^{n(d+1)}.
\eea
This series is necessarily convergent as $0<z\leq\za<z_+$ (see figure \ref{mvzplot}), and can be recovered using ansatz \eqref{pertA} with
\be
b_{n>0}(d)=
\left\{
	\begin{array}{ll}
		\dfrac{1}{2\sqrt{\pi}(n+1-d)}\dfrac{\G(\frac{2n+1}{2})}{\G(n+1)}\dfrac{\G(\frac{n+d+1}{2d})}{\G(\frac{n+1}{2d})}\dfrac{1}{z_+^n} & \mbox{if } \frac{n}{d+1}\in\Z^+\\
		0 & \mbox{otherwise }
	\end{array}
\right.
\ee

\newpage

\section{Extracting the Bulk Metric for a Circular Disk}

Now we have a method for determining the metric function $f(z)$ in \eqref{metric} in the case of a straight belt $A_S$, we can look at other types of shapes where the calculation of the entanglement entropy in the CFT is relatively simple. In particular we consider the circular disk $A_D$ defined by \eqref{cdisk}.

We can rewrite \eqref{metric} in polar coordinates, to respect the symmetry of $A_D$, as
\be
ds^2=R^2\left(\dfrac{-h(z)^2\,dt^2+f(z)^2\,dz^2+dr^2+r^2d\Om^2_{d-1}}{z^2}\right),
\ee
where $r$ is the radial coordinate on the boundary.

Following the same procedure as the straight belt, but using the embedding $z=z(r)$, the area of a general $m$ dimensional static surface $N$, such that $\pd N=\pd A_D$, is given by
\bea
\label{aread}
\Area(N)=\A_N(l)&=&R^m\,\text{Vol}(S^{m-1})\int^l_0dr\,r^{m-1}\dfrac{\sqrt{1+(z'f(z))^2}}{z^m}\nn
&=&R^m\,\text{Vol}(S^{m-1})\int^l_0\!\L(z',z,r)\,dr
\eea
Compared to the straight belt, we have $r$ dependence in the Lagrangian and so the Hamiltonian is not constant. Thus using the Euler-Lagrange equation
\be
\dfrac{d}{dr}\dfrac{d\L}{dz'}=\dfrac{d\L}{dz}
\ee
we have
\bea
\label{eomd}
rf(z)^2zz''&+&(m-1)f(z)^4z(z')^3+mrf(z)^2(z')^2\nn
&+&rf(z)f'(z)z(z')^2+(m-1)f(z)^2zz'+mr=0,
\eea
Solving for $z(r)$ with the boundary conditions $z(l)=0$ and $z'(0)=0$ give us the minimal surface $\g_{A_D}$. For pure AdS where $f(z)=1$, the solution to \eqref{eomd} is given by, $z^2+r^2=l^2$. Thus the static minimal surface $\g_{A_D}$ anchored to the boundary of $A_D$, in pure AdS space, is a semi-circle of radius $l$ for $m=1$ and a hemisphere of radius $l$ for $m>2$.

\subsection{A Perturbative Approach}

A general solution to \eqref{eomd} is intractable, but we can attempt to find $z(r)$ perturbatively, solving order by order in a parameter $\e$ about the known AdS solution.

To proceed, we will find the perturbation equations simplify if we rewrite \eqref{eomd} in the coordinate $v=r^2$, giving
\bea
\label{eomd2}
2f(z)^2[mz'(z+2vz')+2vzz'']&+&8(m-1)vf(z)^4z(z')^3\nn
&+&4vf(z)zf'(z)(z')^2+m=0
\eea
where $z'=\frac{dz}{dv}$ etc.

Now we can ansatz a solution for $z(v)$ in orders of $\e$, which will eventually be set to 1 in a similar way to the method outlined in \S\ref{pertmeth},
\be
\label{antz}
f(z)=1+\sum_{i=1}^\infty\e^ia_iz^{i(d+1)}\quad\text{and}\quad z(v)=\sqrt{b-v}+\sum_{i=1}^\infty\e^iz_{(i)}(v),
\ee
where $v=r^2$ and $b=l^2$.

Substituting ansatz \eqref{antz} into \eqref{eomd2}, the $\O(\e^n)$ equation is given by
\be
\label{peom}
A(v)\,z''_{(n)}(v)+B(v)\,z'_{(n)}(v)+C(v)\,z_{(n)}(v)=D_{(n-1)}(v),
\ee
where
\be
A(v)=4v\sqrt{b-v},\,\,B(v)=\dfrac{2(bm-3v)}{\sqrt{b-v}},\,\,C(v)=-\dfrac{bm}{(b-v)^{3/2}},
\ee
and $D_{(n-1)}(v)$ is determined from the $(n-1)^\text{th}$ solution.

The homogeneous equation has a non-trivial particular solution, 
\be
z^{(1)}(v)=(b-v)^{-1/2}.
\ee
From this one can construct the second solution\cite{Diff}
\bea
z^{(2)}(v)&=&z^{(1)}(v)\int\dfrac{W(v)}{(z^{(1)}(v))^2}\,dv,\quad\text{where}\quad W(v)=e^{-\int\frac{B(v)}{A(v)}\,dv}\nn
&=&\dfrac{1}{\sqrt{b-v}}\int dv\dfrac{(b-v)^{\frac{5-m}{2}}}{v^{m/2}}.
\eea
The general solution of \eqref{peom} is then given by,
\bea
\label{pmin}
z_{(n)}(v)=C_1z^{(1)}(v)&+&C_2z^{(2)}(v)+z^{(2)}(v)\int\!dv\dfrac{D_{(n-1)}(v)\,z^{(1)}(v)}{A(v)\,W(v)}\nn
&-&z^{(1)}(v)\int\!dv\dfrac{D_{(n-1)}(v)\,z^{(2)}(v)}{A(v)\,W(v)},
\eea
where $C_1,C_2$ are determined by the boundary conditions $z_{(n)}(b)=0$ and $\lim_{v\tend0}(\sqrt{v}z_{(n)}'(v))=0$. We see immediately that as $v\tend0$, $\sqrt{v}z'^{(2)}(v)\sim\O(v^{-\frac{m-1}{2}})$, and so we must set $C_2=0$. Expression \eqref{pmin} then becomes
\bea
\label{pmin2}
z_{(n)}(v)&=&\dfrac{C_1}{\sqrt{b-v}}+\sqrt{b-v}\left(\int dv\dfrac{(b-v)^{\frac{m-1}{2}}}{v^{m/2}}\right)\int\!dv\,D_{(n-1)}(v)\dfrac{v^{\frac{m-2}{2}}}{4(b-v)^{\frac{m+1}{2}}}\nn
&&-\sqrt{b-v}\int\!dv\left(D_{(n-1)}(v)\dfrac{v^{\frac{m-2}{2}}}{4(b-v)^{\frac{m+1}{2}}}\int dv\dfrac{(b-v)^{\frac{m-1}{2}}}{v^{m/2}}\right).
\eea
The constant $C_1$ is determined by the condition that $z_{(n)}(v)$ is non-singular at the boundary $v=b$.
 
\begin{figure}[h!t]
\centering
\begin{tabular}{cc}
\begin{minipage}{200pt}
\centerline{\includegraphics[width=200pt]{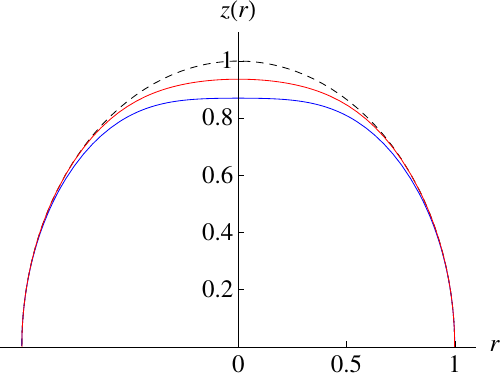}}
\end{minipage}
&
\begin{minipage}{200pt}
\centerline{\includegraphics[width=200pt]{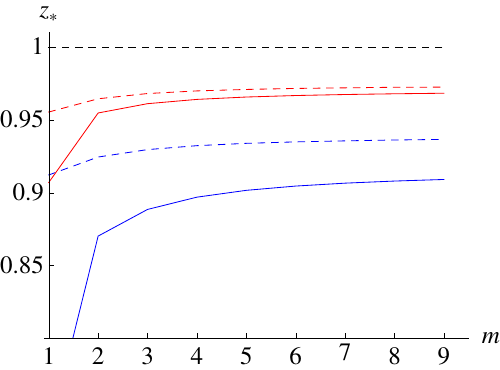}}
\end{minipage}
\end{tabular}
\caption{\label{minsurfdisk}The left diagram plots profile functions $z(r)$ anchored to the the circular disk $A_D$ of radius $l=1$ and dimension $m$, for the metric function $f(z)=(1-z^{10})^{-1/2}$. This spacetime describes a planar black hole in $10+1$ dimensions of horizon depth $z_+=1$. The dashed, red and blue curves show pure AdS, the numerical solution and the first order perturbed solution to \eqref{eomd} respectively for $m=2$. The $r$ axis is the boundary of the spacetime. The right diagram shows how the maximum height $\za$ changes for these curves at $z_+=1,10/9$ as one increases the dimension of the circular disk up to the maximum allowed by the dimension of the spacetime.}
\end{figure}

The integrals is \eqref{pmin2} can be solved analytically at first order given the dimension of the surface $m$ and $D_{(0)}=\frac{2bm-3v}{\sqrt{b-v}}a_1$. In figure \ref{minsurfdisk} we compare the shape of a 2-dimensional minimal surface $\g_{A_D}$ using the perturbative method to first order with the numerical solution for a particular planar black hole. We also compare the numerical and first order solutions for increasing $m$. We note the same result as for the straight belt that $\za(m+1)>\za(m)$, where $z_\ast=z_{(1)}(0)$ is the maximum distance the minimal surface probes the bulk. We also find the expected pattern that the perturbative solution approaches the numerical solution as it nears the pure AdS hemisphere.  

In theory, we have an expression for the minimal surface to any order in the perturbation, we can simply plug \eqref{pmin} and \eqref{antz} into \eqref{aread} and equate orders of $\e$ to determine the perturbed minimal surface area $\Area(\g_{A_D})=\A_\g(l)$. Computing $\A_\g(l)$ at first order in $\e$ by substituting \eqref{antz} into \eqref{aread} with $r=l\sqrt{1-y^2}$, we have
\bea
\A_\g(l)&=&R^m\,\text{Vol}(S^{m-1})\int^1_{a/l}dy\dfrac{(1-y^2)^{\frac{m-2}{2}}}{y^m}\nn
&&+R^m\,\text{Vol}(S^{m-1})\dfrac{\e a_1}{l}\dfrac{d}{2}\int^1_{a/l}dy\dfrac{(1-y^2)^{\frac{m-2}{2}}}{y^{m+2}}z_{(1)}(y)\nn
&&-R^m\,\text{Vol}(S^{m-1})\dfrac{\e a_1}{l}\int^1_{a/l}dy\dfrac{(1-y^2)^{m/2}}{y^m}z'_{(1)}(y)
\eea
One can now relate the expressions for $\A_\g(l)$ and $f(z)$ through the parameter $a_1$ once $\e$ is set to 1. 

\section{Summary}
In this section we extend the ideas and methods of metric extraction explored in previous chapters to higher dimensions using the holographic entanglement entropy proposal \eqref{holent}. We attempt this in two distinct types of subsystem of the boundary CFT where expressions for the entanglement entropy are known (see \cite{Ryu}). These are the straight belt $A_S$ and the circular disk $A_D$ defined by \eqref{infstr} and \eqref{cdisk} respectively. In particular we find expressions for the area of static minimal surfaces $\g_{A_S}$ and $\g_{\A_D}$, such that $\pd\g_A=\pd A$, in terms of the metric function $f(z)$ given in AdS Poincar\'e coordinates \eqref{metric}. This is achieved explicitly in the case of the straight belt, where we exploited the boundary planar symmetry of the area Hamiltonian. A perturbative method is outlined in the case of the circular disk where such symmetry was not manifest. In both cases, motivation was provided for the statement that as one increases the dimension of the minimal surfaces, more of the bulk is recovered. Combined with the fact that we don't live in $2+1$ dimensions, we can justify extending the analysis of chapter 4 to spacetime dimensions greater than 2+1.

\chapter{Discussion}
\label{discussion}
\textit{This chapter is an overview of the material written in this thesis in the author's own words.}\\

In this final chapter, we review what has been achieved in this thesis and provide an overall picture of the results we have presented and their limitations. We can then consider how one can generalise such results to less symmetric cases and potential future areas of interest.

In chapter one, we provided qualitative explanations of why one should study string theory and the AdS/CFT correspondence. Through the holographic nature of AdS/CFT, one can ask questions about how quantities in the bulk AdS space relate to quantities in the boundary CFT and visa versa. More specifically, in the large $\la$ limit of AdS/CFT, as defined in \S\ref{limit}, we investigated CFT correlators and entanglement entropy on the boundary as probes of the bulk geometry. The hope is that one might uncover physics from behind the horizon of black holes from the physics of the boundary. These ideas were made more quantitative in chapter two. It was motivated in \S\ref{probulk} that a boundary correlation function $G(\vec{x},\vec{x}')$ is singular if there exists null geodesics connecting $x$ and $x'$. If the null geodesics travel through the bulk, then $G(\vec{x},\vec{x}')$ has a ``bulk-cone singularity'' \cite{hubeny}. This suggests that if one has knowledge of the locus of its singular boundary correlator operator insertion points, one can begin to determine the geometry of the dual bulk spacetime using boundary-to-boundary null probes. 

In \S\ref{ent2}, we illustrated the holographic interpretation of the entanglement entropy $S_A$ for arbitrary submanifolds $A$ in the boundary CFT. As claimed in \cite{Ryu,Ryu1}, $S_A$ in the $d+1$ dimensional CFT can be determined from the $d$ dimensional static minimal surface $\g_A$ anchored to the boundary such that $\pd\g_A=\pd A$. $S_A$ is then determined by the area of $\g_A$ through \eqref{holent}. This suggests another way one can recover the geometry of the dual bulk description from boundary data. In this case, knowledge of the entanglement entropy for certain subsystems of the boundary CFT can help determine the geometry of the dual bulk spacetime using minimal surface probes. In the specific case of a $2+1$ dimensional bulk, these probes turn out to be zero energy boundary-to-boundary spacelike geodesics.

It was the two holographic relationships illustrated above that allowed us to ask this question which would concern us most in the chapters to follow:

\textit{Given a $d$-dimensional boundary CFT, where one has knowledge of the locus of its singular boundary correlator operator insertion points, and knowledge of the entanglement entropy for certain subsystems of the boundary CFT, how much of the bulk geometry can be determined for the $d+1$-dimensional asymptotically AdS spacetime dual to this CFT?}

In chapter 3, we set out to answer this question for static, $(d+2)-$dimensional, spherically symmetric, asymptotically AdS spacetimes of the form
\be
ds^2 = -f(r)\,dt^2 + \dfrac{dr^2}{f(r)} +r^2\,d\Om^{2}_d.
\ee
Determining the geometry of this spacetime boiled down to finding the metric function $f(r)$. Since we only require the extraction of one function's worth of information, it was observed that one only requires a function's worth of information on the boundary. Utilising the spherical symmetry of the problem, this boundary information took the form of the locus of endpoints of boundary-to-boundary null geodesics. This is encapsulated in the function $\D t(\D\p)$, where $\D t$ and $\D\p$ are the time and angular separation of the endpoints on the boundary respectively. This can then be related to the bulk-cone singularities of the boundary correlators. We found that one can determine the quantities $\D t$ and $\D\p$ in terms of the bulk null geodesic parameter $\a=\frac{E}{J}$ from the function $\D t(\D\p)$ without requiring an expression for the metric function $f(r)$ through equation \eqref{exfr2}. This allowed us to use the expression \eqref{dt} relating $\D\p(\a)$ to $f(r)$ for determining $f(r)$. Via the observation that equation \eqref{dt} was an integral equation with known solution\footnote{It is interesting to observe that all the extractions performed in this Thesis boil down to solving linear Volterra integral equations of the first kind of the form $f(x)=\int_a^x\!\frac{y(t)dt}{\sqrt{g(x)-g(t)}}$ where $g(t)$ and $f(x)$ are known. It would be interesting to investigate if there were some deeper relationship involved in these extractions due to this observation. As mentioned in \cite{hammerphd}, there is a duality between boundary data of null and zero-energy spacelike geodesics. It is interesting to consider if this duality explains the similar extraction methods used here.}, we were able to extract the metric function $f(r)$ in terms of the boundary data $\D\p(\a)$. Unlike the numerical methods presented in \cite{hammer1}, where an analysis of the stability of numerical errors is required, the analytical extraction presented in chapter three provides no errors.

The range of values of $r$, for which one could determine $f(r)$, depended on the region of the bulk the full set of boundary-to-boundary geodesics could cover. Thus for the class of spacetimes not admitting null circular orbits, one can recover the entire metric. However, given a radius of null circular orbits $r=r_m$, such as spacetimes containing spacelike singularities, one can only recover $f(r)$ over the range $r\in(r_m,\infty)$. This is consistent with the work presented in \cite{hammer1}. A unique feature of the analytical extraction is the ability to find analytical expressions for the metric function $f(r)$. This is the purpose of \S\ref{sersol}, where we were able to find a series solution for $f(r)$ by expanding about the pure AdS solution $f(r)=1+\frac{r^2}{R^2}$. Again, we found a limitation on the domain of $f$ for which $f(r)$ can be extracted. In this case, it is the radius of convergence $r_c$ for the series $f(r)$ that provides the lower bound. This is calculated for a toy model of a ``star'' geometry and the AdS-Schwarzschild black hole.

In chapter four, we generalise the analysis to static, $(d+2)-$dimensional, spherically symmetric, asymptotically AdS spacetimes of the form
\be
ds^2 = -g(r)\,dt^2 +h(r)\,dr^2+r^2\,d\Om^{2}_d.
\ee
Since one now requires the extraction of two function's worth of information $g$ and $h$, the boundary function $\D t(\D\p)$ did not provide us with enough information to extract the metric. Thus one must look for a different set of bulk probes. This was provided by the set of boundary-to-boundary zero energy spacelike geodesics. The boundary data was encapsulated by the proper length of these spacelike boundary geodesics as a function of the angular separation of their endpoints $\L(\D\p)$. In the specific case of a $(2+1)-$dimensional bulk, this quantity is related to the entanglement entropy $S_A(l)$ where $A$ is an interval of length $l$ in the dual boundary $(1+1)-$dimensional CFT (see equation \eqref{area3d}). It was shown that one can find an expression (see \eqref{exhr}) for the metric function $h(r)$ in $(2+1)$ dimensions\footnote{However, as pointed out in \cite{Hub1}, geometrical objects such as geodesics or extremal surfaces which are anchored on the boundary, typically correspond to some probe or observable in the dual CFT. Thus it is conceivable that the quantity $\L(\D\p)$ may given in terms of just such an observable, allowing the extraction method outlined in \S\ref{extrhr} to be generalised to any dimension.}, via knowledge of the function $S_A(l)$. 

As in chapter 3, there were limitations on the extractable domain of the metric function $h$, in this case corresponding to the existence of a horizon in the bulk at $r=r_+$. Past this point, zero-energy spacelike geodesics become purely timelike, and so are unable to probe the spatial region inside the horizon (see figure \ref{BHspace}). Thus for static spacetimes admitting a horizon, $h$ can only be extracted in the domain $r\in(r_+,\infty)$.

One can ask the natural question, how can one probe beyond the horizon in these spacetimes? The answer is negative if the spacetime is static with a single asymptotic region, even if we were to consider finite-energy spacelike probes. However, if one were to promote the metric to be time-dependent, the radius of the event horizon can become time-dependent, and so it is possible to probe the geometry before the horizon radius becomes large. We can still find the boundary observables dual to this probe as the holographic entanglement entropy proposal admits a natural covariant extension\cite{Hub}. 

One can see this in Vaidya-AdS geometries of the form
\be
ds^2=-\left(1+r^2-\dfrac{m(v)}{r^{d-2}}\right)\,dv^2+2\,dv\,dr+r^2d\Om_{d-1}^2.
\ee
where $m(v)$ interpolates between $0$ (pure AdS) and $1$ (Schwarzschild-AdS). In these spacetimes, the horizon radius grows with time, thus the set of zero energy spacelike probes cover the region $r\in(r_+(v),\infty)$, and so it may be possible to extract the function $m(v)$ in $(2+1)-$dimensions given the entanglement entropy of a time-dependent interval of length $l$.

Returning to the analysis in chapter four, once we showed how one can extract $h(r)$, we then considered the extraction of the metric function $g(r)$. Using the locus of bulk-cone singularities given by $\D t(\D\p)$, the extraction proceeded in the same manner to chapter three. It was shown that, given $\D t(\D\p)$ and $h(r)$, an analytical expression for $g(r)$ could be recovered. In a similar way to chapter three, this extraction did not depend on the dimension of the spacetime, but provided the same restriction on the domain of the extracted $g$, namely $r\in(r_m,\infty)$ where $r_m$ is the radius of null circular orbits, or $r\in[0,\infty)$ otherwise. 

Could one extend such methods to less symmetric cases? The answer is yes, in principle. Consider promoting the functions $g$ and $h$ to time-dependent functions such that $g(r)=h(r)\tend k(r,t)$. We still have two functions worth of information in the metric, and so knowledge of the boundary functions $\D t(\D\p)$ and $S_A(l)$ should still provide enough information to extract the metric function $k(r,t)$. Using the same logic, one could also consider spacetimes with reduced spherical symmetry. In this case, the metric functions would depend on the angular coordinate $g(r)=h(r)\tend k(r,\p)$. However, the geodesic equations of motion are not as simple, and so finding analytic expressions for $\D t$, $\D\p$ and $\L$ in terms of $k$ prove troublesome.

In chapter five, we considered another set of bulk probes anchored to the boundary, namely multi-dimensional minimal surfaces. As discussed in \cite{Hub1,Hubun}, higher-dimensional surfaces probe deeper for spacetimes satisfying energy conditions, so it makes sense to study such probes. We considered static, $(d+2)-$dimensional planar symmetric, asymptotically AdS spacetimes of the form\footnote{We use the tilde notation to distinguish from the metric functions described earlier in this chapter.}
\be
\label{plmet}
ds^2=R^2\left(\dfrac{-\tilde{h}(z)^2\,dt^2+\tilde{f}(z)^2\,dz^2+d\Sigma^2_d}{z^2}\right).  
\ee
We pick this particular form of the metric since this is the natural generalisation of the pure AdS metric ($\tilde{h}(z)=\tilde{f}(z)=1$) used in \cite{Ryu,Ryu1}, where analytical expressions for the entanglement entropy were found. Since the minimal surfaces must be anchored to some shape on the boundary, we picked two cases. The straight belt given defined in \eqref{infstr} where $d\Sigma^2_d=\sum_{i=1}^d dx_i^2$, and the circular disk defined in \eqref{cdisk} where $d\Sigma^2_d=dr^2+r^2d\Om^2_{d-1}$.

We first considered the straight belt $A_S$. Comparing minimal surfaces of different dimension at fixed width on the boundary, we found for pure AdS, and planar black hole geometries, that higher dimensional surfaces probe further into the bulk, thus agreeing with \cite{Hub1}. We then found an expression for the area of these minimal surfaces $\A_{\g}(z_\ast)$ in terms of the metric function $\tilde{f}(z)$. We showed that the function $\A_{\g}(z_\ast)$ can be determined from the entanglement entropy $S_{A_S}(l)$ of the straight belt $A_S$ of width $l$ in the boundary dual CFT without prior knowledge of the bulk geometry. Via the observation that equation \eqref{marea} was an integral equation with known solution \eqref{invstr}, we were able to extract the metric function $\tilde{f}(z)$, with knowledge of the entanglement entropy $S_{A_S}(l)$ of the straight belt $A_S$ of width $l$. This was confirmed for the known pure AdS result. We were also able to find a series solution for $\tilde{f}(z)$, where one was able to recover the planar black hole.

One may ask the question whether one can extract $\tilde{h}(z)$? The answer is negative if one considers only static surfaces, since the area functional does not depend on the timelike component of the metric. However, if we make the straight belt time-dependent, then using the covariant holographic entanglement entropy proposal \cite{Hub}, one might be able to determine $\tilde{h}(z)$ using extremal surfaces. 

But what about bulk-cone singularities? As we saw in chapter 4, using boundary data of the locus of bulk-cone singularities helped us determine the timelike component of the metric. Could this be applied to metrics with planar symmetry as given in \eqref{plmet}? Unfortunately, the answer is again negative. As pointed out by the authors of \cite{hubeny}, field theories formulated on $\R^{d,1}$, and states respecting the full Poincar\'e symmetry have no bulk-cone singularities. This is because null geodesics emanating from the boundary have no turning point in bulk coordinate $z$, due to the energy condition on the stress tensor, thus cannot return to the boundary.

We then turned our attention to the circular disk shape $A_D$. One finds in this case that the minimal surface equations are not as simple, since the area density (see \eqref{aread}) depends on the boundary coordinate $r$. As such, one is forced to consider a perturbative approach\footnote{A Hamilton-Jacobi method was also attempted. In this case, the Hamilton-Jacobi equation becomes
\be
\left(\dfrac{\pd S}{\pd r}\right)^2+\dfrac{1}{\tilde{f}(z)^2}\left(\dfrac{\pd S}{\pd z}\right)^2=\dfrac{r^{2m-2}}{z^{2m}},
\ee
where $S=S(z,r,C)$ is Hamilton's principle function. If this can be solved for $S$, one can determine $z'(r)$ via the equation $\frac{\pd\L}{\pd z'}=\frac{\pd S}{\pd z}$, where $\L=\L(z,z',r)$ is given in \eqref{area}.}. This was achieved by perturbing about the known pure AdS solution where the minimal surface is a $d-$dimensional hemisphere of radius $l$, given in terms of the profile function $z(r)$ as $z^2+r^2=l^2$. Fortunately, the minimal surface equations reduce to non-linear second order, inhomogeneous differential equations with a known particular solution \eqref{peom}. This allowed us to construct a general solution at each order in the perturbation, and at first order, find an analytical expression for the minimal surface profile function $z(r)$ in any dimension. One could then compare such a solution to the solution obtained by numerically solving the equations of motion. This was illustrated in Figure \ref{minsurfdisk}, where it was observed that as the numerical solution approaches the pure AdS solution, the first order perturbative solution becomes more accurate. Obtaining higher order terms to the profile function proved intractable since the integrals of \eqref{pmin} cannot be solved analytically, and a series solution to the integrals ran into convergence issues.

How could one extend the methods presented in chapter 5 to other cases? The most natural extension is to consider other shapes on the boundary where expressions for the entanglement entropy dual to pure AdS are known. A known example\footnote{See \cite{Ryu3} for a holographic calculation of the entanglement entropy in pure AdS} is the cusp $A_W$ in a $(2+1)-$dimensional CFT. In polar boundary coordinates $(r,\th)$, a cusp of angle $\Om$ is defined in terms of its boundary
\be
\pd A_W=\{(r,\th)|0\leq r<\infty,\th=0\}\cup\{(r,\th)|0\leq r<\infty,\th=\Om\}.
\ee
One could then attempt to solve for the area of the minimal surface anchored to $\pd A_W$ in terms of the metric function $\tilde{f}(z)$. However, one would have to find a suitable regulator for the area of the cusp in the bulk.

Overall, we have attempted to answer the question posed at the start of this chapter, using analytical methods, for 3 classes of highly symmetric spacetimes. However, the ultimate goal of the techniques used in this thesis would be to answer the more encompassing question: \textit{From CFT boundary data, can one determine the holographic dual bulk spacetime in a fully covariant manner?}. Here we have presented the first steps in answering such a question.

\appendix
\chapter{Integral Equations}
\numberwithin{equation}{chapter}
In this section we check the consistency of the integral equations used in this thesis. We do this by considering the most general form of integral equation
\be
\label{intf}
f(x)=\int_a^x\!\frac{y(t)dt}{\sqrt{g(x)-g(t)}},
\ee
with solution,
\be
\label{inty}
y(x)=\dfrac{1}{\pi}\dfrac{d}{dx}\int_a^x\!dt\,\dfrac{f(t)g'(t)}{\sqrt{g(x)-g(t)}},
\ee
where $g(x)$ is a strictly monotonic function of $x$.

We can check the solution by substituting in the expression \eqref{intf} for $f(t)$ into \eqref{inty}, giving us
\be
y(x)=\dfrac{1}{\pi}\dfrac{d}{dx}\int_a^x\left(\int_a^t\dfrac{y(u)}{\sqrt{g(t)-g(u)}}\,du\right)\dfrac{g'(t)}{\sqrt{g(x)-g(t)}}\,dt.
\ee
Changing the order of integration gives us,
\bea
\label{intI}
y(x)&=&-\dfrac{1}{\pi}\dfrac{d}{dx}\int_a^xy(u)\left(\int_x^u\frac{g'(t)}{\sqrt{(g(t)-g(u))(g(x)-g(t))}}\,dt\right)du\nn
&\equiv&\dfrac{1}{\pi}\dfrac{d}{dx}\int_a^xy(u)I(u,x)\,du.
\eea
By letting $v=g(t)$, and integrating over $v$, we have
\bea
I(u,x)&=&-\int_{g(x)}^{g(u)}\dfrac{dv}{\sqrt{-v^2+(g(u)+g(x))v-g(x)g(u)}}\nn
&=&\left[\arcsin\left(\dfrac{g(u)+g(x)-2v}{g(u)-g(x)}\right)\right]^{g(u)}_{g(x)}\nn
&=&\arcsin(-1)-\arcsin(1)\nn
&=&3\pi/2-\pi/2\nn
&=&\pi.
\eea
Substituting this result into the RHS of \eqref{intI}, we recover the LHS, and so \eqref{inty} is indeed a solution to \eqref{intf}.

To recover Abel's integral equation \eqref{abelint} used in chapters 3 and 4, we take $g(x)=x$. To recover the modified integral equation \eqref{modAbel} used in chapter 4, we take $g(x)=x^2$. Finally, to recover the integral equation \eqref{genAbel} used in chapter 5, we take $g(x)=x^{2d}$.
\newpage


\bibliographystyle{utphys}
\bibliography{Thesis}

\end{document}